\documentclass[prd,twocolumn,showpacs,superscriptaddress,nofootinbib,floatfix,showkeys,10pt]{revtex4-2}
\usepackage{bm}
\usepackage{times}
\usepackage{braket}
\usepackage{amsfonts,amssymb,stmaryrd,latexsym,amsmath}
\usepackage[usenames,dvipsnames]{color}
\usepackage{epsfig}
\usepackage{slashed}
\usepackage{hyperref}
\usepackage{subfigure}
\usepackage{graphics}
\usepackage{orcidlink}
\usepackage{multirow}
\usepackage{dashrule}
\allowdisplaybreaks[1]

\definecolor{nicered}{rgb}{0.7,0.1,0.1}
\definecolor{nicegreen}{rgb}{0.1,0.5,0.1}
\definecolor{emph}{rgb}{1,0,0}
\definecolor{doub}{rgb}{0.7,0.2,1.0}
\definecolor{navyblue}{RGB}{0, 110, 184}
\hypersetup{colorlinks,citecolor=nicegreen,linkcolor=nicered,urlcolor=navyblue}

\newcommand{\eq}{Eq.~\eqref}
\newcommand{\Tcc}{{T_{cc}(3875)}}
\newcommand{\Tcs}{{T_{cs0}(2870)}}

\newcommand{\clabel}[2][]{#2}

\begin{document}


\title{Existence of the $DD^*\bar{K}^*$ and $BB^*K^*$ three-body molecular states}

\author{Yan-Ke Chen \orcidlink{0000-0002-9984-163X}}
\affiliation{School of Physics and Center of High Energy Physics, Peking University, Beijing 100871, China}

\author{Lu Meng \orcidlink{0000-0001-9791-7138}}\email{lmeng@seu.edu.cn}
\affiliation{School of Physics, Southeast University, Nanjing 211189, China}

\author{Jun-Zhang Wang \orcidlink{0000-0002-3404-8569}}\email{wangjzh@cqu.edu.cn}
\affiliation{Department of Physics and Chongqing Key Laboratory for Strongly Coupled Physics, Chongqing University, Chongqing 401331, China}

\author{Shi-Lin Zhu\,\orcidlink{0000-0002-4055-6906}}\email{zhusl@pku.edu.cn}
\affiliation{School of Physics and Center of High Energy Physics, Peking University, Beijing 100871, China}

\begin{abstract}
We investigate the existence of the three-body molecular state composed of $DD^*\bar{K}^*$ within the one-boson-exchange model. A major challenge is that while the pseudoscalar-meson couplings are well determined, the couplings for scalar- and vector-meson exchanges render significant model dependence. To ensure the reliability of our predictions and reduce model dependence, we recalibrate the coupling constants of the one-boson-exchange model. We treat the pole position of $Z_c(3900)$, or equivalently the scalar $\sigma$-exchange coupling constant, as the only unknown parameter. The coupling constants for the vector $\rho$ and $\omega$ exchanges are determined by the pole positions of the well-established states $X(3872)$ and $\Tcc$. We demonstrate that these parameter sets also successfully describe the $\Tcs$ without further tuning. For the three-body system, our results indicate that an $I\left(J^P\right)=1 / 2\left(0^{-}\right)$ three-body molecular bound state exists when $Z_c(3900)$ is a virtual state located within approximately $-10~\text{MeV}$ of the $D\bar{D}^*$ threshold. Furthermore, we extend our analysis to the complex energy plane using the complex scaling method to search for molecular resonances, though no evidence of resonances is found in considered channels. We also apply this formalism to the bottom analog  $BB^*K^*$ system. In this sector, the conditions for the existence of a three-body bound state are more relaxed, as a $Z_c(3900)$ virtual state located within $-25~\text{MeV}$ below the threshold suffices, although three-body molecular resonances remain absent. We suggest that future experiments precisely measure the pole position of $Z_c(3900)$ or search for the three-body bound state in $DD\bar{K}\pi\pi$ and $DD\bar{K}$ channels, as these efforts would mutually illuminate the nature of the associated states.
\end{abstract}

\maketitle

\section{Introduction} 
\label{sec:introduction}

Over the past two decades, the experimental discovery of numerous exotic hadrons that cannot be simply described as quark-antiquark mesons or three-quark baryons within the conventional quark model has revitalized the field. These discoveries have sparked significant advancements in hadron spectroscopy and provide an excellent platform for probing the non-perturbative dynamics of quantum chromodynamics (QCD), as reviewed in Refs.~\cite{Chen:2016qju,Esposito:2016noz,Lebed:2016hpi,Ali:2017jda,Guo:2017jvc,Olsen:2017bmm,Karliner:2017qhf,Liu:2019zoy,Brambilla:2019esw,Chen:2022asf,Meng:2022ozq,Wu:2022ftm}. Several of these exotic states, such as the $X(3872)$ (also referred to as $\chi_{c1}(3872)$)~\cite{Belle:2003nnu}, $Z_c(3900)$~\cite{BESIII:2013ris,Belle:2013yex}, $T_{cs0}(2870)$~\cite{LHCb:2020bls,LHCb:2020pxc,LHCb:2024vfz}, and $T_{cc}(3875)$~\cite{LHCb:2021vvq,LHCb:2021auc} have attracted particular attention due to their unique properties. As illustrated in Table~\ref{tab:molecule_example}, these states are located close to $S$-wave meson-meson thresholds; therefore it is natural to interpret them as hadronic molecular states of the corresponding meson pairs.

The concept of a hadronic molecular state composed of two color-singlet hadrons is not unique to the heavy-flavor sector. Indeed, the deuteron stands as the archetype of a two-body hadronic molecular state. Just as atomic nuclei with various mass numbers, such as$~^3\mathrm{H}$ and$~^3\mathrm{He}$, exist in nature, anticipating the existence of heavy-flavor molecular states composed of three or more hadrons is logical. Theoretical studies of three-body hadronic molecules have been conducted on various systems, including three light mesons~\cite{Shen:2022etd,Zhang:2021hcl}, $D^{(*)}D^{(*)}D^{(*)}$~\cite{Wu:2021kbu,Luo:2021ggs,Bayar:2022bnc,Pan:2022whr,Ortega:2024ecy,Fu:2025joa}, $D^{(*)}D^{(*)}\bar{D}^{(*)}$~\cite{Tan:2024omp,Valderrama:2018sap}, $D^{(*)}B^{(*)}\bar{B}^{(*)}$~\cite{Dias:2018iuy}, $BBB^*$~\cite{Ma:2018vhp}, and systems where kaons replace one or more heavy mesons~\cite{Ma:2017ery,Ren:2018pcd,Ren:2018qhr,Wu:2019vsy,Di:2019jsx,Ikeno:2022jbb,Zhang:2024yfj,Zhai:2024luy,Ren:2024mjh,Pan:2025xvq}. Furthermore, three-body universality in the charmed and bottom sectors was investigated in Refs.~\cite{Canham:2009zq, Lin:2017dbo}. For a comprehensive review of three-body heavy-flavor systems, we refer the reader to Refs.~\cite{MartinezTorres:2020hus,Liu:2024uxn}.

A system of particular interest that remains relatively unexplored is the $DD^*\bar{K}^*$. In this configuration, the $DD^*$ subsystem can form the molecular state $\Tcc$ , and the $T_{cs0}(2870)$ was proposed as the $D^*\bar{K}^*$ molecular state ~\cite{Molina:2010tx,Chen:2020aos,He:2020btl,Liu:2020nil,Hu:2020mxp,Agaev:2020nrc,Wang:2021lwy,Ortega:2023azl,Wang:2023hpp}. Consequently, determining the existence of a three-body molecular state in this system is a compelling question.

In this work, we focus on the $DD^*\bar{K}^*$ three-body system and its bottom analog, the $BB^*K^*$ system. We describe the interactions between hadrons using the one-boson-exchange (OBE) model, which incorporates the exchange of mesons such as $\pi$, $\eta$, $\rho$, $\omega$, and $\sigma$. This model has achieved significant success in elucidating heavy-flavor hadronic molecules~\cite{Tornqvist:1991ks,Tornqvist:1993ng,Gamermann:2006nm,Liu:2008fh,Liu:2008xz,Thomas:2008ja,Liu:2009qhy,Ding:2009vj,Lee:2009hy,Sun:2011uh,Li:2012cs,Li:2012ss,Chen:2017vai,Liu:2019stu,Chen:2020yvq, Chen:2021vhg,Dong:2021bvy, Dong:2021juy,Lin:2022wmj,Cheng:2022qcm,Peng:2023lfw,Lin:2024qcq,Wang:2024ukc,Wang:2025kpm,Cheng:2026cgo}. Notably, the existence of the $DD^*$ bound state was predicted theoretically within the OBE model prior to the experimental discovery of the $\Tcc$~\cite{Li:2012cs,Li:2012ss}. Recently, the OBE model has been extended to investigate $P$-wave $DD^*$, $D\bar{D}^*/\bar{D}D^*$~\cite{Lin:2024qcq,Cheng:2026cgo}, $D^*\bar{K}^*$~\cite{Wang:2024ukc}, and $B^{(*)}\bar{B}^{(*)}$~\cite{Wang:2025kpm} systems. 

A potential issue within the OBE model is that while the pseudoscalar-meson couplings are well determined, the couplings for scalar- and vector-meson exchanges introduce significant model dependence. For example, the $\sigma$-exchange potential in Refs.~\cite{Liu:2019stu,Wu:2021kbu} is nearly 20 times stronger than that in Refs.~\cite{Li:2012cs,Li:2012ss}. To address the model dependence and make theoretical predictions more reliable, we adopt the approach from Ref.~\cite{Zhu:2024hgm} to recalibrate the model parameters. The key strategy involves utilizing the $Z_c(3900)$ state, where the $\rho$ and $\omega$ exchanges interactions are expected to largely cancel, to constrain the strength of the scalar $\sigma$-exchange interaction. The pole position of the $Z_c(3900)$, which is equivalent to the scalar $\sigma$-exchange coupling constant, is treated as the only varying parameter. Then the coupling constants for the vector $\rho$- and $\omega$-exchanges are fixed by the pole positions of the well-established states $X(3872)$ and $T_{cc}(3875)$. Based on heavy quark symmetry, we apply the same model parameters from the $DD^*\bar{K}^*$ system to the $BB^*K^*$ system.

Regarding the solution of the three-body problem, we employ the Gaussian expansion method (GEM)~\cite{Hiyama:2003cu} to solve the Schr\"odinger equation. Furthermore, we utilize the complex scaling method (CSM)~\cite{Aguilar:1971ve,Balslev:1971vb,Moiseyev:1998gjp,Aoyama:2006hrz} to search for three-body molecular resonances, as this technique allows bound states and resonances to be treated on an equal footing via square-integrable basis expansion.

The paper is organized as follows. In Sec.~\ref{sec:formalism}, we introduce the formalism. We construct the effective Lagrangians and derive the OBE effective potentials in Sec.~\ref{sub:model}. In Sec.~\ref{sub:methodology}, we present the methodology for studying the three-body systems, specifically the GEM and CSM. We discuss the numerical results in Sec.~\ref{sec:results_and_discussions}. Finally, a brief summary is provided in Sec.~\ref{sec:summary}. 

\begin{table*}[htbp]
\centering
\caption{Examples for hadronic molecular states. The experimental data are taken from Ref.~\cite{ParticleDataGroup:2024cfk}. $\Delta E$ represents the mass difference between the molecular state and the threshold of the corresponding meson pairs. The symbol $\star$ indicates that the $I(J^P)$ quantum numbers have not been determined experimentally, and the listed values represent theoretically favored assignments.}
\label{tab:molecule_example}
\setlength{\tabcolsep}{3.5mm}
\begin{tabular}{cccccc}
\hline 
\hline
State & $I^G\left(J^{P C}\right)$ & $M[\text{MeV}]$ & $\Gamma[\text{MeV}]$ & $S$-wave threshold & $\Delta E[\text{MeV}]$ \tabularnewline
\hline \multirow[t]{3}{*}{$X(3872)$} & \multirow[t]{3}{*}{$0\left(1^{++}\right)$} & \multirow[t]{3}{*}{$3871.64 \pm 0.06$} & \multirow[t]{3}{*}{$1.19 \pm 0.21$} & $D^{*+} D^{-}+\mathrm{c.c}$ & $-8.1 \pm 0.4$ \tabularnewline
& & & & $D^{* 0} \bar{D}^0+\mathrm{c.c}$ & $-0.05 \pm 0.09$ \tabularnewline
& & & & $\left[D \bar{D}^* / \bar{D} D^*\right]^{C=+1}$ & $-4.09 \pm 0.21$ \tabularnewline
\hline
\multirow[t]{3}{*}{$Z_{c}(3900)$} & $1^{+}\left(1^{+-}\right)$ & $3887.1 \pm 2.6$ & $28.4 \pm 2.6$ & $D^{*+} D^{-}-\mathrm{c.c}$ . & $7.3 \pm 2.6$ \tabularnewline
& & & & $D^{* 0} \bar{D}^0-\mathrm{c.c}$ & $15.4 \pm 2.6$ \tabularnewline
& & & & $\left[D \bar{D}^* / \bar{D} D^*\right]^{C=-1}$ & $11.4 \pm 2.6$ \tabularnewline
\hline
\multirow[t]{3}{*}{$T_{c c}(3875)$} & \multirow[t]{3}{*}{$0\left(1^{+}\right)^{\star}$} & \multirow[t]{3}{*}{$3874.74 \pm 0.10$} & \multirow[t]{3}{*}{$0.048_{-0.0141}^{+0.0020}$} & $D^{* 0} D^{+}$ & $-1.6 \pm 0.4$ \tabularnewline
 & & & & $D^{*+} D^0$ & $-0.36 \pm 0.12$ \tabularnewline
& & & & $D^* D$ & $-0.99 \pm 0.23$ \tabularnewline
\hline
$T_{c s 0}(2870)$ & $0\left(0^{+}\right)^{\star}$ & $2872 \pm 16$ & $67 \pm 24$ & $D^* \bar{K}^*$ & $-30 \pm 17$ \tabularnewline
\hline\hline
\end{tabular}
\end{table*}

\section{formalism} 
\label{sec:formalism}
\subsection{Effective Lagrangians and potentials} 
\label{sub:model}

For the near-threshold hadronic molecular states considered in this work, the typical center-of-mass momenta are significantly smaller than the constituent masses of the charm and strange quarks. Therefore, it is reasonable to treat these heavy quarks as spectators and employ heavy meson effective Lagrangians to describe the interactions between hadrons. The Lagrangians in the OBE model are constructed based on heavy quark symmetry, chiral symmetry, and SU(2) isospin symmetry~\cite{Georgi:1990cx,Mannel:1990vg,Falk:1991nq,Wise:1992hn,Yan:1992gz,Casalbuoni:1996pg,Liu:2008fh,Liu:2009qhy,Li:2012cs,Li:2012ss,Wang:2024ukc,Zhu:2024hgm}:
\begin{align}
\mathcal{L}_{D^{(*)} D^{(*)} \sigma} = & -2 g_s D_b^{\dagger} D_b \sigma+2 g_s D_b^* \cdot D_b^{* \dagger} \sigma,\\
\mathcal{L}_{D^{(*)} D^{(*)} P} = & +\frac{2 g_a}{f_\pi}\left(D_b D_{a \lambda}^{* \dagger}+D_{b \lambda}^* D_a^{\dagger}\right) \partial^\lambda P_{b a} \nonumber \\
& + i \frac{2 g_a}{f_\pi} v^\alpha \varepsilon_{\alpha \mu \nu \lambda} D_b^{* \mu} D_a^{* \lambda \dagger} \partial^\nu P_{b a},\\
\mathcal{L}_{D^{(*)} D^{(*)} V}= & +\sqrt{2} \beta g_V D_b D_a^{\dagger} v \cdot V_{b a}-2 \sqrt{2} \lambda g_V \nonumber\\
& \times \epsilon_{\lambda \mu \alpha \beta} v^\lambda\left(D_b D_a^{* \mu \dagger}+D_b^{* \mu} D_a^{\dagger}\right)\left(\partial^\alpha V_{b a}^\beta\right) \nonumber\\
& -\sqrt{2} \beta g_V D_b^* \cdot D_a^{* \dagger} v \cdot V_{b a} \nonumber \\
& -i 2 \sqrt{2} \lambda g_V D_b^{* \mu} D_a^{* \nu \dagger}\left(\partial_\mu V_\nu-\partial_\nu V_\mu\right)_{b a},\\
\mathcal{L}_{K^{(*)} K^{(*)} \sigma} = & -2 g_s^{\prime} K_b^{\dagger} K_b \sigma+2 g_s^{\prime} K_b^* \cdot K_b^{* \dagger} \sigma,\\
\mathcal{L}_{K^{(*)} K^{(*)} P}= & +\frac{2 g_a^{\prime}}{f_\pi}\left(K_b K_{a \lambda}^{* \dagger}+K_{b \lambda}^* K_a^{\dagger}\right) \partial^\lambda P_{b a} \nonumber \\
& +i \frac{2 g_a^{\prime}}{f_\pi} v^{\prime \alpha} \varepsilon_{\alpha \mu \nu \lambda} K_b^{* \mu} K_a^{* \lambda \dagger} \partial^\nu P_{b a},\\
\mathcal{L}_{K^{(*)} K^{(*) V}}= & +\sqrt{2} \beta^{\prime} g_V^{\prime} K_b K_a^{\dagger} v^{\prime} \cdot V_{b a}-2 \sqrt{2} \lambda^{\prime} g_V^{\prime} \nonumber \\
& \times \epsilon_{\lambda \mu \alpha \beta} v^{\prime \lambda}\left(K_b K_a^{* \mu \dagger}+K_b^{* \mu} K_a^{\dagger}\right)\left(\partial^\alpha V_{b a}^\beta\right) \nonumber \\
& -\sqrt{2} \beta^{\prime} g_V^{\prime} K_b^* \cdot K_a^{* \dagger} v^{\prime} \cdot V_{b a} \nonumber \\
& -i 2 \sqrt{2} \lambda^{\prime} g_V^{\prime} K_b^{* \mu} K_a^{* \nu \dagger}\left(\partial_\mu V_\nu-\partial_\nu V_\mu\right)_{b a},
\end{align}
where $g_a^{(\prime)}$, $g_s^{(\prime)}$, $g_V^{(\prime)}$, $\beta^{(\prime)}$, and $\lambda^{(\prime)}$ are coupling constants. $f_\pi$ is the pion decay constant. $v=(1,0,0,0)$ is the velocity of the heavy meson. The matrices $P$ and $V$ represent the pseudoscalar and vector meson fields, respectively:
\begin{align}
&P=\left(\begin{array}{cc}
\frac{\pi^0}{\sqrt{2}}+\frac{\eta}{\sqrt{6}} & \pi^{+} \\
\pi^{-} & -\frac{\pi^0}{\sqrt{2}}+\frac{\eta}{\sqrt{6}}
\end{array}\right),\\
&V=\left(\begin{array}{cc}
\frac{\rho^0}{\sqrt{2}}+\frac{\omega}{\sqrt{2}} & \rho^{+} \\
\rho^{-} & -\frac{\rho^0}{\sqrt{2}}+\frac{\omega}{\sqrt{2}}.
\end{array}\right).
\end{align}
Here, we group the isotriplet $\pi/\rho$ and the isosinglet $\eta/\omega$ into the same matrix according to SU(3) flavor symmetry to reduce the number of coupling constants.

\begin{figure}
    \centering
    \includegraphics[width=0.48\textwidth]{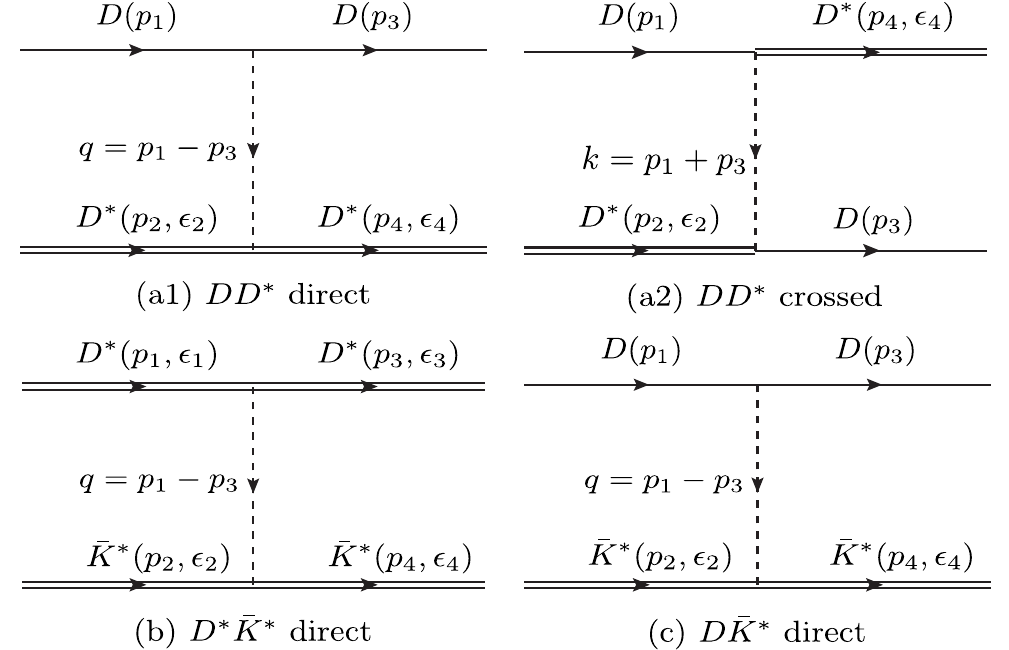}
    \caption{Feynman diagrams to derive OBE potentials: (a1) and (a2) for $DD^*$, (b) for $D^*\bar{K}^*$, and (c) for $D\bar{K}^*$.}
    \label{fig:scattering_fm_diagram}
\end{figure}

Based on the OBE Lagrangians, the effective potentials are derived from the scattering Feynman diagrams shown in Fig.~\ref{fig:scattering_fm_diagram}, including contributions from both direct and crossed diagrams. The effective potentials for the $D\left(p_1\right) D^*\left(p_2,\epsilon_2\right) \rightarrow D\left(p_3\right) D^*\left(p_4,\epsilon_4\right)$ are:
\begin{align}
	V^{DD^*}_\sigma=&-g_s^2\frac{1}{\bm{q}^2+u_{\sigma}^2}\mathcal{O}_{\sigma}^D, \label{eq:V_D_Dast_sigma}\\
	V^{DD^*}_{\pi/\eta}=& + \frac{g_a^2}{f_\pi^2}\frac{(\bm{k}\cdot \bm{\epsilon}_4^{\dag})(\bm{k}\cdot \bm{\epsilon}_2)}{\bm{k}^2+u_{\pi/\eta}^2}\mathcal{O}_{\pi/\eta}^C,\\
	V^{DD^*}_{\rho/\omega}= & + \frac{\beta^2g_v^2}{2}\frac{(\bm{\epsilon}_2\cdot \bm{\epsilon}_4^{\dag})}{\bm{q}^2+u_{\rho/\omega}^2}\mathcal{O}_{\rho/\omega}^D \nonumber \\
	&+2\lambda^2g_v^2\frac{\bm{k}^2(\bm{\epsilon}_2\cdot \bm{\epsilon}_4^{\dag})-(\bm{k}\cdot \bm{\epsilon}_4^{\dag})(\bm{k}\cdot \bm{\epsilon}_2)}{\bm{k}^2+u_{\rho/\omega}^2}\mathcal{O}_{\rho/\omega}^C.
\end{align}
The effective potentials for the $D^*\left(p_1,\epsilon_1\right) \bar K^*\left(p_2,\epsilon_2\right) \rightarrow D^*\left(p_3,\epsilon_3\right) \bar K^*\left(p_4,\epsilon_4\right)$ are:
\begin{align}
\mathcal{V}^{D^*\bar K^{*}}_{\sigma} = & -g_sg_{s}^{\prime}\frac{(\bm{\epsilon}_1\cdot \bm{\epsilon}_3^{\dag})(\bm{\epsilon}_2\cdot \bm{\epsilon}_4^{\dag})}{\bm{q}^2+u_{\sigma}^2}\mathcal{O}_{\sigma}^D,\\
\mathcal{V}^{D^*\bar K^{*}}_{\pi/\eta}= & + \frac{g_ag_a^{\prime}}{f_{\pi}^2}\frac{((\bm{\epsilon}_1\times \bm{\epsilon}_3^{ \dag})\cdot\bm{q})((\bm{\epsilon}_2\times \bm{\epsilon}_4^{ \dag})\cdot\bm{q})}{\bm{q}^2+u_{\pi/\eta}^2}\mathcal{O}_{\pi/\eta}^D,\\
\mathcal{V}^{D^*\bar K^{*}}_{\rho/\omega} = & \{ + \frac{\beta\beta^{\prime}g_Vg_V^{\prime}}{2}\frac{(\bm{\epsilon}_1\cdot \bm{\epsilon}_3^{\dag})(\bm{\epsilon}_2\cdot \bm{\epsilon}_4^{\dag})}{\bm{q}^2+u_{\rho/\omega}^2} \nonumber\\
&~+\frac{2\lambda\lambda^{\prime}g_Vg_V^{\prime}}{\bm{q}^2+u_{\rho/\omega}^2} [ (\bm{\epsilon}_1\cdot \bm{\epsilon}_2)(\bm{\epsilon}_3^{\dag}\cdot \bm{q})(\bm{\epsilon}_4^{\dag}\cdot \bm{q}) \nonumber \\
& ~~~~~-(\bm{\epsilon}_1\cdot \bm{\epsilon}_4^{\dag})(\bm{\epsilon}_3^{\dag}\cdot \bm{q})(\bm{\epsilon}_2\cdot \bm{q}) \nonumber \\
& ~~~~~-(\bm{\epsilon}_2\cdot \bm{\epsilon}_3^{\dag})(\bm{\epsilon}_4^{\dag}\cdot \bm{q})(\bm{\epsilon}_1\cdot \bm{q}) \nonumber \\
& ~~~~~ +(\bm{\epsilon}_3^{\dag} \cdot \bm{\epsilon}_4^{\dag})(\bm{\epsilon}_1\cdot \bm{q})(\bm{\epsilon}_2\cdot \bm{q}) ]\}\mathcal{O}_{\rho/\omega}^D.
\end{align}
The effective potentials for the $D\left(p_1\right) \bar K^*\left(p_2,\epsilon_2\right) \rightarrow D\left(p_3\right) \bar K^*\left(p_4,\epsilon_4\right)$ are
\begin{align}
\mathcal{V}^{D \bar K^{*}}_{\sigma} = & -g_sg_{s}^{\prime}\frac{1}{\bm{q}^2+u_{\sigma}^2}\mathcal{O}_{\sigma}^D, \\
\mathcal{V}^{D \bar K^{*}}_{\rho/\omega} = & +\frac{\beta\beta^{\prime}g_Vg_V^{\prime}}{2}\frac{\bm{\epsilon}_2\cdot \bm{\epsilon}_4^{ \dag}}{\bm{q}^2+u_{\rho/\omega}^2}\mathcal{O}_{\rho/\omega}^D. \label{eq:V_D_Kast_vector}
\end{align}
The effective potentials for the hidden-charm $D\bar{D}^*/\bar{D}D^*$ systems can be found in Ref.~\cite{Zhu:2024hgm}. The isospin operators $\mathcal{O}$ are listed in Table~\ref{tab:isospin_operator}. In the direct diagram, the momentum of the propagator is $q=p_1-p_3$, whereas in the crossed diagram, it is $k=p_1+p_3$. As discussed in Refs.~\cite{Zhu:2024hgm,Lin:2024qcq}, this distinction significantly affects systems with odd partial waves. $u_{\mathrm{exc}}$ represents the effective mass of the exchanged meson, defined as
\begin{align}
	\frac{-1}{k_{0}^{2} - \bm{k}^{2} - m_{\mathrm{exc}}^{2}} = \frac{1}{\bm{k}^{2} + (m_{\mathrm{exc}}^{2} - k_{0}^{2})} = \frac{1}{\bm{k}^{2} + u_{\mathrm{exc}}^{2}},\\
	\frac{-1}{q_{0}^{2} - \bm{q}^{2} - m_{\mathrm{exc}}^{2}} = \frac{1}{\bm{q}^{2} + (m_{\mathrm{exc}}^{2} - q_{0}^{2})} = \frac{1}{\bm{q}^{2} + u_{\mathrm{exc}}^{2}}.
\end{align}
In the static approximation, the energy transfers $k_0$ and $q_0$ are the mass differences between the initial and final states. Notably, we have $q_0=0$ and $k_{DD^*,0}^2\approx m_\pi^2$. Therefore, we set $u=0$ for the crossed pion exchange in the $DD^*$ system and $u=m_{\mathrm{exc}}$ for all other cases.

\begin{table*}
\centering
\caption{The isospin operators $\mathcal{O}_{\mathrm{exc}}$ in Eqs.~\eqref{eq:V_D_Dast_sigma}-\eqref{eq:V_D_Kast_vector}. $\mathcal{O}^D$ and $\mathcal{O}^C$ represent operators for the direct and crossed diagram, respectively. $\mathbb{I}$ is the $2\times2$ identity matrix and $\bm{\tau}=(\tau_1,\tau_2,\tau_3)$ denotes the SU(2) isospin Pauli matrix acting on the heavy meson fields.}
\label{tab:isospin_operator}
\begin{tabular*}{\hsize}{@{}@{\extracolsep{\fill}}cccccc@{}}
\hline 
\hline
Meson & $\sigma$ & $\pi$ & $\eta$ & $\rho$ & $\omega$ \tabularnewline
\hline
$\mathcal{O}^D$ & $\mathbb{I}$ & $\frac{\bm{\tau}_1 \cdot \bm{\tau}_2}{2}$ & $\frac{1}{6}\mathbb{I}$ & $\frac{\boldsymbol{\tau}_1 \cdot \boldsymbol{\tau}_2}{2}$ & $\frac{1}{2}\mathbb{I}$ \tabularnewline
$\mathcal{O}^C$ & & $-\frac{1}{4}\left(3 \mathbb{I}-\boldsymbol{\tau}_1 \cdot \boldsymbol{\tau}_2\right)$ & $-\frac{1}{12}\left(\mathbb{I}+\boldsymbol{\tau}_1 \cdot \boldsymbol{\tau}_2\right)$ & $-\frac{1}{4}\left(3 \mathbb{I}-\boldsymbol{\tau}_1 \cdot \boldsymbol{\tau}_2\right)$ & $-\frac{1}{4}\left(\mathbb{I}+\boldsymbol{\tau}_1 \cdot \boldsymbol{\tau}_2\right)$ \tabularnewline
\hline \hline
\end{tabular*}
\end{table*}

To regularize potential short-range singularities, we introduce a form factor to suppress high-momentum contributions:
\begin{equation}
	\begin{aligned}
	&V(\bm{p},u)\to V(\bm p, u)F^2(u,\Lambda,\bm p^2) \\
	&F\left(u, \Lambda, \bm{p}^2\right)=\frac{\Lambda^2-u^2}{\Lambda^2+\bm{p}^2}
\end{aligned}
\end{equation}
where $\Lambda$ denotes the cutoff parameter. In this work, we vary the cutoff from $1.10~\text{GeV}$ to $1.35~\text{GeV}$ in steps of $0.05~\text{GeV}$ to address the cutoff dependence. The coordinate space potential can be readily obtained via Fourier transformation, as presented in the Appendix~\ref{app:fourier}.

The values of the OBE parameters are listed in Table~\ref{tab:lag_para}. The pion decay constant $f_\pi$ is a well-known quantity, while the axial coupling constants $g_a$ and $g_a^\prime$ are extracted from the widths of the $D^*\to D\pi$ and $K^*\to K\pi$, respectively. However, the remaining three vector coupling constants $g_V$, $\beta$, and $\lambda$, as well as the scalar coupling $g_s$, can only be determined through model-dependent methods. Three of these are independent, as $g_V$ acts as an overall factor in the vector sector. The values in the table are adopted from Refs.~\cite{Li:2012cs,Li:2012ss} as a baseline, which are determined using vector meson dominance combined with results from lattice QCD and light cone sum rules~\cite{Isola:2003fh}. Although the coupling constants in the charm and strange sectors may differ in principle, we set $g_s^\prime=g_s$, $g_V^\prime=g_V$, $\beta^\prime=\beta$, and $\lambda^\prime=\lambda$ to reduce the number of the parameters.

\begin{table*}[]
    \centering
        \caption{Hadron masses and coupling constants in the OBE Lagrangians. The constants $g_a$ and $g_a^\prime$ are extracted from the widths of the $D^*\to D\pi$ and $K^*\to K\pi$, respectively. The experimental data are taken from Ref.~\cite{ParticleDataGroup:2024cfk}. The values for $g_s$, $\lambda$, and $\beta$ are adopted from Refs.~\cite{Li:2012cs,Li:2012ss,Isola:2003fh} and serve as baseline parameters, where three rescaling factors in \eq{eq:coupling-ratio} are applied.}
    \label{tab:lag_para}
\begin{tabular*}{\hsize}{@{}@{\extracolsep{\fill}}ccccccccc@{}}
\hline 
\hline 
\multirow{2}{*}{Mass } & $m_{D}[\text{GeV}]$ & $m_{D^{*}}[\text{GeV}]$ & $m_{\pi}[\text{GeV}]$ & $m_{\eta}[\text{GeV}]$ & $m_{\rho}[\text{GeV}]$ & $m_{\omega}[\text{GeV}]$ & $m_{\sigma}[\text{GeV}]$ & $m_{K^*}[\text{GeV}]$ \tabularnewline
\cline{2-9} 
 & $1.867$ & $2.009$ & $0.137$ & $0.548$ & $0.775$ & $0.783$ & $0.600$ & $0.894$\tabularnewline
\hline 
\multirow{2}{*}{Coupling} & $f_{\pi}~[\text{GeV}]$ & $g_{a}$ & $g_{a}^{\prime}$ & $g_{V}=g_V^\prime$ & $\beta=\beta^\prime$ & $\lambda=\lambda^\prime[\text{GeV}^{-1}]$  & $g_{s}=g_s^\prime$ & \tabularnewline
\cline{2-9} 
 & $0.13025$ & $0.57$ & $0.88$ & $5.8$ & $0.9$ & $0.56$  & $0.76$ & \tabularnewline
\hline 
\hline 
\end{tabular*}
\end{table*}

\begin{figure}
    \centering
    \includegraphics[width=0.4\textwidth]{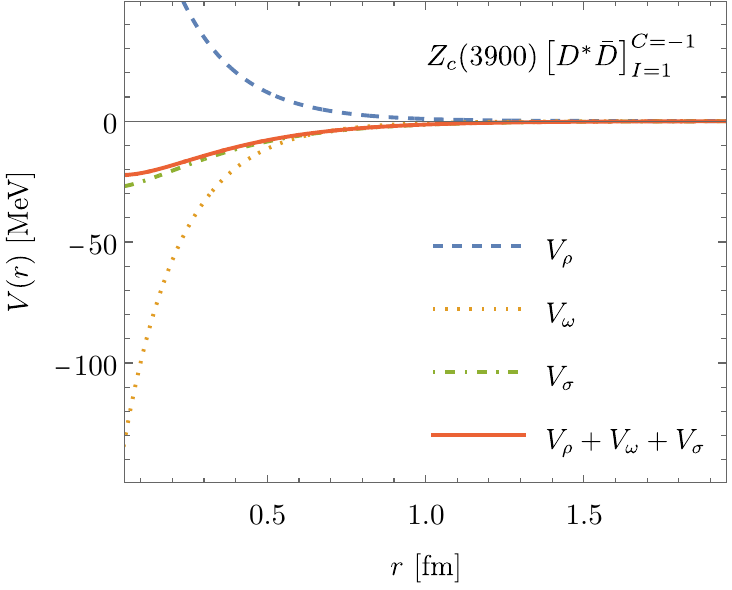}
    \caption{Effective potentials of the scalar and vector meson exchange in the $Z_c(3900)~[D\bar{D}^*]^{C=-1}_{I=1}$ system. The blue dashed, yellow dotted, and green dot-dashed lines represent the contributions from the $\rho$, $\omega$, and $\sigma$ exchanges, respectively. The red solid line denotes the total of them. The cutoff $\Lambda$ is set to $1.10~\text{GeV}$, and the parameters $(R_\beta, R_\lambda, R_s)$ are set to $(1.0, 1.0, 2.0)$ as an example.} 
    \label{fig:Zc3900_potential}
\end{figure}

\begin{figure*}
    \centering
    \includegraphics[width=0.9\textwidth]{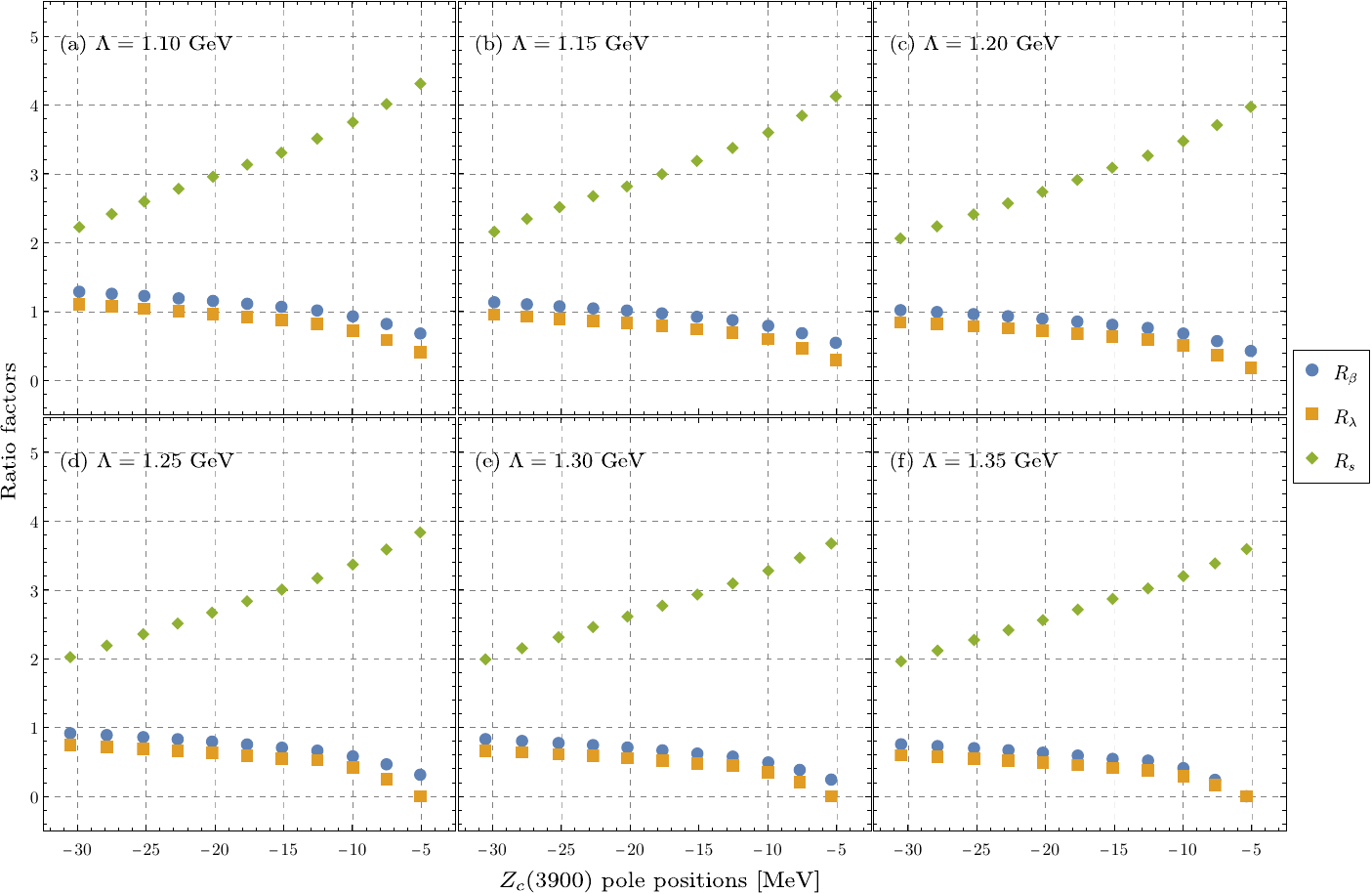}
    \caption{The ratio factors for the scalar- and vector-meson-exchange interactions vary with the $Z_c(3900)$ virtual state pole position across different cutoffs. The blue circles, yellow squares, and green diamonds represent $R_\beta$, $R_\lambda$, and $R_s$, respectively. These ratios are defined relative to the baseline values as detailed in Eq.~\eqref{eq:coupling-ratio} and Table~\ref{tab:lag_para}.}
    \label{fig:coupling}
\end{figure*}

Following the approach in Ref.~\cite{Zhu:2024hgm}, we introduce three ratio factors:
\begin{equation}\label{eq:coupling-ratio}
    \lambda\to\lambda R_{\lambda},\quad\beta\to\beta R_{\beta},\quad g_{s}\to g_{s}R_{s},
\end{equation}
to investigate the impact of variations in the scalar and vector coupling constants and to address model dependence. As demonstrated in Ref.~\cite{Zhu:2024hgm} and illustrated in Fig.~\ref{fig:Zc3900_potential}, for a special state $Z_c(3900)$, which is considered as a two-body $\left[D^* \bar{D}\right]_{I=1}^{C=-1}$ system, the exchange potentials of the $\rho$ and $\omega$ mesons are nearly equal in magnitude and effectively cancel each other. This cancellation renders the pole position of the $Z_c(3900)$ primarily sensitive to $\sigma$ exchange $R_s$. Considering the physical nature of the $Z_c(3900)$ remains a subject of debate~\cite{Albaladejo:2015lob,Pilloni:2016obd,Nakamura:2023obk,Yu:2024sqv,Chen:2023def}, we assume $Z_c(3900)$ is a virtual state and treat its pole position, or equivalently $R_s$, as a free parameter. We vary the pole position of the $Z_c(3900)$ virtual state from $-5~\text{MeV}$ to $-30~\text{MeV}$ in increments of $2.5~\text{MeV}$. Subsequently, for each cutoff $\Lambda$ and $Z_c(3900)$ pole position, we fix the $R_\lambda$ and $R_\beta$ by reproducing the isospin averaged binding energies of the $X(3872)$ ($\Delta E=-4.09~\text{MeV}$) and $\Tcc$ ($\Delta E=-1.6~\text{MeV}$) as listed in Table~\ref{tab:molecule_example}. The resulting coupling constants are presented in Fig.~\ref{fig:coupling}. It is evident that $R_s$ correlates strongly with the pole position of the $Z_c(3900)$, whereas $R_\beta$ and $R_\lambda$ remain close to unity with minimal variation.

\clabel[change_1]{Three-body forces may also exist in the system. However, such interactions remain elusive even in nucleon systems~\cite{Kalantar-Nayestanaki:2011rzs}. In chiral effective field theory, the three-nucleon force first appears at next-to-next-to-leading order~\cite{Epelbaum:2009sd,Machleidt:2011zz}. Power counting in this framework places the typical size of three-nucleon terms at only a few percent (roughly $5\%$) of the two-nucleon contribution~\cite{vanKolck:1994yi}. Phenomenologically, light nuclei show a similar hierarchy between three- and two-nucleon binding-energy contributions~\cite{Hammer:2012id}. Recently, Ref.~\cite{Pan:2025vyg} showed that, in hadronic systems with definite $C$ parity, a $C$-parity-dependent transition term can act as an intrinsic three-body force. The present $DD^*\bar{K}^*$ system is different since it contains three isospin-$1/2$ hadrons whose light-flavor degrees of freedom are all antiquarks and has no definite charge parity. Its leading dynamics is therefore expected to be closer to that of the three-nucleon system, where pairwise interactions dominate. We expect the three-body-force effect in the present system to be of the same order as in the three-nucleon case and smaller than the present uncertainty associated with the two-body OBE interaction. Moreover, isolating three-body forces through physical observables is challenging. For these reasons, we neglect the three-body force in this work.}

\subsection{Methodology} 
\label{sub:methodology}

\begin{figure}
    \centering
    \includegraphics[width=0.475\textwidth]{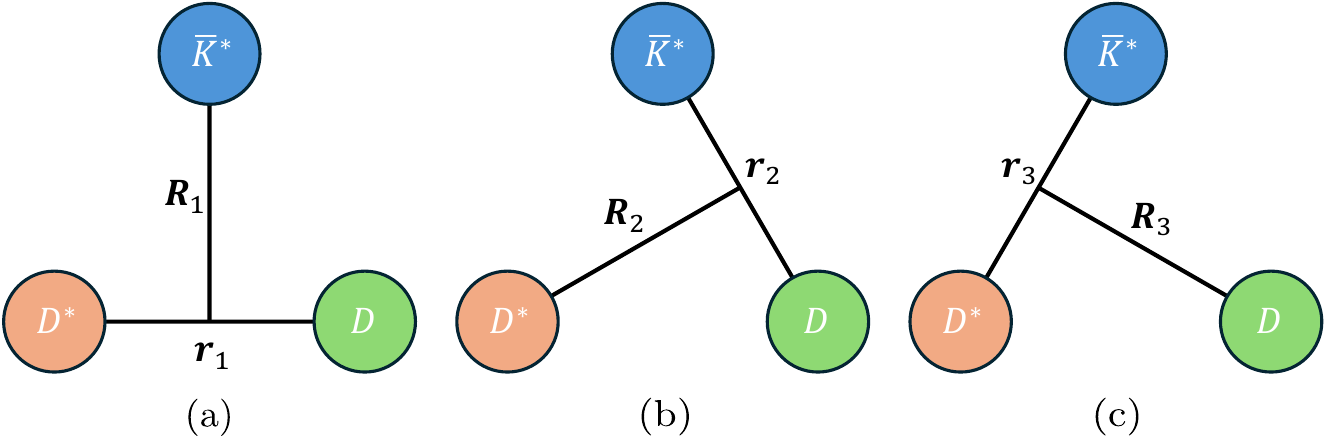}
    \caption{The three sets of Jacobi coordinates for the $DD^*\bar{K}^*$ three-body system.} 
    \label{fig:jac}
\end{figure}

The three-body system is investigated using a nonrelativistic Hamiltonian
\begin{equation}
	H=\sum_{i=1}^3\left(m_i+\frac{p_i^2}{2 m_i}\right)+\sum_{i<j} V_{i j},
\end{equation}
where $V_{ij}$ is the two-body effective potential of the OBE model defined in Eqs.~\eqref{eq:V_D_Dast_sigma}-\eqref{eq:V_D_Kast_vector}. The $S$-wave $DD^*\bar{K}^*$ three-body wave function can be expressed as
\begin{equation}
	\begin{aligned}
		\Psi_{I,J}=&\sum_{\alpha,n_{i},n_{i}^\prime,i}C_{\alpha,n_{i},n_{i}^\prime,i}\chi_{\alpha}^{I,J}\\
		&\qquad\times\left[\phi^{G}_{n_ilm}(\bm r_i)\phi^G_{n^\prime_i l^\prime m^\prime}(\bm R_i)\right]_{L=0},
	\end{aligned}
\end{equation}
where $R_{i}$ and $r_{i}$ represent the independent Jacobi coordinates  shown in Fig.~\ref{fig:jac}. The expansion coefficients $C_{\alpha,n_{i},n^\prime,i}$ are determined via the Rayleigh-Ritz variational method. $\chi_{\alpha}^{I,J}$ denotes the spin-isospin wave functions, given by
\begin{equation}
	\begin{aligned}
	& \chi^{\frac{1}{2},J}_{1}=\left\{D \left[D^*\bar{K^*}\right]^{0,J}\right\}^{\frac{1}{2},J},\\
	& \chi^{\frac{1}{2},J}_{2}=\left\{D \left[D^*\bar{K^*}\right]^{1,J}\right\}^{\frac{1}{2},J},\\
	&\chi^{\frac{3}{2},J}=\left\{D \left[D^*\bar{K^*}\right]^{1,J}\right\}^{\frac{3}{2},J},
\end{aligned}
\end{equation}
where $J=0,1,2$. 

We employ the GEM to solve the three-body Schr\"odinger equation. This approach is widely used in few-body calculations and has proven to be highly effective~\cite{Hiyama:2003cu}. In GEM, the spatial wave functions take the form
\begin{equation}
	\phi_{n l m}(\boldsymbol{r})=\sqrt{\frac{2^{l+5 / 2}}{\Gamma\left(l+\frac{3}{2}\right) r_n^3}}\left(\frac{r}{r_n}\right)^l e^{-\frac{r^2}{r_n^2}} Y_{l m}(\hat{r}),
\end{equation}
where $Y_{lm}$ represents the spherical harmonic function. It is worth noting that any single set of Jacobi coordinates $(\bm r_i,\bm R_i)$ suffices when considering an infinite number of radially and orbitally excited basis functions. However, a more efficient approach involves restricting the calculation to the $S$-wave $\phi_{n00}(\bm r_i)\phi_{n^\prime 00}(\bm R_i)$, while utilizing all three sets of Jacobi coordinates ($i=1$ to $3$)~\cite{Hiyama:2003cu,Meng:2020knc,Meng:2023jqk}. The inclusion of $S$-wave functions from all Jacobi coordinates also partially compensates for contributions from higher partial wave angular momentum components within a single coordinate. The Gaussian size parameter $r_n$ follows a geometric progression:
\begin{equation}
r_n=r_1 a^{n-1}, \quad a=\left(\frac{r_1}{r_{\max }}\right)^{\left(n_{\max }-1\right)}.
\end{equation}
In this work, the size parameters are taken as
\begin{equation}
	r_1=0.01~\mathrm{fm},\quad r_{\mathrm{max}}=10~\mathrm{fm},\quad n_{\mathrm{max}}=20,
\end{equation}
which cover both long-range and short-range correlations and provide an efficient approximation for the radial component of the wave functions.

\begin{figure}
  \centering
  \includegraphics[width=0.9\linewidth]{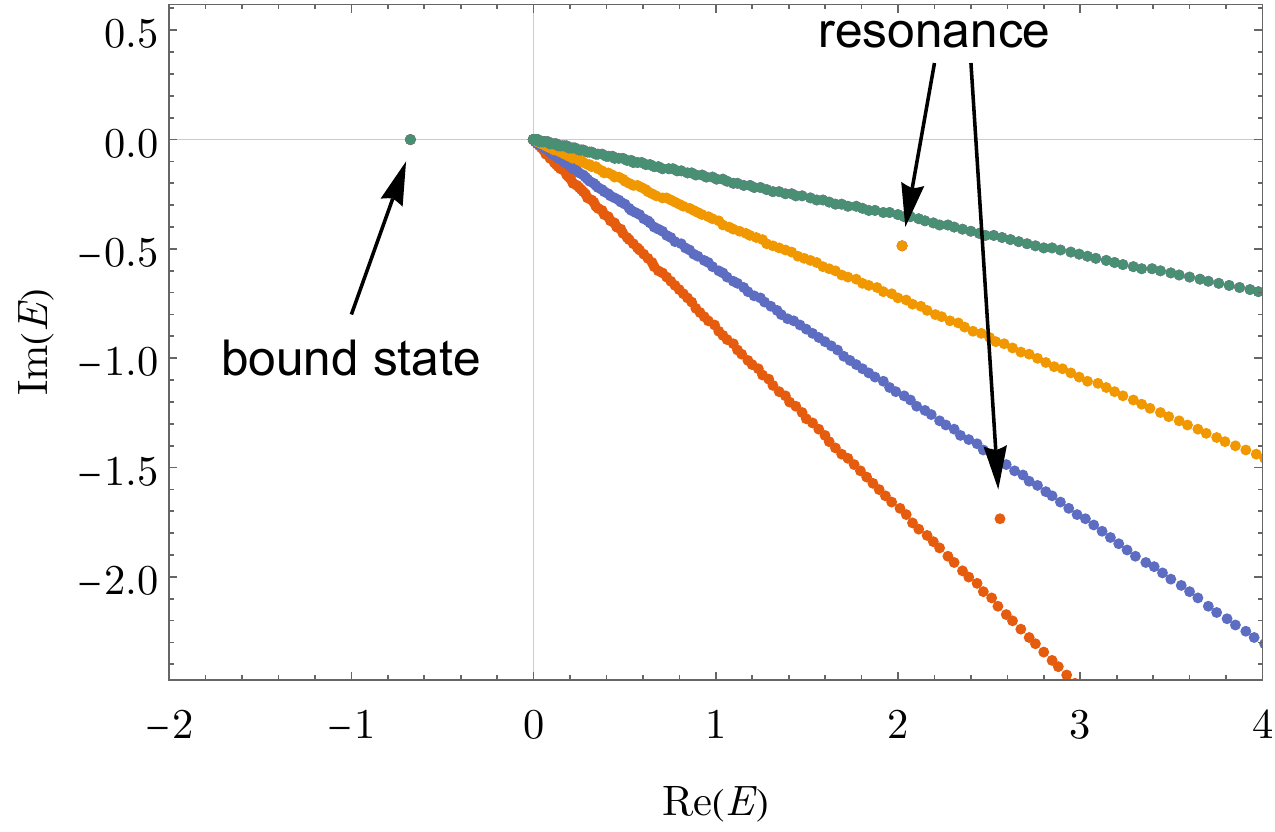}
  \caption{A typical solution of the complex-scaled Schr\"odinger equation. The scattering states align along $\operatorname{Arg}(E)=-2 \theta$, whereas bound states and resonances do not shift as $\theta$ changes.The resonance with energy $E_R=m_R-i \frac{\Gamma_R}{2}$ can be obtained when $2 \theta>\left|\operatorname{Arg}\left(E_R\right)\right|$.}
  \label{fig:csm_typical}
\end{figure}

The approach described above is highly effective for solving bound states. However, the wave functions of resonances are not square integrable and cannot be obtained by directly solving the eigenequation of the Hermitian Hamiltonian. To identify three-body resonant solutions, we employ the CSM~\cite{Aguilar:1971ve,Balslev:1971vb,Moiseyev:1998gjp,Aoyama:2006hrz}. This technique introduces a complex rotation $U(\theta)$ to the radial coordinate $r$ and the conjugate momentum $p$:
\begin{align}
&U(\theta)r=re^{i\theta},\quad U(\theta)p=pe^{-i\theta},\\
&H(\theta)=\sum_{i=1}^3(m_i+\frac{p_i^2e^{-2i\theta}}{2m_i})+\sum_{i<j}V_{ij}(r_{ij}e^{i\theta}),
\end{align}
This transformation allows resonances to be treated similarly to bound states through square-integrable basis expansion. Solving the eigenvalue equation of the complex-scaled Hamiltonian yields the eigenenergies of bound, scattering, and resonant states simultaneously. Figure~\ref{fig:csm_typical} illustrates a typical distribution of eigenenergies obtained by the CSM. The scattering states align along rays starting from threshold energies with $\operatorname{Arg}(E)=-2 \theta$, whereas bound states and resonances remain unchanged as $\theta$ varies. The resonance with energy $E_R=m_R-i \Gamma_R/2$ appears when $2 \theta>\left|\operatorname{Arg}\left(E_R\right)\right|$, where $m_R$ and $\Gamma_R$ denote the mass and width of the resonance, respectively. The combination of the GEM and the CSM has proven to be a powerful tool for investigating few-body resonances~\cite{Aoyama:2006hrz,Carbonell:2013ywa,Hiyama:2016nwn,Dote:2017wkk,Happ:2023kcc,Chen:2023eri,Chen:2023syh,Meng:2024yhu,Ma:2024vsi,Wu:2024euj,Wu:2024zbx}.

\section{Results and discussions} 
\label{sec:results_and_discussions}

{\subsection{$D^*\bar{K}^*$ system}}

\begin{figure*}
    \centering
    \includegraphics[width=0.81\textwidth]{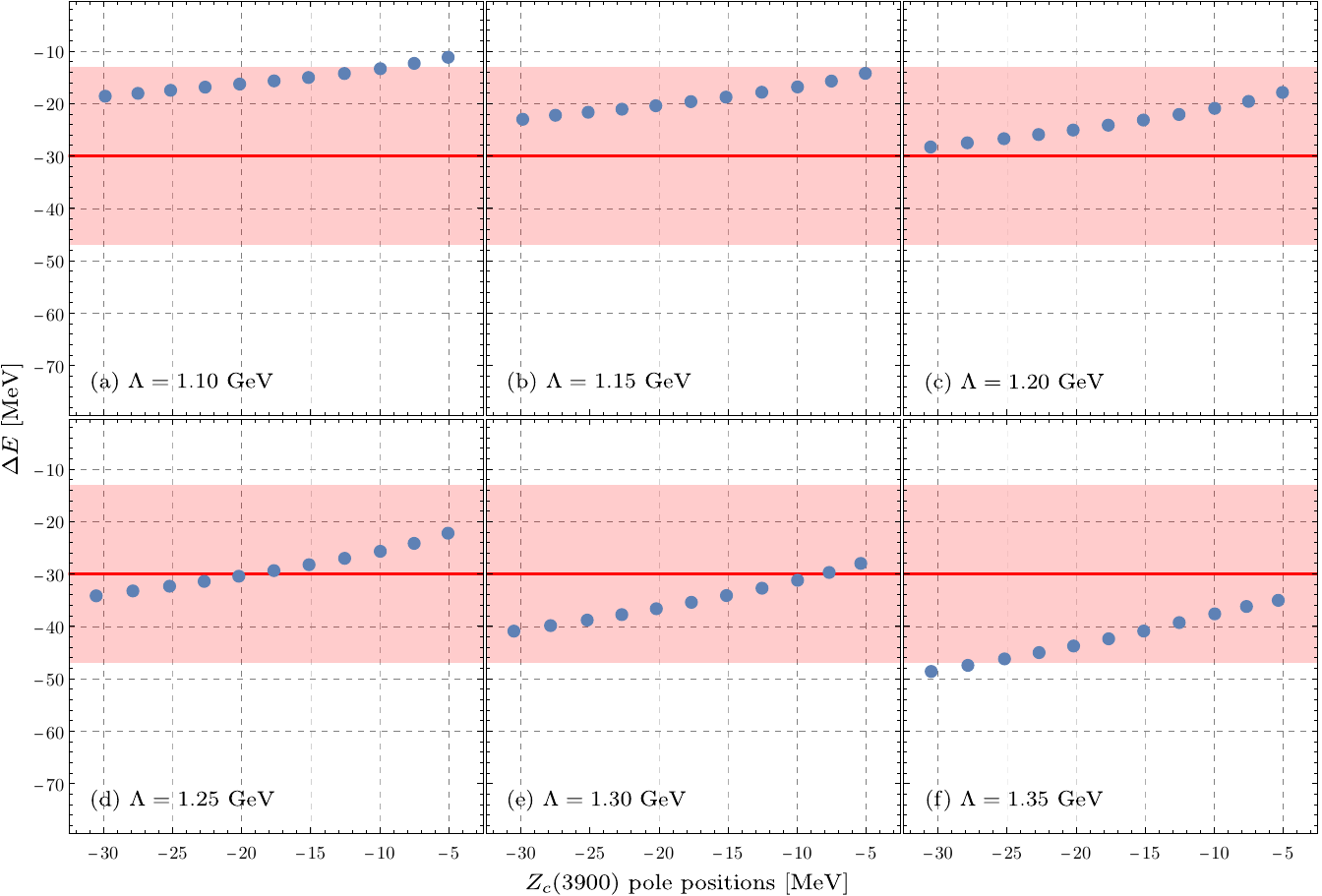}
    \caption{The binding energies of the $I(J^P)=0(0^+)$ $D^*\bar{K}^*$ two-body system vary with the $Z_c(3900)$ virtual state pole position across different cutoffs.  The red solid line indicates the experimental central value of the mass difference between the $T_{cs0}(2870)$ and the $D^*\bar{K}^*$ threshold, and the red shaded region represents the experimental uncertainty. The experimental data are taken from Ref.~\cite{ParticleDataGroup:2024cfk}.}
    \label{fig:tcs_pole_position}
\end{figure*}

Before investigating the three-body systems, we first calculate the binding energies of the $I(J^P)=0(0^+)$ $D^*\bar{K}^*$ two-body system using the parameters presented in Fig.~\ref{fig:coupling}, with the results illustrated in Fig.~\ref{fig:tcs_pole_position}. For each parameter set, corresponding to various cutoff values and $Z_c(3900)$ virtual state pole positions, we consistently obtain a bound state with a binding energy of approximately $30~\text{MeV}$. The slight cutoff dependence in the results arises from our assumption that the parameters $\Lambda, \beta, \lambda$, and $g_s$ are identical for the charm and strange sectors, with they are calibrated solely by the charm sector. Nevertheless, each parameter set provides a good description of the experimentally observed $T_{cs0}(2870)$ without additional tuning, which lies $30\pm17~\text{MeV}$ below the $D^*\bar{K}^*$ threshold~\cite{LHCb:2020bls,LHCb:2020pxc,LHCb:2024vfz,ParticleDataGroup:2024cfk}.  This agreement indicates that treating the $\bar{K}^*$ as a heavy meson, combined with the OBE effective potential and the recalibrated parameters, provides a reliable description of the $D^{(*)}\bar{K}^*$ interactions. Furthermore, these results validate the application of this framework and these parameter sets to the investigation of other systems.

\subsection{$DD^*\bar{K}^*$ system} 
\label{sub:DDastKbarast_system}

\begin{figure*}
    \centering
    \includegraphics[width=0.81\textwidth]{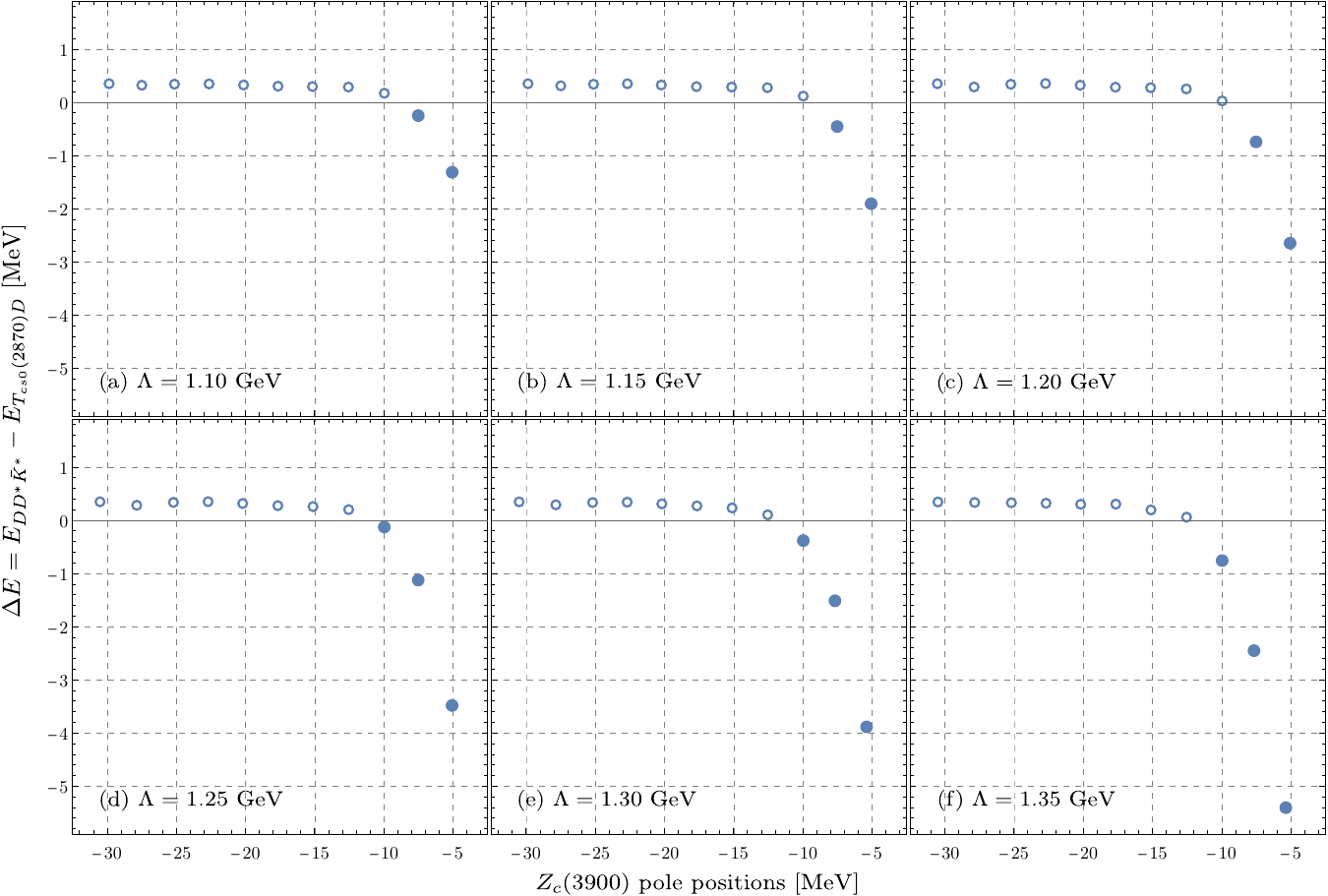}
    \caption{The binding energies $\Delta E$ of the $I(J^P)=1/2(0^-)$ $DD^*\bar{K}^*$ three-body system vary with the $Z_c(3900)$ virtual state pole position across different cutoffs. $\Delta E=E_{DD^*\bar{K}^*}-E_{T_{cs0}(2870)D}$ is defined as the energy difference between the three-body ground state and the theoretical $T_{cs0}(2870)D$ threshold, where the $T_{cs0}(2870)$ corresponds to the $I(J^P)=0(0^+)$ $D^*\bar{K}^*$ two-body bound state with binding energies shown in Fig.~\ref{fig:tcs_pole_position}. Solid points indicate bound states ($\Delta E < 0$), while hollow points indicate unbound states ($\Delta E > 0$).}
    \label{fig:DDastKbarast}
\end{figure*}

We subsequently investigate the $DD^*\bar{K}^*$ three-body system. Using the parameter sets presented in Fig.~\ref{fig:coupling}, we calculate the ground state energy of the $DD^*\bar{K}^*$ system. The results for the $I(J^P)=1/2(0^-)$ configuration are displayed in Fig.~\ref{fig:DDastKbarast}. The energy difference $\Delta E$ is defined relative to the theoretical $T_{cs0}(2870)D$ threshold, where the $T_{cs0}(2870)$ is treated as the $I(J^P)=0(0^+)$ $D^*\bar{K}^*$ two-body bound state, with binding energies shown in Fig.~\ref{fig:tcs_pole_position}. Negative values of $\Delta E$ indicate the formation of a three-body molecular bound state, while positive values indicate unbound cases.

As shown in Fig.~\ref{fig:DDastKbarast}, the existence of a three-body bound state is strongly correlated with the pole position of the $Z_c(3900)$. As the $Z_c(3900)$ virtual state pole approaches the threshold, corresponding to a stronger attractive $\sigma$ exchange, the binding energy of the three-body system increases. Conversely, as the $Z_c(3900)$ pole moves away from the threshold, indicating a weaker $\sigma$-exchange attraction, the three-body binding energy decreases until the bound state disappears. Our results indicate that a three-body molecular bound state emerges when the $Z_c(3900)$ virtual state pole lies within approximately $-10~\text{MeV}$ of the threshold. The binding energies are a few MeV. Notably, varying the cutoff does not qualitatively alter these results, implying that the connection between the existence of the $DD^*\bar{K}^*$ molecule and the $Z_c(3900)$ virtual state pole position is robust. At the quark level, this three-body bound state would correspond to a hexaquark state that has not yet been observed in experiment. Experimentally, this three-body bound state can be searched for in decay channels such as $DD\bar{K}\pi\pi$ and $DD\bar{K}$. Observation of such a state would constrain the position of the $Z_c(3900)$ pole. Meanwhile, a precise measurement of the $Z_c(3900)$ pole position would determine the existence or nonexistence of this three-body molecule.

We calculated the binding energies for all $S$-wave quantum numbers, specifically $I=(1/2, 3/2)$ with $J^P=(0, 1, 2)^-$. However, with the exception of the $1/2(0^-)$ channel, no three-body bound states were found in other configurations.

\clabel[change_2]{In our calculations, we neglect isospin breaking and coupled channels involving $D^* D^* \bar{K}^*$, $D^{(*)} D^{(*)} \bar{K}$, and baryon-antibaryon channels. Coupled-channel dynamics can be important for near-threshold systems, but its impact is expected to be limited for the present three-body bound state. For the lower mesonic channels, the nearest threshold is $D^*D^*\bar{K}$, which lies about $(m_{K^*}-m_K)-(m_{D^*}-m_D)\simeq 260~\text{MeV}$ below the channel of interest. Such a large energy separation suppresses its contribution to the near-threshold eigenenergy. This expectation is also supported by the two-body coupled-channel calculation within the same OBE framework in Ref.~\cite{Wang:2024ukc}, where both $T_{cs0}(2870)$ and $T_{cs1}(2900)$ were found to be almost pure $D^*\bar{K}^*$ molecular states. In addition, the $K$ meson is a light pseudo-Goldstone boson, so treating the $D^{(*)}D^{(*)}\bar{K}$ channels consistently would require dynamics beyond the heavy-meson formalism and nonrelativistic framework adopted for the present $D^{(*)}\bar{K}^*$ interactions. For the higher mesonic channel, the nearest $D^*D^*\bar{K}^*$ threshold lies about $m_{D^*}-m_D\simeq 140~\text{MeV}$ above the states of interest. According to the variational principle, including this higher coupled channel would lower the ground-state energy and thus make the three-body bound state slightly more deeply bound. Owing to the sizable threshold gap, however, this correction should be moderate and would not change the qualitative conclusion on the existence of the $DD^*\bar{K}^*$ bound state. The baryon-antibaryon channels involve short-range dynamics beyond the molecular basis and are also not expected to dominate the near-threshold solution. For isospin breaking, Ref.~\cite{Wu:2021kbu} included effects from mass splittings and Coulomb interactions in the $DDD^*$ three-body system. In Ref.~\cite{Zhu:2024hgm}, the isospin-breaking effects in the $X(3872)$ and $T_{cc}(3875)$ were examined in detail. The binding-energy shifts of the $X(3872)$ and $T_{cc}(3875)$ were found to mainly follow the mass differences between the physical $D^0\bar{D}^{*0}$ and $D^{*+}D^0$ thresholds and their corresponding isospin-averaged thresholds. Applying this threshold estimate to the $DD^*\bar{K}^*$ system gives an effect of order $1~\text{MeV}$. This is much smaller than the uncertainty from the strong interaction already present in the isospin-symmetric calculation. We therefore do not include the minor isospin-violation and Coulomb effects in this work.}

\begin{figure*}
    \centering
    \includegraphics[width=0.81\textwidth]{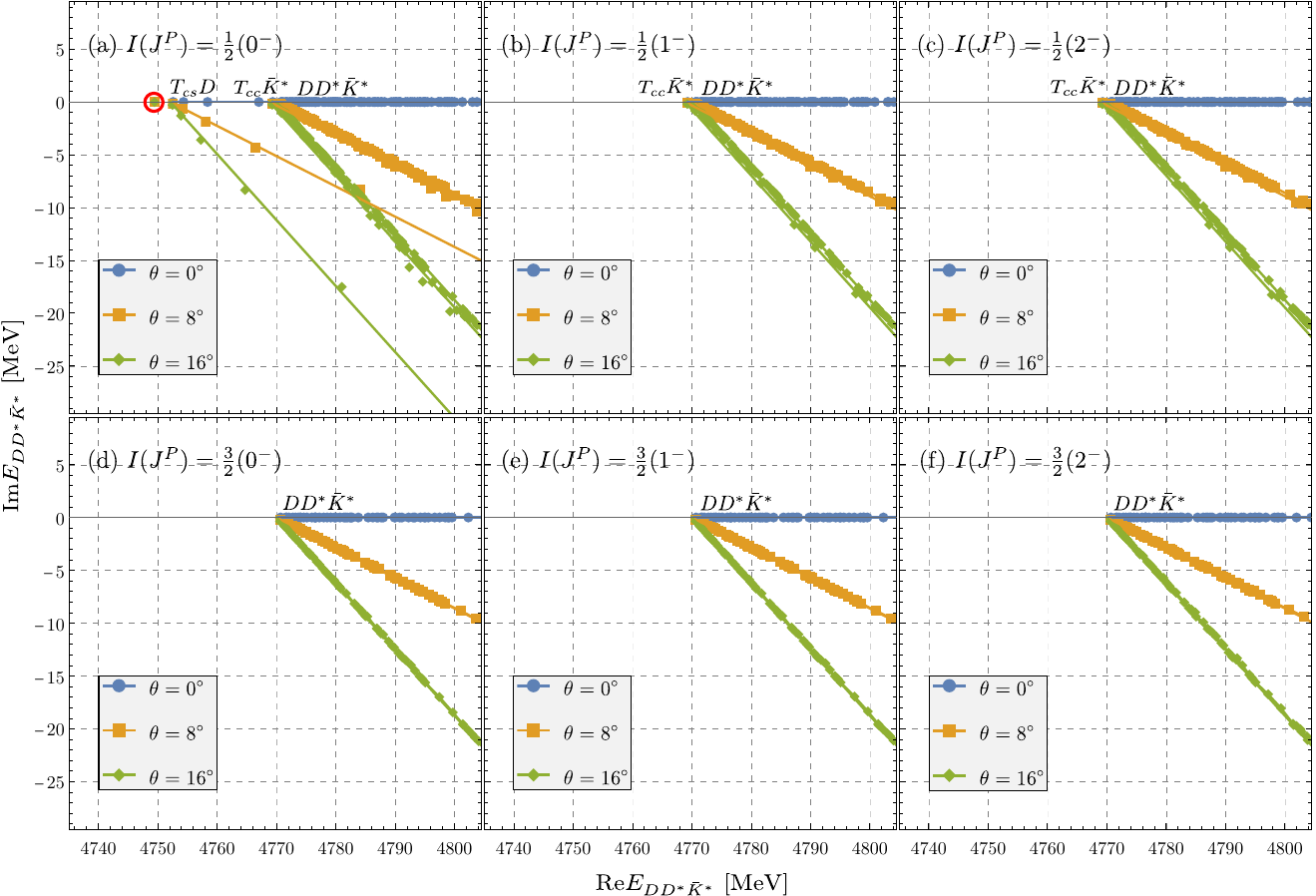}
    \caption{Complex energy spectrum of the $DD^*\bar{K}^*$ three-body system calculated using the CSM with varying complex scaling angles $\theta$. The parameter set corresponds to a cutoff $\Lambda=1.20$ GeV and a $Z_{c}(3900)$ virtual state pole at $-5$ MeV. With the exception of the bound state marked by the red circle, all other states represent scattering states aligned along rays.}
    \label{fig:DDastKbarast_csm}
\end{figure*}

In addition to bound states, we employ the CSM to search for resonances within the three-body system. Figure~\ref{fig:DDastKbarast_csm} illustrates a representative result. With the exception of a bound state pole, all other eigenenergies align along rays that rotate with the scaling angle $\theta$, indicating that they correspond to scattering states. The results for other parameter sets are similar; we do not identify any three-body resonances in any of the considered channels.

\subsection{$BB^*K^*$ system} 
\label{sub:BBastKast_system}

Regarding the $BB^*K^*$ system, which serves as the bottom analog of the $DD^*\bar{K}^*$ system, the lack of experimental data prevents the extraction of the coupling constants.  Consequently, based on heavy quark symmetry, we employ the same parameters as used for the $DD^*\bar{K}^*$ system, replacing only the charm meson masses $(m_D,m_{D^*})$ with the bottom meson masses $(m_B,m_{B^*})$. While this approach introduces theoretical uncertainties, it allows for a qualitative discussion.

\begin{figure*}
    \centering
    \includegraphics[width=0.9\textwidth]{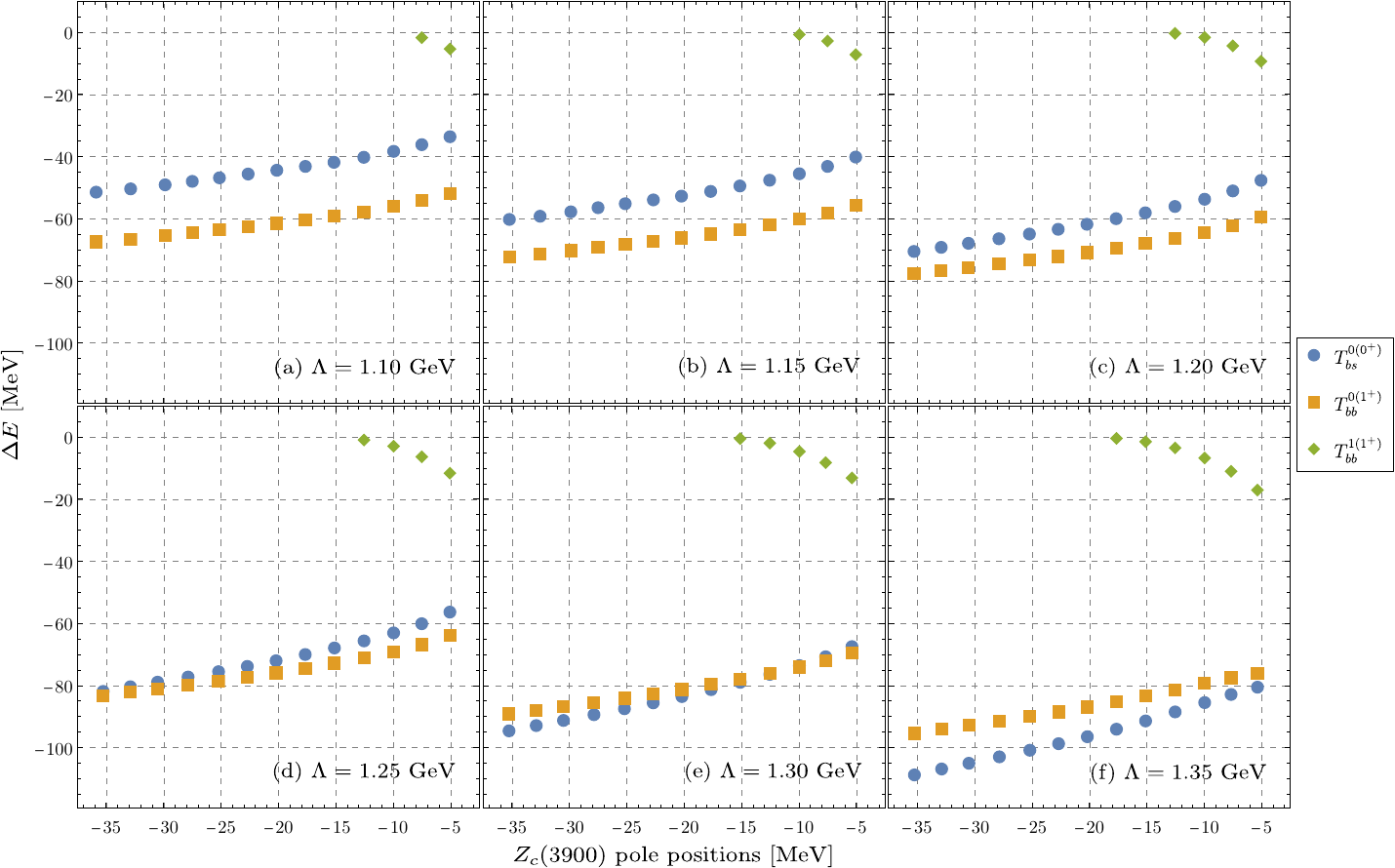}
    \caption{The binding energies of the two-body subsystem bound states within the $BB^*K^*$ three-body system vary with the $Z_c(3900)$ virtual state pole position across different cutoffs. The blue circles, yellow squares, and green diamonds represent the $0(0^+)$ $B^*\bar{K}^*$ bound state $T_{bs}^{0(0^+)}$, the $0(1^+)$ $BB^*$ bound state $T_{bb}^{0(1^+)}$, and the $1(1^+)$ $BB^*$ bound state $T_{bb}^{1(1^+)}$, respectively. }
    \label{fig:BBastKast_two_body}
\end{figure*}

The binding energies of the two-body subsystem bound states within the $BB^*K^*$ three-body system are presented in Fig.~\ref{fig:BBastKast_two_body}. Evidently, the various two-body subsystems in the $BB^*K^*$ sector exhibit deeper binding due to the larger mass of the bottom mesons. In particular, the $BB^*$ system with $I(J^P)=1(1^+)$ forms a shallow bound state, denoted as $T_{bb}^{1(1^+)}$, when the $\sigma$-exchange interaction is strong, or equivalently, when the virtual state pole of the $Z_c(3900)$ lies close to the threshold. Furthermore, at a cutoff $\Lambda \approx 1.30$ GeV, the hierarchy of binding energies between the $0(1^+)$ $BB^*$ bound state $T_{bb}^{0(1^+)}$ and the $0(0^+)$ $B^*K^*$ bound state $T_{bs}^{0(0^+)}$ is reversed. The lowest two-body threshold of the three-body system changes at this point.

\begin{figure*}
    \centering
    \includegraphics[width=0.81\textwidth]{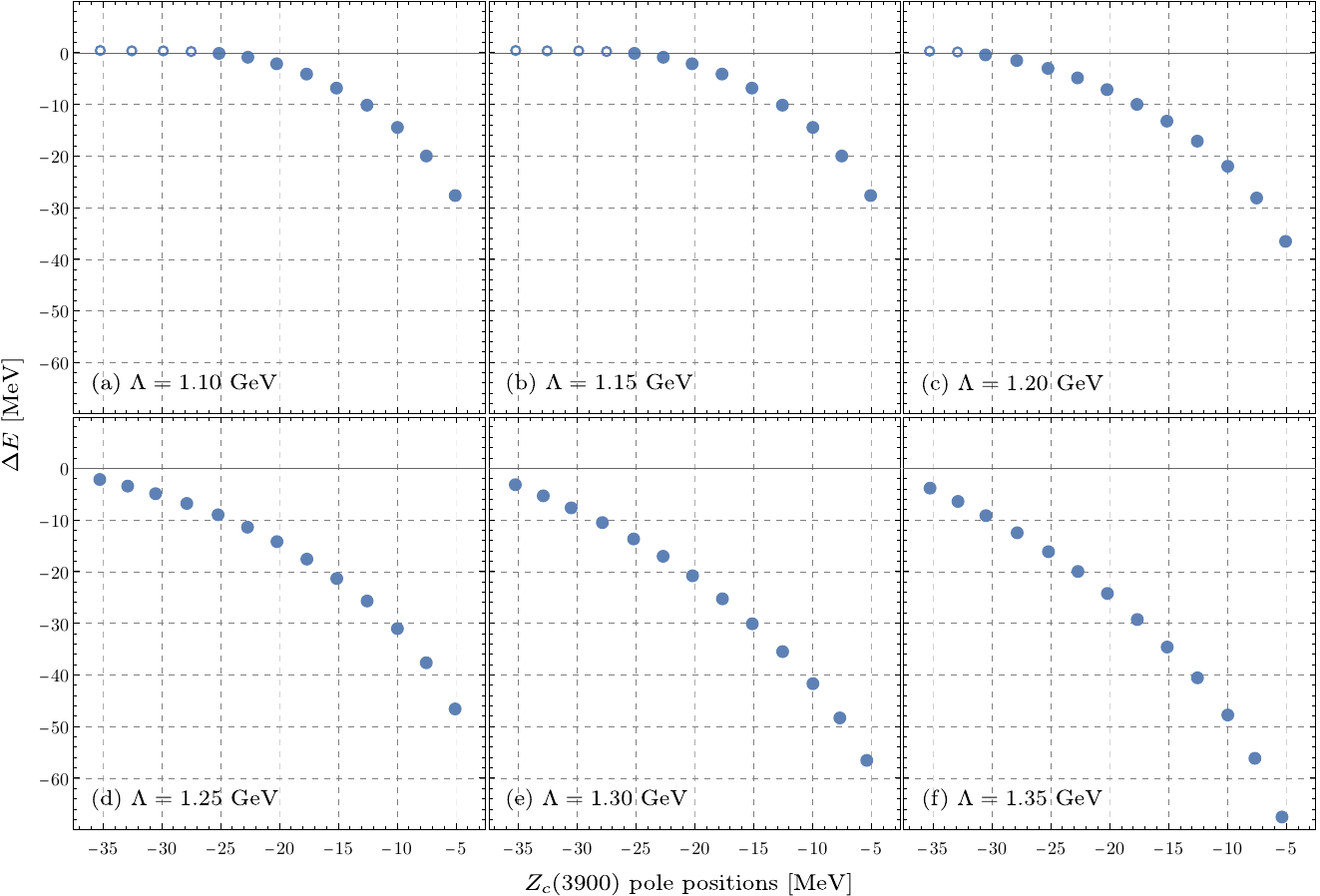}
    \caption{The binding energies $\Delta E$ of the $I(J^P)=1/2(0^-)$ $BB^*K^*$ three-body system vary with the $Z_c(3900)$ virtual state pole position across different cutoffs. $\Delta E$ is defined as the energy difference between the three-body ground state and the lowest two-body threshold of the three-body system. Solid points indicate bound states ($\Delta E < 0$), while hollow points indicate unbound states ($\Delta E > 0$).}
    \label{fig:BBastKast_three_body}
\end{figure*}

The ground state energies of the $I(J^P)=1/2(0^-)$ $BB^*K^*$ system are displayed in Fig.~\ref{fig:BBastKast_three_body}. The energy difference $\Delta E$ is defined relative to the lowest two-body threshold of the three-body system. The results indicate that the conditions for forming a three-body bound state are significantly more relaxed in the $BB^*K^*$ system. Depending on the specific cutoff, a $Z_c(3900)$ virtual state located between $-25$ and $-35$ MeV below the threshold suffices to form a three-body bound state. Moreover, the three-body state is more deeply bound, with binding energies reaching several tens of MeV. Apart from the $1/2(0^-)$ channel, no three-body bound states are found in other configurations.

\begin{figure*}
    \centering
    \includegraphics[width=0.81\textwidth]{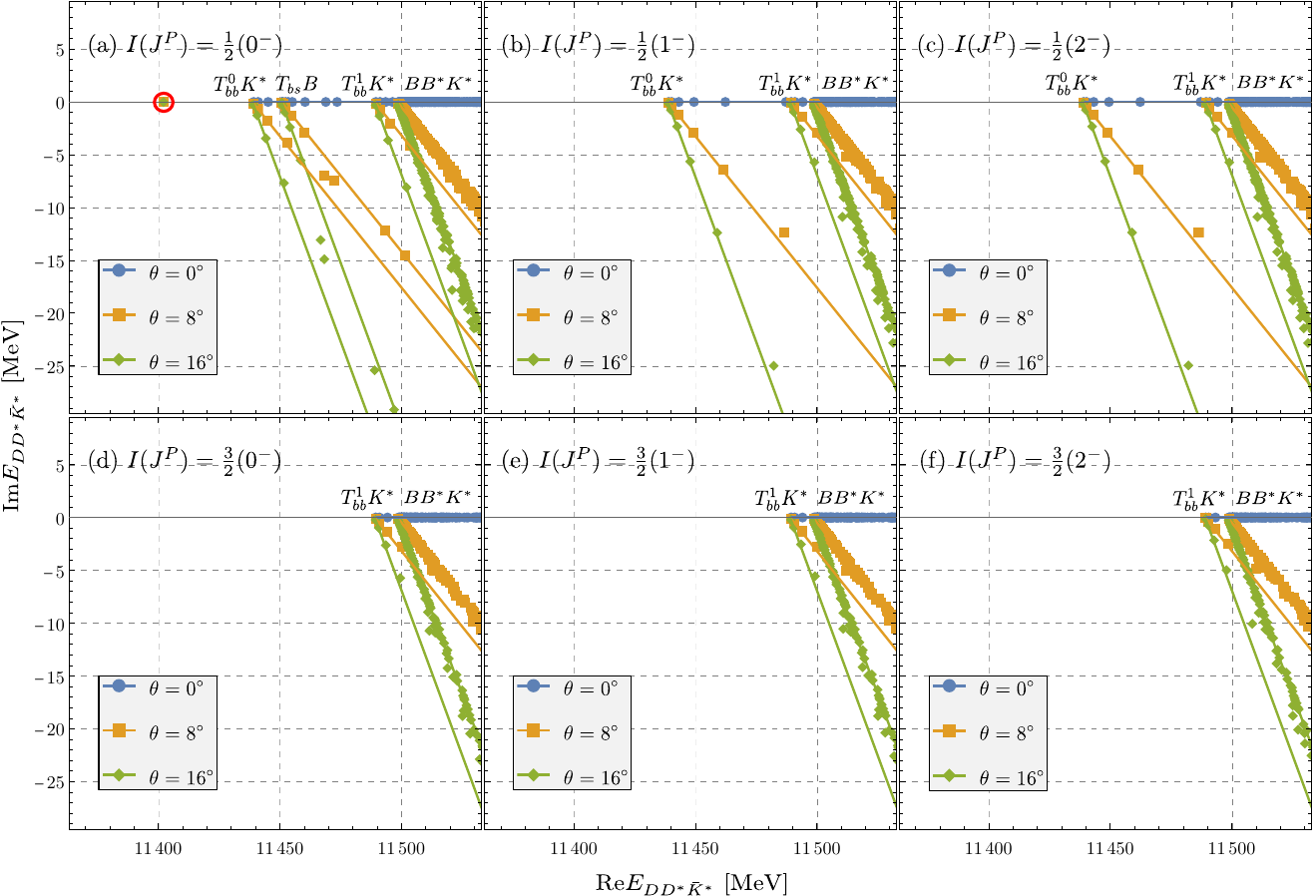}
    \caption{Complex energy spectrum of the $BB^*K^*$ three-body system calculated using the CSM with varying complex scaling angles $\theta$. The parameter set corresponds to a cutoff $\Lambda=1.20$ GeV and a $Z_{c}(3900)$ virtual state pole at $-5$ MeV. With the exception of the bound state marked by the red circle, all other states represent scattering states aligned along rays.}
    \label{fig:BBastKast_csm}
\end{figure*}

A typical complex energy spectrum of the $BB^*K^*$ system calculated using the CSM is illustrated in Fig.~\ref{fig:BBastKast_csm} Similar to the $DD^*\bar{K}^*$ system, all solutions other than the bound state correspond to scattering states. We do not identify any three-body resonances in any of the considered channels.

\section{Summary} 
\label{sec:summary}

In this work, we investigated the existence of the $DD^*\bar{K}^*$ three-body molecular state and its bottom analog, the $BB^*K^*$ system, within the OBE model. To address the inherent model dependence associated with scalar- and vector-meson-exchange interactions, we recalibrated their coupling constants. We assumed the $Z_c(3900)$ is a virtual state and treated its pole position as a free parameter, thereby constraining the strength of the scalar $\sigma$-exchange interaction. The vector couplings were determined using the well-established $X(3872)$ and $T_{cc}(3875)$ states. The resulting parameter sets successfully reproduce the mass of the $T_{cs0}(2870)$ without further tuning, validating the application of this framework and these parameter sets to the investigation of $D^{(*)}\bar{K}^*$ and other systems.

We employed the Gaussian expansion method to solve the three-body Schr\"odinger equation. Our results reveal a strong correlation between the existence of the three-body molecular bound state and the pole position of the $Z_c(3900)$ virtual state.  Specifically, an $I(J^P)=1/2(0^-)$ $DD^*\bar{K}^*$ bound state exists when the $Z_c(3900)$ virtual state is located within approximately $-10~\text{MeV}$ of the $D\bar{D}^*$ threshold, with a binding energy of a few MeV. For the $BB^*K^*$ system, the conditions for forming a three-body bound state are more relaxed; a $Z_c(3900)$ virtual state located within $-25$ MeV of the threshold suffices. Furthermore, this state is more deeply bound, with a binding energy reaching several tens of MeV.

We also employ the complex scaling method to search for three-body resonances, but no such states are identified in either the charm or bottom sectors for any of the considered channels.

At the quark level, the predicted three-body bound states correspond to hexaquark states that have not yet been observed experimentally. We propose that future experiments search for the $DD^*\bar{K}^*$ three-body bound state in the $DD\bar{K}\pi\pi$ and $DD\bar{K}$ decay channels or precisely measure the pole position of the $Z_c(3900)$, as these efforts would mutually illuminate the nature of the associated states.


\section*{ACKNOWLEDGMENTS}
	
	This project was supported by the National Natural Science Foundation of China (Grants No. 12475137 and No. 12405088) and  Start-up Funds of Southeast University (Grant No. 4007022506) and Start-up Funds of Chongqing University. The computational resources were supported by High-performance Computing Platform of Peking University.

\section*{DATA AVAILABILITY}

The data supporting this study's findings are available within the article.

\appendix
\section{Fourier transformation}\label{app:fourier}
Taking the direct diagrams as an example, the analytical expressions for the coordinate space potentials obtained via the Fourier transform are:
\begin{align}
	& \frac{1}{u^2+\bm{q}^2} F\left(u, \Lambda, q^2\right)^2 \rightarrow H_0(u, \Lambda, r),\\
	& \frac{\bm{q}^2}{u^2+\bm{q}^2} F\left(u, \Lambda, q^2\right)^2 \rightarrow-H_1(u, \Lambda, r),\\
	& \frac{q_i q_j}{u^2+\bm{q}^2} F\left(u, \Lambda, q^2\right)^2 \nonumber \\
	& \qquad \rightarrow-\left[H_3(u, \Lambda, r) T_{i j}+H_1(u, \Lambda, r) \frac{\delta_{i j}}{3}\right],
\end{align}
where $T_{i j}=\frac{3 r_i r_j}{r^2}-\delta_{i j}$, and the expressions of $H_{0,1,3}$ are
\begin{align}
	 H_0(u, \Lambda, r)= & \frac{u}{4 \pi}\left[\frac{e^{-u r}-e^{-\Lambda r}}{u r}-\frac{\Lambda^2-u^2}{2 u \Lambda} e^{-\Lambda r}\right],\\
	 H_1(u, \Lambda, r)= & \frac{u^3}{4 \pi}\left[\frac{e^{-u r}-e^{-\Lambda r}}{u r}\right.\nonumber \\
	 & \qquad \left.-\frac{\left(\Lambda^2-u^2\right) \Lambda^2}{2 u^3 \Lambda} e^{-\Lambda r}\right],\\
	H_3(u, \Lambda, r) = & \frac{u^3}{12 \pi}\left[-\frac{e^{-\Lambda r} \Lambda^2\left(\frac{3}{\Lambda^2 r^2}+\frac{3}{\Lambda r}+1\right)}{r u^3}\right.\\
	&\qquad ~-\frac{e^{-\Lambda r}(\Lambda r+1)\left(\Lambda^2-u^2\right)}{2 r u^3}\nonumber\\
	&\qquad ~+\left.\frac{e^{-u r}\left(\frac{3}{r^2 u^2}+\frac{3}{r u}+1\right)}{r u}\right].
\end{align}
The transformation for $V(\bm k)$ follows a similar procedure. As discussed in Refs.~\cite{Zhu:2024hgm,Lin:2024qcq}, the coordinate space potentials derived from $V(\bm k)$ and $V(\bm q)$ are identical for even partial waves but differ by a sign for odd partial waves.
\bibliography{ref}

\begin{thebibliography}{117}%
\makeatletter
\providecommand \@ifxundefined [1]{%
 \@ifx{#1\undefined}
}%
\providecommand \@ifnum [1]{%
 \ifnum #1\expandafter \@firstoftwo
 \else \expandafter \@secondoftwo
 \fi
}%
\providecommand \@ifx [1]{%
 \ifx #1\expandafter \@firstoftwo
 \else \expandafter \@secondoftwo
 \fi
}%
\providecommand \natexlab [1]{#1}%
\providecommand \enquote  [1]{``#1''}%
\providecommand \bibnamefont  [1]{#1}%
\providecommand \bibfnamefont [1]{#1}%
\providecommand \citenamefont [1]{#1}%
\providecommand \href@noop [0]{\@secondoftwo}%
\providecommand \href [0]{\begingroup \@sanitize@url \@href}%
\providecommand \@href[1]{\@@startlink{#1}\@@href}%
\providecommand \@@href[1]{\endgroup#1\@@endlink}%
\providecommand \@sanitize@url [0]{\catcode `\\12\catcode `\$12\catcode `\&12\catcode `\#12\catcode `\^12\catcode `\_12\catcode `\%12\relax}%
\providecommand \@@startlink[1]{}%
\providecommand \@@endlink[0]{}%
\providecommand \url  [0]{\begingroup\@sanitize@url \@url }%
\providecommand \@url [1]{\endgroup\@href {#1}{\urlprefix }}%
\providecommand \urlprefix  [0]{URL }%
\providecommand \Eprint [0]{\href }%
\providecommand \doibase [0]{https://doi.org/}%
\providecommand \selectlanguage [0]{\@gobble}%
\providecommand \bibinfo  [0]{\@secondoftwo}%
\providecommand \bibfield  [0]{\@secondoftwo}%
\providecommand \translation [1]{[#1]}%
\providecommand \BibitemOpen [0]{}%
\providecommand \bibitemStop [0]{}%
\providecommand \bibitemNoStop [0]{.\EOS\space}%
\providecommand \EOS [0]{\spacefactor3000\relax}%
\providecommand \BibitemShut  [1]{\csname bibitem#1\endcsname}%
\let\auto@bib@innerbib\@empty
\bibitem [{\citenamefont {Chen}\ \emph {et~al.}(2016)\citenamefont {Chen}, \citenamefont {Chen}, \citenamefont {Liu},\ and\ \citenamefont {Zhu}}]{Chen:2016qju}%
  \BibitemOpen
  \bibfield  {author} {\bibinfo {author} {\bibfnamefont {H.-X.}\ \bibnamefont {Chen}}, \bibinfo {author} {\bibfnamefont {W.}~\bibnamefont {Chen}}, \bibinfo {author} {\bibfnamefont {X.}~\bibnamefont {Liu}},\ and\ \bibinfo {author} {\bibfnamefont {S.-L.}\ \bibnamefont {Zhu}},\ }\bibfield  {title} {\bibinfo {title} {{The hidden-charm pentaquark and tetraquark states}},\ }\href {https://doi.org/10.1016/j.physrep.2016.05.004} {\bibfield  {journal} {\bibinfo  {journal} {Phys. Rept.}\ }\textbf {\bibinfo {volume} {639}},\ \bibinfo {pages} {1} (\bibinfo {year} {2016})},\ \Eprint {https://arxiv.org/abs/1601.02092} {arXiv:1601.02092 [hep-ph]} \BibitemShut {NoStop}%
\bibitem [{\citenamefont {Esposito}\ \emph {et~al.}(2017)\citenamefont {Esposito}, \citenamefont {Pilloni},\ and\ \citenamefont {Polosa}}]{Esposito:2016noz}%
  \BibitemOpen
  \bibfield  {author} {\bibinfo {author} {\bibfnamefont {A.}~\bibnamefont {Esposito}}, \bibinfo {author} {\bibfnamefont {A.}~\bibnamefont {Pilloni}},\ and\ \bibinfo {author} {\bibfnamefont {A.~D.}\ \bibnamefont {Polosa}},\ }\bibfield  {title} {\bibinfo {title} {{Multiquark Resonances}},\ }\href {https://doi.org/10.1016/j.physrep.2016.11.002} {\bibfield  {journal} {\bibinfo  {journal} {Phys. Rept.}\ }\textbf {\bibinfo {volume} {668}},\ \bibinfo {pages} {1} (\bibinfo {year} {2017})},\ \Eprint {https://arxiv.org/abs/1611.07920} {arXiv:1611.07920 [hep-ph]} \BibitemShut {NoStop}%
\bibitem [{\citenamefont {Lebed}\ \emph {et~al.}(2017)\citenamefont {Lebed}, \citenamefont {Mitchell},\ and\ \citenamefont {Swanson}}]{Lebed:2016hpi}%
  \BibitemOpen
  \bibfield  {author} {\bibinfo {author} {\bibfnamefont {R.~F.}\ \bibnamefont {Lebed}}, \bibinfo {author} {\bibfnamefont {R.~E.}\ \bibnamefont {Mitchell}},\ and\ \bibinfo {author} {\bibfnamefont {E.~S.}\ \bibnamefont {Swanson}},\ }\bibfield  {title} {\bibinfo {title} {{Heavy-Quark QCD Exotica}},\ }\href {https://doi.org/10.1016/j.ppnp.2016.11.003} {\bibfield  {journal} {\bibinfo  {journal} {Prog. Part. Nucl. Phys.}\ }\textbf {\bibinfo {volume} {93}},\ \bibinfo {pages} {143} (\bibinfo {year} {2017})},\ \Eprint {https://arxiv.org/abs/1610.04528} {arXiv:1610.04528 [hep-ph]} \BibitemShut {NoStop}%
\bibitem [{\citenamefont {Ali}\ \emph {et~al.}(2017)\citenamefont {Ali}, \citenamefont {Lange},\ and\ \citenamefont {Stone}}]{Ali:2017jda}%
  \BibitemOpen
  \bibfield  {author} {\bibinfo {author} {\bibfnamefont {A.}~\bibnamefont {Ali}}, \bibinfo {author} {\bibfnamefont {J.~S.}\ \bibnamefont {Lange}},\ and\ \bibinfo {author} {\bibfnamefont {S.}~\bibnamefont {Stone}},\ }\bibfield  {title} {\bibinfo {title} {{Exotics: Heavy Pentaquarks and Tetraquarks}},\ }\href {https://doi.org/10.1016/j.ppnp.2017.08.003} {\bibfield  {journal} {\bibinfo  {journal} {Prog. Part. Nucl. Phys.}\ }\textbf {\bibinfo {volume} {97}},\ \bibinfo {pages} {123} (\bibinfo {year} {2017})},\ \Eprint {https://arxiv.org/abs/1706.00610} {arXiv:1706.00610 [hep-ph]} \BibitemShut {NoStop}%
\bibitem [{\citenamefont {Guo}\ \emph {et~al.}(2018)\citenamefont {Guo}, \citenamefont {Hanhart}, \citenamefont {Mei{\ss}ner}, \citenamefont {Wang}, \citenamefont {Zhao},\ and\ \citenamefont {Zou}}]{Guo:2017jvc}%
  \BibitemOpen
  \bibfield  {author} {\bibinfo {author} {\bibfnamefont {F.-K.}\ \bibnamefont {Guo}}, \bibinfo {author} {\bibfnamefont {C.}~\bibnamefont {Hanhart}}, \bibinfo {author} {\bibfnamefont {U.-G.}\ \bibnamefont {Mei{\ss}ner}}, \bibinfo {author} {\bibfnamefont {Q.}~\bibnamefont {Wang}}, \bibinfo {author} {\bibfnamefont {Q.}~\bibnamefont {Zhao}},\ and\ \bibinfo {author} {\bibfnamefont {B.-S.}\ \bibnamefont {Zou}},\ }\bibfield  {title} {\bibinfo {title} {{Hadronic molecules}},\ }\href {https://doi.org/10.1103/RevModPhys.90.015004} {\bibfield  {journal} {\bibinfo  {journal} {Rev. Mod. Phys.}\ }\textbf {\bibinfo {volume} {90}},\ \bibinfo {pages} {015004} (\bibinfo {year} {2018})},\ \bibinfo {note} {[Erratum: Rev.Mod.Phys. 94, 029901 (2022)]},\ \Eprint {https://arxiv.org/abs/1705.00141} {arXiv:1705.00141 [hep-ph]} \BibitemShut {NoStop}%
\bibitem [{\citenamefont {Olsen}\ \emph {et~al.}(2018)\citenamefont {Olsen}, \citenamefont {Skwarnicki},\ and\ \citenamefont {Zieminska}}]{Olsen:2017bmm}%
  \BibitemOpen
  \bibfield  {author} {\bibinfo {author} {\bibfnamefont {S.~L.}\ \bibnamefont {Olsen}}, \bibinfo {author} {\bibfnamefont {T.}~\bibnamefont {Skwarnicki}},\ and\ \bibinfo {author} {\bibfnamefont {D.}~\bibnamefont {Zieminska}},\ }\bibfield  {title} {\bibinfo {title} {{Nonstandard heavy mesons and baryons: Experimental evidence}},\ }\href {https://doi.org/10.1103/RevModPhys.90.015003} {\bibfield  {journal} {\bibinfo  {journal} {Rev. Mod. Phys.}\ }\textbf {\bibinfo {volume} {90}},\ \bibinfo {pages} {015003} (\bibinfo {year} {2018})},\ \Eprint {https://arxiv.org/abs/1708.04012} {arXiv:1708.04012 [hep-ph]} \BibitemShut {NoStop}%
\bibitem [{\citenamefont {Karliner}\ \emph {et~al.}(2018)\citenamefont {Karliner}, \citenamefont {Rosner},\ and\ \citenamefont {Skwarnicki}}]{Karliner:2017qhf}%
  \BibitemOpen
  \bibfield  {author} {\bibinfo {author} {\bibfnamefont {M.}~\bibnamefont {Karliner}}, \bibinfo {author} {\bibfnamefont {J.~L.}\ \bibnamefont {Rosner}},\ and\ \bibinfo {author} {\bibfnamefont {T.}~\bibnamefont {Skwarnicki}},\ }\bibfield  {title} {\bibinfo {title} {{Multiquark States}},\ }\href {https://doi.org/10.1146/annurev-nucl-101917-020902} {\bibfield  {journal} {\bibinfo  {journal} {Ann. Rev. Nucl. Part. Sci.}\ }\textbf {\bibinfo {volume} {68}},\ \bibinfo {pages} {17} (\bibinfo {year} {2018})},\ \Eprint {https://arxiv.org/abs/1711.10626} {arXiv:1711.10626 [hep-ph]} \BibitemShut {NoStop}%
\bibitem [{\citenamefont {Liu}\ \emph {et~al.}(2019{\natexlab{a}})\citenamefont {Liu}, \citenamefont {Chen}, \citenamefont {Chen}, \citenamefont {Liu},\ and\ \citenamefont {Zhu}}]{Liu:2019zoy}%
  \BibitemOpen
  \bibfield  {author} {\bibinfo {author} {\bibfnamefont {Y.-R.}\ \bibnamefont {Liu}}, \bibinfo {author} {\bibfnamefont {H.-X.}\ \bibnamefont {Chen}}, \bibinfo {author} {\bibfnamefont {W.}~\bibnamefont {Chen}}, \bibinfo {author} {\bibfnamefont {X.}~\bibnamefont {Liu}},\ and\ \bibinfo {author} {\bibfnamefont {S.-L.}\ \bibnamefont {Zhu}},\ }\bibfield  {title} {\bibinfo {title} {{Pentaquark and Tetraquark states}},\ }\href {https://doi.org/10.1016/j.ppnp.2019.04.003} {\bibfield  {journal} {\bibinfo  {journal} {Prog. Part. Nucl. Phys.}\ }\textbf {\bibinfo {volume} {107}},\ \bibinfo {pages} {237} (\bibinfo {year} {2019}{\natexlab{a}})},\ \Eprint {https://arxiv.org/abs/1903.11976} {arXiv:1903.11976 [hep-ph]} \BibitemShut {NoStop}%
\bibitem [{\citenamefont {Brambilla}\ \emph {et~al.}(2020)\citenamefont {Brambilla}, \citenamefont {Eidelman}, \citenamefont {Hanhart}, \citenamefont {Nefediev}, \citenamefont {Shen}, \citenamefont {Thomas}, \citenamefont {Vairo},\ and\ \citenamefont {Yuan}}]{Brambilla:2019esw}%
  \BibitemOpen
  \bibfield  {author} {\bibinfo {author} {\bibfnamefont {N.}~\bibnamefont {Brambilla}}, \bibinfo {author} {\bibfnamefont {S.}~\bibnamefont {Eidelman}}, \bibinfo {author} {\bibfnamefont {C.}~\bibnamefont {Hanhart}}, \bibinfo {author} {\bibfnamefont {A.}~\bibnamefont {Nefediev}}, \bibinfo {author} {\bibfnamefont {C.-P.}\ \bibnamefont {Shen}}, \bibinfo {author} {\bibfnamefont {C.~E.}\ \bibnamefont {Thomas}}, \bibinfo {author} {\bibfnamefont {A.}~\bibnamefont {Vairo}},\ and\ \bibinfo {author} {\bibfnamefont {C.-Z.}\ \bibnamefont {Yuan}},\ }\bibfield  {title} {\bibinfo {title} {{The $XYZ$ states: experimental and theoretical status and perspectives}},\ }\href {https://doi.org/10.1016/j.physrep.2020.05.001} {\bibfield  {journal} {\bibinfo  {journal} {Phys. Rept.}\ }\textbf {\bibinfo {volume} {873}},\ \bibinfo {pages} {1} (\bibinfo {year} {2020})},\ \Eprint {https://arxiv.org/abs/1907.07583} {arXiv:1907.07583 [hep-ex]} \BibitemShut {NoStop}%
\bibitem [{\citenamefont {Chen}\ \emph {et~al.}(2023)\citenamefont {Chen}, \citenamefont {Chen}, \citenamefont {Liu}, \citenamefont {Liu},\ and\ \citenamefont {Zhu}}]{Chen:2022asf}%
  \BibitemOpen
  \bibfield  {author} {\bibinfo {author} {\bibfnamefont {H.-X.}\ \bibnamefont {Chen}}, \bibinfo {author} {\bibfnamefont {W.}~\bibnamefont {Chen}}, \bibinfo {author} {\bibfnamefont {X.}~\bibnamefont {Liu}}, \bibinfo {author} {\bibfnamefont {Y.-R.}\ \bibnamefont {Liu}},\ and\ \bibinfo {author} {\bibfnamefont {S.-L.}\ \bibnamefont {Zhu}},\ }\bibfield  {title} {\bibinfo {title} {{An updated review of the new hadron states}},\ }\href {https://doi.org/10.1088/1361-6633/aca3b6} {\bibfield  {journal} {\bibinfo  {journal} {Rept. Prog. Phys.}\ }\textbf {\bibinfo {volume} {86}},\ \bibinfo {pages} {026201} (\bibinfo {year} {2023})},\ \Eprint {https://arxiv.org/abs/2204.02649} {arXiv:2204.02649 [hep-ph]} \BibitemShut {NoStop}%
\bibitem [{\citenamefont {Meng}\ \emph {et~al.}(2023{\natexlab{a}})\citenamefont {Meng}, \citenamefont {Wang}, \citenamefont {Wang},\ and\ \citenamefont {Zhu}}]{Meng:2022ozq}%
  \BibitemOpen
  \bibfield  {author} {\bibinfo {author} {\bibfnamefont {L.}~\bibnamefont {Meng}}, \bibinfo {author} {\bibfnamefont {B.}~\bibnamefont {Wang}}, \bibinfo {author} {\bibfnamefont {G.-J.}\ \bibnamefont {Wang}},\ and\ \bibinfo {author} {\bibfnamefont {S.-L.}\ \bibnamefont {Zhu}},\ }\bibfield  {title} {\bibinfo {title} {{Chiral perturbation theory for heavy hadrons and chiral effective field theory for heavy hadronic molecules}},\ }\href {https://doi.org/10.1016/j.physrep.2023.04.003} {\bibfield  {journal} {\bibinfo  {journal} {Phys. Rept.}\ }\textbf {\bibinfo {volume} {1019}},\ \bibinfo {pages} {1} (\bibinfo {year} {2023}{\natexlab{a}})},\ \Eprint {https://arxiv.org/abs/2204.08716} {arXiv:2204.08716 [hep-ph]} \BibitemShut {NoStop}%
\bibitem [{\citenamefont {Wu}\ \emph {et~al.}(2022{\natexlab{a}})\citenamefont {Wu}, \citenamefont {Pan}, \citenamefont {Liu},\ and\ \citenamefont {Geng}}]{Wu:2022ftm}%
  \BibitemOpen
  \bibfield  {author} {\bibinfo {author} {\bibfnamefont {T.-W.}\ \bibnamefont {Wu}}, \bibinfo {author} {\bibfnamefont {Y.-W.}\ \bibnamefont {Pan}}, \bibinfo {author} {\bibfnamefont {M.-Z.}\ \bibnamefont {Liu}},\ and\ \bibinfo {author} {\bibfnamefont {L.-S.}\ \bibnamefont {Geng}},\ }\bibfield  {title} {\bibinfo {title} {{Multi-hadron molecules: status and prospect}},\ }\href {https://doi.org/10.1016/j.scib.2022.08.007} {\bibfield  {journal} {\bibinfo  {journal} {Sci. Bull.}\ }\textbf {\bibinfo {volume} {67}},\ \bibinfo {pages} {1735} (\bibinfo {year} {2022}{\natexlab{a}})},\ \Eprint {https://arxiv.org/abs/2208.00882} {arXiv:2208.00882 [hep-ph]} \BibitemShut {NoStop}%
\bibitem [{\citenamefont {Choi}\ \emph {et~al.}(2003)\citenamefont {Choi} \emph {et~al.}}]{Belle:2003nnu}%
  \BibitemOpen
  \bibfield  {author} {\bibinfo {author} {\bibfnamefont {S.~K.}\ \bibnamefont {Choi}} \emph {et~al.} (\bibinfo {collaboration} {Belle}),\ }\bibfield  {title} {\bibinfo {title} {{Observation of a narrow charmonium-like state in exclusive $B^\pm \to K^\pm \pi^+ \pi^- J/\psi$ decays}},\ }\href {https://doi.org/10.1103/PhysRevLett.91.262001} {\bibfield  {journal} {\bibinfo  {journal} {Phys. Rev. Lett.}\ }\textbf {\bibinfo {volume} {91}},\ \bibinfo {pages} {262001} (\bibinfo {year} {2003})},\ \Eprint {https://arxiv.org/abs/hep-ex/0309032} {arXiv:hep-ex/0309032} \BibitemShut {NoStop}%
\bibitem [{\citenamefont {Ablikim}\ \emph {et~al.}(2013)\citenamefont {Ablikim} \emph {et~al.}}]{BESIII:2013ris}%
  \BibitemOpen
  \bibfield  {author} {\bibinfo {author} {\bibfnamefont {M.}~\bibnamefont {Ablikim}} \emph {et~al.} (\bibinfo {collaboration} {BESIII}),\ }\bibfield  {title} {\bibinfo {title} {{Observation of a Charged Charmoniumlike Structure in $e^+e^- \to \pi^+\pi^- J/\psi$ at $\sqrt{s}$ =4.26 GeV}},\ }\href {https://doi.org/10.1103/PhysRevLett.110.252001} {\bibfield  {journal} {\bibinfo  {journal} {Phys. Rev. Lett.}\ }\textbf {\bibinfo {volume} {110}},\ \bibinfo {pages} {252001} (\bibinfo {year} {2013})},\ \Eprint {https://arxiv.org/abs/1303.5949} {arXiv:1303.5949 [hep-ex]} \BibitemShut {NoStop}%
\bibitem [{\citenamefont {Liu}\ \emph {et~al.}(2013)\citenamefont {Liu} \emph {et~al.}}]{Belle:2013yex}%
  \BibitemOpen
  \bibfield  {author} {\bibinfo {author} {\bibfnamefont {Z.~Q.}\ \bibnamefont {Liu}} \emph {et~al.} (\bibinfo {collaboration} {Belle}),\ }\bibfield  {title} {\bibinfo {title} {{Study of $e^+e^- \to \pi^+ \pi^- J/\psi$ and Observation of a Charged Charmoniumlike State at Belle}},\ }\href {https://doi.org/10.1103/PhysRevLett.110.252002} {\bibfield  {journal} {\bibinfo  {journal} {Phys. Rev. Lett.}\ }\textbf {\bibinfo {volume} {110}},\ \bibinfo {pages} {252002} (\bibinfo {year} {2013})},\ \bibinfo {note} {[Erratum: Phys.Rev.Lett. 111, 019901 (2013)]},\ \Eprint {https://arxiv.org/abs/1304.0121} {arXiv:1304.0121 [hep-ex]} \BibitemShut {NoStop}%
\bibitem [{\citenamefont {Aaij}\ \emph {et~al.}(2020{\natexlab{a}})\citenamefont {Aaij} \emph {et~al.}}]{LHCb:2020bls}%
  \BibitemOpen
  \bibfield  {author} {\bibinfo {author} {\bibfnamefont {R.}~\bibnamefont {Aaij}} \emph {et~al.} (\bibinfo {collaboration} {LHCb}),\ }\bibfield  {title} {\bibinfo {title} {{A model-independent study of resonant structure in $B^+\to D^+D^-K^+$ decays}},\ }\href {https://doi.org/10.1103/PhysRevLett.125.242001} {\bibfield  {journal} {\bibinfo  {journal} {Phys. Rev. Lett.}\ }\textbf {\bibinfo {volume} {125}},\ \bibinfo {pages} {242001} (\bibinfo {year} {2020}{\natexlab{a}})},\ \Eprint {https://arxiv.org/abs/2009.00025} {arXiv:2009.00025 [hep-ex]} \BibitemShut {NoStop}%
\bibitem [{\citenamefont {Aaij}\ \emph {et~al.}(2020{\natexlab{b}})\citenamefont {Aaij} \emph {et~al.}}]{LHCb:2020pxc}%
  \BibitemOpen
  \bibfield  {author} {\bibinfo {author} {\bibfnamefont {R.}~\bibnamefont {Aaij}} \emph {et~al.} (\bibinfo {collaboration} {LHCb}),\ }\bibfield  {title} {\bibinfo {title} {{Amplitude analysis of the $B^+\to D^+D^-K^+$ decay}},\ }\href {https://doi.org/10.1103/PhysRevD.102.112003} {\bibfield  {journal} {\bibinfo  {journal} {Phys. Rev. D}\ }\textbf {\bibinfo {volume} {102}},\ \bibinfo {pages} {112003} (\bibinfo {year} {2020}{\natexlab{b}})},\ \Eprint {https://arxiv.org/abs/2009.00026} {arXiv:2009.00026 [hep-ex]} \BibitemShut {NoStop}%
\bibitem [{\citenamefont {Aaij}\ \emph {et~al.}(2024)\citenamefont {Aaij} \emph {et~al.}}]{LHCb:2024vfz}%
  \BibitemOpen
  \bibfield  {author} {\bibinfo {author} {\bibfnamefont {R.}~\bibnamefont {Aaij}} \emph {et~al.} (\bibinfo {collaboration} {LHCb}),\ }\bibfield  {title} {\bibinfo {title} {{Observation of New Charmonium or Charmoniumlike States in B+{\textrightarrow}D*{\ensuremath{\pm}}D{\ensuremath{\mp}}K+ Decays}},\ }\href {https://doi.org/10.1103/PhysRevLett.133.131902} {\bibfield  {journal} {\bibinfo  {journal} {Phys. Rev. Lett.}\ }\textbf {\bibinfo {volume} {133}},\ \bibinfo {pages} {131902} (\bibinfo {year} {2024})},\ \Eprint {https://arxiv.org/abs/2406.03156} {arXiv:2406.03156 [hep-ex]} \BibitemShut {NoStop}%
\bibitem [{\citenamefont {Aaij}\ \emph {et~al.}(2022{\natexlab{a}})\citenamefont {Aaij} \emph {et~al.}}]{LHCb:2021vvq}%
  \BibitemOpen
  \bibfield  {author} {\bibinfo {author} {\bibfnamefont {R.}~\bibnamefont {Aaij}} \emph {et~al.} (\bibinfo {collaboration} {LHCb}),\ }\bibfield  {title} {\bibinfo {title} {{Observation of an exotic narrow doubly charmed tetraquark}},\ }\href {https://doi.org/10.1038/s41567-022-01614-y} {\bibfield  {journal} {\bibinfo  {journal} {Nature Phys.}\ }\textbf {\bibinfo {volume} {18}},\ \bibinfo {pages} {751} (\bibinfo {year} {2022}{\natexlab{a}})},\ \Eprint {https://arxiv.org/abs/2109.01038} {arXiv:2109.01038 [hep-ex]} \BibitemShut {NoStop}%
\bibitem [{\citenamefont {Aaij}\ \emph {et~al.}(2022{\natexlab{b}})\citenamefont {Aaij} \emph {et~al.}}]{LHCb:2021auc}%
  \BibitemOpen
  \bibfield  {author} {\bibinfo {author} {\bibfnamefont {R.}~\bibnamefont {Aaij}} \emph {et~al.} (\bibinfo {collaboration} {LHCb}),\ }\bibfield  {title} {\bibinfo {title} {{Study of the doubly charmed tetraquark $T_{cc}^{+}$}},\ }\href {https://doi.org/10.1038/s41467-022-30206-w} {\bibfield  {journal} {\bibinfo  {journal} {Nature Commun.}\ }\textbf {\bibinfo {volume} {13}},\ \bibinfo {pages} {3351} (\bibinfo {year} {2022}{\natexlab{b}})},\ \Eprint {https://arxiv.org/abs/2109.01056} {arXiv:2109.01056 [hep-ex]} \BibitemShut {NoStop}%
\bibitem [{\citenamefont {Shen}\ and\ \citenamefont {Xie}(2023)}]{Shen:2022etd}%
  \BibitemOpen
  \bibfield  {author} {\bibinfo {author} {\bibfnamefont {Q.-H.}\ \bibnamefont {Shen}}\ and\ \bibinfo {author} {\bibfnamefont {J.-J.}\ \bibnamefont {Xie}},\ }\bibfield  {title} {\bibinfo {title} {{Faddeev fixed-center approximation to the {\ensuremath{\eta}}K*K{\textasciimacron}*, {\ensuremath{\pi}}K*K{\textasciimacron}*, and KK*K{\textasciimacron}* systems}},\ }\href {https://doi.org/10.1103/PhysRevD.107.034019} {\bibfield  {journal} {\bibinfo  {journal} {Phys. Rev. D}\ }\textbf {\bibinfo {volume} {107}},\ \bibinfo {pages} {034019} (\bibinfo {year} {2023})},\ \Eprint {https://arxiv.org/abs/2211.04911} {arXiv:2211.04911 [hep-ph]} \BibitemShut {NoStop}%
\bibitem [{\citenamefont {Zhang}\ \emph {et~al.}(2022)\citenamefont {Zhang}, \citenamefont {Hanhart}, \citenamefont {Mei{\ss}ner},\ and\ \citenamefont {Xie}}]{Zhang:2021hcl}%
  \BibitemOpen
  \bibfield  {author} {\bibinfo {author} {\bibfnamefont {X.}~\bibnamefont {Zhang}}, \bibinfo {author} {\bibfnamefont {C.}~\bibnamefont {Hanhart}}, \bibinfo {author} {\bibfnamefont {U.-G.}\ \bibnamefont {Mei{\ss}ner}},\ and\ \bibinfo {author} {\bibfnamefont {J.-J.}\ \bibnamefont {Xie}},\ }\bibfield  {title} {\bibinfo {title} {{Remarks on non-perturbative three{\textendash}body dynamics and its application to the $KK{\bar{K}}$ system}},\ }\href {https://doi.org/10.1140/epja/s10050-021-00661-y} {\bibfield  {journal} {\bibinfo  {journal} {Eur. Phys. J. A}\ }\textbf {\bibinfo {volume} {58}},\ \bibinfo {pages} {20} (\bibinfo {year} {2022})},\ \Eprint {https://arxiv.org/abs/2107.03168} {arXiv:2107.03168 [hep-ph]} \BibitemShut {NoStop}%
\bibitem [{\citenamefont {Wu}\ \emph {et~al.}(2022{\natexlab{b}})\citenamefont {Wu}, \citenamefont {Pan}, \citenamefont {Liu}, \citenamefont {Luo}, \citenamefont {Geng},\ and\ \citenamefont {Liu}}]{Wu:2021kbu}%
  \BibitemOpen
  \bibfield  {author} {\bibinfo {author} {\bibfnamefont {T.-W.}\ \bibnamefont {Wu}}, \bibinfo {author} {\bibfnamefont {Y.-W.}\ \bibnamefont {Pan}}, \bibinfo {author} {\bibfnamefont {M.-Z.}\ \bibnamefont {Liu}}, \bibinfo {author} {\bibfnamefont {S.-Q.}\ \bibnamefont {Luo}}, \bibinfo {author} {\bibfnamefont {L.-S.}\ \bibnamefont {Geng}},\ and\ \bibinfo {author} {\bibfnamefont {X.}~\bibnamefont {Liu}},\ }\bibfield  {title} {\bibinfo {title} {{Discovery of the doubly charmed Tcc+ state implies a triply charmed Hccc hexaquark state}},\ }\href {https://doi.org/10.1103/PhysRevD.105.L031505} {\bibfield  {journal} {\bibinfo  {journal} {Phys. Rev. D}\ }\textbf {\bibinfo {volume} {105}},\ \bibinfo {pages} {L031505} (\bibinfo {year} {2022}{\natexlab{b}})},\ \Eprint {https://arxiv.org/abs/2108.00923} {arXiv:2108.00923 [hep-ph]} \BibitemShut {NoStop}%
\bibitem [{\citenamefont {Luo}\ \emph {et~al.}(2022)\citenamefont {Luo}, \citenamefont {Wu}, \citenamefont {Liu}, \citenamefont {Geng},\ and\ \citenamefont {Liu}}]{Luo:2021ggs}%
  \BibitemOpen
  \bibfield  {author} {\bibinfo {author} {\bibfnamefont {S.-Q.}\ \bibnamefont {Luo}}, \bibinfo {author} {\bibfnamefont {T.-W.}\ \bibnamefont {Wu}}, \bibinfo {author} {\bibfnamefont {M.-Z.}\ \bibnamefont {Liu}}, \bibinfo {author} {\bibfnamefont {L.-S.}\ \bibnamefont {Geng}},\ and\ \bibinfo {author} {\bibfnamefont {X.}~\bibnamefont {Liu}},\ }\bibfield  {title} {\bibinfo {title} {{Triple-charm molecular states composed of D*D*D and D*D*D*}},\ }\href {https://doi.org/10.1103/PhysRevD.105.074033} {\bibfield  {journal} {\bibinfo  {journal} {Phys. Rev. D}\ }\textbf {\bibinfo {volume} {105}},\ \bibinfo {pages} {074033} (\bibinfo {year} {2022})},\ \Eprint {https://arxiv.org/abs/2111.15079} {arXiv:2111.15079 [hep-ph]} \BibitemShut {NoStop}%
\bibitem [{\citenamefont {Bayar}\ \emph {et~al.}(2023)\citenamefont {Bayar}, \citenamefont {Martinez~Torres}, \citenamefont {Khemchandani}, \citenamefont {Molina},\ and\ \citenamefont {Oset}}]{Bayar:2022bnc}%
  \BibitemOpen
  \bibfield  {author} {\bibinfo {author} {\bibfnamefont {M.}~\bibnamefont {Bayar}}, \bibinfo {author} {\bibfnamefont {A.}~\bibnamefont {Martinez~Torres}}, \bibinfo {author} {\bibfnamefont {K.~P.}\ \bibnamefont {Khemchandani}}, \bibinfo {author} {\bibfnamefont {R.}~\bibnamefont {Molina}},\ and\ \bibinfo {author} {\bibfnamefont {E.}~\bibnamefont {Oset}},\ }\bibfield  {title} {\bibinfo {title} {{Exotic states with triple charm}},\ }\href {https://doi.org/10.1140/epjc/s10052-023-11207-5} {\bibfield  {journal} {\bibinfo  {journal} {Eur. Phys. J. C}\ }\textbf {\bibinfo {volume} {83}},\ \bibinfo {pages} {46} (\bibinfo {year} {2023})},\ \Eprint {https://arxiv.org/abs/2211.09294} {arXiv:2211.09294 [hep-ph]} \BibitemShut {NoStop}%
\bibitem [{\citenamefont {Pan}\ \emph {et~al.}(2022)\citenamefont {Pan}, \citenamefont {Wu}, \citenamefont {Liu},\ and\ \citenamefont {Geng}}]{Pan:2022whr}%
  \BibitemOpen
  \bibfield  {author} {\bibinfo {author} {\bibfnamefont {Y.-W.}\ \bibnamefont {Pan}}, \bibinfo {author} {\bibfnamefont {T.-W.}\ \bibnamefont {Wu}}, \bibinfo {author} {\bibfnamefont {M.-Z.}\ \bibnamefont {Liu}},\ and\ \bibinfo {author} {\bibfnamefont {L.-S.}\ \bibnamefont {Geng}},\ }\bibfield  {title} {\bibinfo {title} {{Hadronic molecules composed of a doubly charmed tetraquark state and a charmed meson}},\ }\href {https://doi.org/10.1140/epjc/s10052-022-10881-1} {\bibfield  {journal} {\bibinfo  {journal} {Eur. Phys. J. C}\ }\textbf {\bibinfo {volume} {82}},\ \bibinfo {pages} {908} (\bibinfo {year} {2022})},\ \Eprint {https://arxiv.org/abs/2208.05385} {arXiv:2208.05385 [hep-ph]} \BibitemShut {NoStop}%
\bibitem [{\citenamefont {Ortega}(2024)}]{Ortega:2024ecy}%
  \BibitemOpen
  \bibfield  {author} {\bibinfo {author} {\bibfnamefont {P.~G.}\ \bibnamefont {Ortega}},\ }\bibfield  {title} {\bibinfo {title} {{Exploring the Efimov effect in the D*D*D* system}},\ }\href {https://doi.org/10.1103/PhysRevD.110.034015} {\bibfield  {journal} {\bibinfo  {journal} {Phys. Rev. D}\ }\textbf {\bibinfo {volume} {110}},\ \bibinfo {pages} {034015} (\bibinfo {year} {2024})},\ \Eprint {https://arxiv.org/abs/2403.10244} {arXiv:2403.10244 [hep-ph]} \BibitemShut {NoStop}%
\bibitem [{\citenamefont {Fu}\ \emph {et~al.}(2025)\citenamefont {Fu}, \citenamefont {Lin}, \citenamefont {Guo}, \citenamefont {Hammer}, \citenamefont {Mei{\ss}ner}, \citenamefont {Rusetsky},\ and\ \citenamefont {Zhang}}]{Fu:2025joa}%
  \BibitemOpen
  \bibfield  {author} {\bibinfo {author} {\bibfnamefont {H.-L.}\ \bibnamefont {Fu}}, \bibinfo {author} {\bibfnamefont {Y.-H.}\ \bibnamefont {Lin}}, \bibinfo {author} {\bibfnamefont {F.-K.}\ \bibnamefont {Guo}}, \bibinfo {author} {\bibfnamefont {H.-W.}\ \bibnamefont {Hammer}}, \bibinfo {author} {\bibfnamefont {U.-G.}\ \bibnamefont {Mei{\ss}ner}}, \bibinfo {author} {\bibfnamefont {A.}~\bibnamefont {Rusetsky}},\ and\ \bibinfo {author} {\bibfnamefont {X.}~\bibnamefont {Zhang}},\ }\bibfield  {title} {\bibinfo {title} {{Exploring Efimov states in D$^{*}$D$^{*}$D$^{*}$ and DD$^{*}$D$^{*}$ three-body systems}},\ }\href {https://doi.org/10.1007/JHEP07(2025)081} {\bibfield  {journal} {\bibinfo  {journal} {JHEP}\ }\textbf {\bibinfo {volume} {07}},\ \bibinfo {pages} {081}},\ \Eprint {https://arxiv.org/abs/2503.19709} {arXiv:2503.19709 [hep-ph]} \BibitemShut {NoStop}%
\bibitem [{\citenamefont {Tan}\ \emph {et~al.}(2024)\citenamefont {Tan}, \citenamefont {Liu}, \citenamefont {Chen}, \citenamefont {Yang}, \citenamefont {Huang},\ and\ \citenamefont {Ping}}]{Tan:2024omp}%
  \BibitemOpen
  \bibfield  {author} {\bibinfo {author} {\bibfnamefont {Y.}~\bibnamefont {Tan}}, \bibinfo {author} {\bibfnamefont {X.}~\bibnamefont {Liu}}, \bibinfo {author} {\bibfnamefont {X.}~\bibnamefont {Chen}}, \bibinfo {author} {\bibfnamefont {Y.}~\bibnamefont {Yang}}, \bibinfo {author} {\bibfnamefont {H.}~\bibnamefont {Huang}},\ and\ \bibinfo {author} {\bibfnamefont {J.}~\bibnamefont {Ping}},\ }\bibfield  {title} {\bibinfo {title} {{Dynamical study of D*DK and D*DD{\textasciimacron} systems at quark level}},\ }\href {https://doi.org/10.1103/PhysRevD.110.016005} {\bibfield  {journal} {\bibinfo  {journal} {Phys. Rev. D}\ }\textbf {\bibinfo {volume} {110}},\ \bibinfo {pages} {016005} (\bibinfo {year} {2024})},\ \Eprint {https://arxiv.org/abs/2404.02048} {arXiv:2404.02048 [hep-ph]} \BibitemShut {NoStop}%
\bibitem [{\citenamefont {Valderrama}(2018)}]{Valderrama:2018sap}%
  \BibitemOpen
  \bibfield  {author} {\bibinfo {author} {\bibfnamefont {M.~P.}\ \bibnamefont {Valderrama}},\ }\bibfield  {title} {\bibinfo {title} {{$D^* D^* \bar{D}$ and $D^* D^* \bar{D}^*$ three-body systems}},\ }\href {https://doi.org/10.1103/PhysRevD.98.034017} {\bibfield  {journal} {\bibinfo  {journal} {Phys. Rev. D}\ }\textbf {\bibinfo {volume} {98}},\ \bibinfo {pages} {034017} (\bibinfo {year} {2018})},\ \Eprint {https://arxiv.org/abs/1805.10584} {arXiv:1805.10584 [hep-ph]} \BibitemShut {NoStop}%
\bibitem [{\citenamefont {Dias}\ \emph {et~al.}(2018)\citenamefont {Dias}, \citenamefont {Roca},\ and\ \citenamefont {Sakai}}]{Dias:2018iuy}%
  \BibitemOpen
  \bibfield  {author} {\bibinfo {author} {\bibfnamefont {J.~M.}\ \bibnamefont {Dias}}, \bibinfo {author} {\bibfnamefont {L.}~\bibnamefont {Roca}},\ and\ \bibinfo {author} {\bibfnamefont {S.}~\bibnamefont {Sakai}},\ }\bibfield  {title} {\bibinfo {title} {{Prediction of new states from $D^{(*)}B^{(*)}\bar{B}^{(*)}$ three-body interactions}},\ }\href {https://doi.org/10.1103/PhysRevD.97.056019} {\bibfield  {journal} {\bibinfo  {journal} {Phys. Rev. D}\ }\textbf {\bibinfo {volume} {97}},\ \bibinfo {pages} {056019} (\bibinfo {year} {2018})},\ \Eprint {https://arxiv.org/abs/1801.03504} {arXiv:1801.03504 [hep-ph]} \BibitemShut {NoStop}%
\bibitem [{\citenamefont {Ma}\ \emph {et~al.}(2019{\natexlab{a}})\citenamefont {Ma}, \citenamefont {Wang},\ and\ \citenamefont {Mei{\ss}ner}}]{Ma:2018vhp}%
  \BibitemOpen
  \bibfield  {author} {\bibinfo {author} {\bibfnamefont {L.}~\bibnamefont {Ma}}, \bibinfo {author} {\bibfnamefont {Q.}~\bibnamefont {Wang}},\ and\ \bibinfo {author} {\bibfnamefont {U.-G.}\ \bibnamefont {Mei{\ss}ner}},\ }\bibfield  {title} {\bibinfo {title} {{Trimeson bound state BBB* via a delocalized {\ensuremath{\pi}} bond}},\ }\href {https://doi.org/10.1103/PhysRevD.100.014028} {\bibfield  {journal} {\bibinfo  {journal} {Phys. Rev. D}\ }\textbf {\bibinfo {volume} {100}},\ \bibinfo {pages} {014028} (\bibinfo {year} {2019}{\natexlab{a}})},\ \Eprint {https://arxiv.org/abs/1812.09750} {arXiv:1812.09750 [hep-ph]} \BibitemShut {NoStop}%
\bibitem [{\citenamefont {Ma}\ \emph {et~al.}(2019{\natexlab{b}})\citenamefont {Ma}, \citenamefont {Wang},\ and\ \citenamefont {Mei{\ss}ner}}]{Ma:2017ery}%
  \BibitemOpen
  \bibfield  {author} {\bibinfo {author} {\bibfnamefont {L.}~\bibnamefont {Ma}}, \bibinfo {author} {\bibfnamefont {Q.}~\bibnamefont {Wang}},\ and\ \bibinfo {author} {\bibfnamefont {U.-G.}\ \bibnamefont {Mei{\ss}ner}},\ }\bibfield  {title} {\bibinfo {title} {{Double heavy tri-hadron bound state via delocalized $\pi$ bond}},\ }\href {https://doi.org/10.1088/1674-1137/43/1/014102} {\bibfield  {journal} {\bibinfo  {journal} {Chin. Phys. C}\ }\textbf {\bibinfo {volume} {43}},\ \bibinfo {pages} {014102} (\bibinfo {year} {2019}{\natexlab{b}})},\ \Eprint {https://arxiv.org/abs/1711.06143} {arXiv:1711.06143 [hep-ph]} \BibitemShut {NoStop}%
\bibitem [{\citenamefont {Ren}\ \emph {et~al.}(2018)\citenamefont {Ren}, \citenamefont {Malabarba}, \citenamefont {Geng}, \citenamefont {Khemchandani},\ and\ \citenamefont {Mart{\'\i}nez~Torres}}]{Ren:2018pcd}%
  \BibitemOpen
  \bibfield  {author} {\bibinfo {author} {\bibfnamefont {X.-L.}\ \bibnamefont {Ren}}, \bibinfo {author} {\bibfnamefont {B.~B.}\ \bibnamefont {Malabarba}}, \bibinfo {author} {\bibfnamefont {L.-S.}\ \bibnamefont {Geng}}, \bibinfo {author} {\bibfnamefont {K.~P.}\ \bibnamefont {Khemchandani}},\ and\ \bibinfo {author} {\bibfnamefont {A.}~\bibnamefont {Mart{\'\i}nez~Torres}},\ }\bibfield  {title} {\bibinfo {title} {{$K^*$ mesons with hidden charm arising from $KX(3872)$ and $KZ_c(3900)$ dynamics}},\ }\href {https://doi.org/10.1016/j.physletb.2018.08.034} {\bibfield  {journal} {\bibinfo  {journal} {Phys. Lett. B}\ }\textbf {\bibinfo {volume} {785}},\ \bibinfo {pages} {112} (\bibinfo {year} {2018})},\ \Eprint {https://arxiv.org/abs/1805.08330} {arXiv:1805.08330 [hep-ph]} \BibitemShut {NoStop}%
\bibitem [{\citenamefont {Ren}\ and\ \citenamefont {Sun}(2019)}]{Ren:2018qhr}%
  \BibitemOpen
  \bibfield  {author} {\bibinfo {author} {\bibfnamefont {X.-L.}\ \bibnamefont {Ren}}\ and\ \bibinfo {author} {\bibfnamefont {Z.-F.}\ \bibnamefont {Sun}},\ }\bibfield  {title} {\bibinfo {title} {{Possible bound states with hidden bottom from $\bar{K}^{(*)}B^{(*)}\bar{B}^{(*)}$ systems}},\ }\href {https://doi.org/10.1103/PhysRevD.99.094041} {\bibfield  {journal} {\bibinfo  {journal} {Phys. Rev. D}\ }\textbf {\bibinfo {volume} {99}},\ \bibinfo {pages} {094041} (\bibinfo {year} {2019})},\ \Eprint {https://arxiv.org/abs/1812.09931} {arXiv:1812.09931 [hep-ph]} \BibitemShut {NoStop}%
\bibitem [{\citenamefont {Wu}\ \emph {et~al.}(2019)\citenamefont {Wu}, \citenamefont {Liu}, \citenamefont {Geng}, \citenamefont {Hiyama},\ and\ \citenamefont {Valderrama}}]{Wu:2019vsy}%
  \BibitemOpen
  \bibfield  {author} {\bibinfo {author} {\bibfnamefont {T.-W.}\ \bibnamefont {Wu}}, \bibinfo {author} {\bibfnamefont {M.-Z.}\ \bibnamefont {Liu}}, \bibinfo {author} {\bibfnamefont {L.-S.}\ \bibnamefont {Geng}}, \bibinfo {author} {\bibfnamefont {E.}~\bibnamefont {Hiyama}},\ and\ \bibinfo {author} {\bibfnamefont {M.~P.}\ \bibnamefont {Valderrama}},\ }\bibfield  {title} {\bibinfo {title} {{$DK$, $DDK$, and $DDDK$ molecules{\textendash}understanding the nature of the $D_{s0}^*(2317)$}},\ }\href {https://doi.org/10.1103/PhysRevD.100.034029} {\bibfield  {journal} {\bibinfo  {journal} {Phys. Rev. D}\ }\textbf {\bibinfo {volume} {100}},\ \bibinfo {pages} {034029} (\bibinfo {year} {2019})},\ \Eprint {https://arxiv.org/abs/1906.11995} {arXiv:1906.11995 [hep-ph]} \BibitemShut {NoStop}%
\bibitem [{\citenamefont {Di}\ and\ \citenamefont {Wang}(2019)}]{Di:2019jsx}%
  \BibitemOpen
  \bibfield  {author} {\bibinfo {author} {\bibfnamefont {Z.-Y.}\ \bibnamefont {Di}}\ and\ \bibinfo {author} {\bibfnamefont {Z.-G.}\ \bibnamefont {Wang}},\ }\bibfield  {title} {\bibinfo {title} {{Analysis of the $D\bar{D}^*K$ system with QCD sum rules}},\ }\href {https://doi.org/10.1155/2019/8958079} {\bibfield  {journal} {\bibinfo  {journal} {Adv. High Energy Phys.}\ }\textbf {\bibinfo {volume} {2019}},\ \bibinfo {pages} {8958079} (\bibinfo {year} {2019})},\ \Eprint {https://arxiv.org/abs/1901.05196} {arXiv:1901.05196 [hep-ph]} \BibitemShut {NoStop}%
\bibitem [{\citenamefont {Ikeno}\ \emph {et~al.}(2023)\citenamefont {Ikeno}, \citenamefont {Bayar},\ and\ \citenamefont {Oset}}]{Ikeno:2022jbb}%
  \BibitemOpen
  \bibfield  {author} {\bibinfo {author} {\bibfnamefont {N.}~\bibnamefont {Ikeno}}, \bibinfo {author} {\bibfnamefont {M.}~\bibnamefont {Bayar}},\ and\ \bibinfo {author} {\bibfnamefont {E.}~\bibnamefont {Oset}},\ }\bibfield  {title} {\bibinfo {title} {{Molecular states of D*D*K{\textasciimacron}* nature}},\ }\href {https://doi.org/10.1103/PhysRevD.107.034006} {\bibfield  {journal} {\bibinfo  {journal} {Phys. Rev. D}\ }\textbf {\bibinfo {volume} {107}},\ \bibinfo {pages} {034006} (\bibinfo {year} {2023})},\ \Eprint {https://arxiv.org/abs/2208.03698} {arXiv:2208.03698 [hep-ph]} \BibitemShut {NoStop}%
\bibitem [{\citenamefont {Zhang}\ \emph {et~al.}(2025)\citenamefont {Zhang}, \citenamefont {Hu}, \citenamefont {He}, \citenamefont {Liu}, \citenamefont {Shi}, \citenamefont {Lu},\ and\ \citenamefont {Wang}}]{Zhang:2024yfj}%
  \BibitemOpen
  \bibfield  {author} {\bibinfo {author} {\bibfnamefont {Z.}~\bibnamefont {Zhang}}, \bibinfo {author} {\bibfnamefont {X.-Y.}\ \bibnamefont {Hu}}, \bibinfo {author} {\bibfnamefont {G.}~\bibnamefont {He}}, \bibinfo {author} {\bibfnamefont {J.}~\bibnamefont {Liu}}, \bibinfo {author} {\bibfnamefont {J.-A.}\ \bibnamefont {Shi}}, \bibinfo {author} {\bibfnamefont {B.-N.}\ \bibnamefont {Lu}},\ and\ \bibinfo {author} {\bibfnamefont {Q.}~\bibnamefont {Wang}},\ }\bibfield  {title} {\bibinfo {title} {{Binding of the three-hadron DD*K system from the lattice effective field theory}},\ }\href {https://doi.org/10.1103/PhysRevD.111.036002} {\bibfield  {journal} {\bibinfo  {journal} {Phys. Rev. D}\ }\textbf {\bibinfo {volume} {111}},\ \bibinfo {pages} {036002} (\bibinfo {year} {2025})},\ \Eprint {https://arxiv.org/abs/2409.01325} {arXiv:2409.01325 [hep-ph]} \BibitemShut {NoStop}%
\bibitem [{\citenamefont {Zhai}\ \emph {et~al.}(2025)\citenamefont {Zhai}, \citenamefont {Molina}, \citenamefont {Oset},\ and\ \citenamefont {Geng}}]{Zhai:2024luy}%
  \BibitemOpen
  \bibfield  {author} {\bibinfo {author} {\bibfnamefont {Q.-Y.}\ \bibnamefont {Zhai}}, \bibinfo {author} {\bibfnamefont {R.}~\bibnamefont {Molina}}, \bibinfo {author} {\bibfnamefont {E.}~\bibnamefont {Oset}},\ and\ \bibinfo {author} {\bibfnamefont {L.-S.}\ \bibnamefont {Geng}},\ }\bibfield  {title} {\bibinfo {title} {{Study of the exotic three-body ND*K{\textasciimacron}* system}},\ }\href {https://doi.org/10.1103/PhysRevD.111.034039} {\bibfield  {journal} {\bibinfo  {journal} {Phys. Rev. D}\ }\textbf {\bibinfo {volume} {111}},\ \bibinfo {pages} {034039} (\bibinfo {year} {2025})},\ \Eprint {https://arxiv.org/abs/2411.18285} {arXiv:2411.18285 [hep-ph]} \BibitemShut {NoStop}%
\bibitem [{\citenamefont {Ren}\ \emph {et~al.}(2024)\citenamefont {Ren}, \citenamefont {Khemchandani},\ and\ \citenamefont {Mart{\'\i}nez~Torres}}]{Ren:2024mjh}%
  \BibitemOpen
  \bibfield  {author} {\bibinfo {author} {\bibfnamefont {X.-L.}\ \bibnamefont {Ren}}, \bibinfo {author} {\bibfnamefont {K.~P.}\ \bibnamefont {Khemchandani}},\ and\ \bibinfo {author} {\bibfnamefont {A.}~\bibnamefont {Mart{\'\i}nez~Torres}},\ }\bibfield  {title} {\bibinfo {title} {{Heavy $K^*$ mesons with open charm from $KD^{(*)}D^*$ interactions}},\ }\href {https://doi.org/10.1140/epjc/s10052-024-13683-9} {\bibfield  {journal} {\bibinfo  {journal} {Eur. Phys. J. C}\ }\textbf {\bibinfo {volume} {84}},\ \bibinfo {pages} {1297} (\bibinfo {year} {2024})},\ \Eprint {https://arxiv.org/abs/2409.16281} {arXiv:2409.16281 [hep-ph]} \BibitemShut {NoStop}%
\bibitem [{\citenamefont {Pan}\ \emph {et~al.}(2025)\citenamefont {Pan}, \citenamefont {Lu}, \citenamefont {Hiyama}, \citenamefont {Geng},\ and\ \citenamefont {Hosaka}}]{Pan:2025xvq}%
  \BibitemOpen
  \bibfield  {author} {\bibinfo {author} {\bibfnamefont {Y.-W.}\ \bibnamefont {Pan}}, \bibinfo {author} {\bibfnamefont {J.-X.}\ \bibnamefont {Lu}}, \bibinfo {author} {\bibfnamefont {E.}~\bibnamefont {Hiyama}}, \bibinfo {author} {\bibfnamefont {L.-S.}\ \bibnamefont {Geng}},\ and\ \bibinfo {author} {\bibfnamefont {A.}~\bibnamefont {Hosaka}},\ }\bibfield  {title} {\bibinfo {title} {{Effect of a repulsive three-body interaction on the DD(*)K molecule}},\ }\href {https://doi.org/10.1103/jk8x-qv2p} {\bibfield  {journal} {\bibinfo  {journal} {Phys. Rev. D}\ }\textbf {\bibinfo {volume} {111}},\ \bibinfo {pages} {114006} (\bibinfo {year} {2025})},\ \Eprint {https://arxiv.org/abs/2502.00438} {arXiv:2502.00438 [nucl-th]} \BibitemShut {NoStop}%
\bibitem [{\citenamefont {Canham}\ \emph {et~al.}(2009)\citenamefont {Canham}, \citenamefont {Hammer},\ and\ \citenamefont {Springer}}]{Canham:2009zq}%
  \BibitemOpen
  \bibfield  {author} {\bibinfo {author} {\bibfnamefont {D.~L.}\ \bibnamefont {Canham}}, \bibinfo {author} {\bibfnamefont {H.~W.}\ \bibnamefont {Hammer}},\ and\ \bibinfo {author} {\bibfnamefont {R.~P.}\ \bibnamefont {Springer}},\ }\bibfield  {title} {\bibinfo {title} {{On the scattering of D and D* mesons off the X(3872)}},\ }\href {https://doi.org/10.1103/PhysRevD.80.014009} {\bibfield  {journal} {\bibinfo  {journal} {Phys. Rev. D}\ }\textbf {\bibinfo {volume} {80}},\ \bibinfo {pages} {014009} (\bibinfo {year} {2009})},\ \Eprint {https://arxiv.org/abs/0906.1263} {arXiv:0906.1263 [hep-ph]} \BibitemShut {NoStop}%
\bibitem [{\citenamefont {Lin}\ \emph {et~al.}(2025)\citenamefont {Lin}, \citenamefont {Wilbring}, \citenamefont {Fu}, \citenamefont {Hammer},\ and\ \citenamefont {Mei{\ss}ner}}]{Lin:2017dbo}%
  \BibitemOpen
  \bibfield  {author} {\bibinfo {author} {\bibfnamefont {Y.-H.}\ \bibnamefont {Lin}}, \bibinfo {author} {\bibfnamefont {E.}~\bibnamefont {Wilbring}}, \bibinfo {author} {\bibfnamefont {H.-L.}\ \bibnamefont {Fu}}, \bibinfo {author} {\bibfnamefont {H.-W.}\ \bibnamefont {Hammer}},\ and\ \bibinfo {author} {\bibfnamefont {U.-G.}\ \bibnamefont {Mei{\ss}ner}},\ }\bibfield  {title} {\bibinfo {title} {{Three-body universality in the B meson sector}},\ }\href {https://doi.org/10.1088/1361-6471/ae06be} {\bibfield  {journal} {\bibinfo  {journal} {J. Phys. G}\ }\textbf {\bibinfo {volume} {52}},\ \bibinfo {pages} {105005} (\bibinfo {year} {2025})},\ \Eprint {https://arxiv.org/abs/1705.06176} {arXiv:1705.06176 [hep-ph]} \BibitemShut {NoStop}%
\bibitem [{\citenamefont {Martinez~Torres}\ \emph {et~al.}(2020)\citenamefont {Martinez~Torres}, \citenamefont {Khemchandani}, \citenamefont {Roca},\ and\ \citenamefont {Oset}}]{MartinezTorres:2020hus}%
  \BibitemOpen
  \bibfield  {author} {\bibinfo {author} {\bibfnamefont {A.}~\bibnamefont {Martinez~Torres}}, \bibinfo {author} {\bibfnamefont {K.~P.}\ \bibnamefont {Khemchandani}}, \bibinfo {author} {\bibfnamefont {L.}~\bibnamefont {Roca}},\ and\ \bibinfo {author} {\bibfnamefont {E.}~\bibnamefont {Oset}},\ }\bibfield  {title} {\bibinfo {title} {{Few-body systems consisting of mesons}},\ }\href {https://doi.org/10.1007/s00601-020-01568-y} {\bibfield  {journal} {\bibinfo  {journal} {Few Body Syst.}\ }\textbf {\bibinfo {volume} {61}},\ \bibinfo {pages} {35} (\bibinfo {year} {2020})},\ \Eprint {https://arxiv.org/abs/2005.14357} {arXiv:2005.14357 [nucl-th]} \BibitemShut {NoStop}%
\bibitem [{\citenamefont {Liu}\ \emph {et~al.}(2025)\citenamefont {Liu}, \citenamefont {Pan}, \citenamefont {Liu}, \citenamefont {Wu}, \citenamefont {Lu},\ and\ \citenamefont {Geng}}]{Liu:2024uxn}%
  \BibitemOpen
  \bibfield  {author} {\bibinfo {author} {\bibfnamefont {M.-Z.}\ \bibnamefont {Liu}}, \bibinfo {author} {\bibfnamefont {Y.-W.}\ \bibnamefont {Pan}}, \bibinfo {author} {\bibfnamefont {Z.-W.}\ \bibnamefont {Liu}}, \bibinfo {author} {\bibfnamefont {T.-W.}\ \bibnamefont {Wu}}, \bibinfo {author} {\bibfnamefont {J.-X.}\ \bibnamefont {Lu}},\ and\ \bibinfo {author} {\bibfnamefont {L.-S.}\ \bibnamefont {Geng}},\ }\bibfield  {title} {\bibinfo {title} {{Three ways to decipher the nature of exotic hadrons: Multiplets, three-body hadronic molecules, and correlation functions}},\ }\href {https://doi.org/10.1016/j.physrep.2024.12.001} {\bibfield  {journal} {\bibinfo  {journal} {Phys. Rept.}\ }\textbf {\bibinfo {volume} {1108}},\ \bibinfo {pages} {1} (\bibinfo {year} {2025})},\ \Eprint {https://arxiv.org/abs/2404.06399} {arXiv:2404.06399 [hep-ph]} \BibitemShut {NoStop}%
\bibitem [{\citenamefont {Molina}\ \emph {et~al.}(2010)\citenamefont {Molina}, \citenamefont {Branz},\ and\ \citenamefont {Oset}}]{Molina:2010tx}%
  \BibitemOpen
  \bibfield  {author} {\bibinfo {author} {\bibfnamefont {R.}~\bibnamefont {Molina}}, \bibinfo {author} {\bibfnamefont {T.}~\bibnamefont {Branz}},\ and\ \bibinfo {author} {\bibfnamefont {E.}~\bibnamefont {Oset}},\ }\bibfield  {title} {\bibinfo {title} {{A new interpretation for the $D^*_{s2}(2573)$ and the prediction of novel exotic charmed mesons}},\ }\href {https://doi.org/10.1103/PhysRevD.82.014010} {\bibfield  {journal} {\bibinfo  {journal} {Phys. Rev. D}\ }\textbf {\bibinfo {volume} {82}},\ \bibinfo {pages} {014010} (\bibinfo {year} {2010})},\ \Eprint {https://arxiv.org/abs/1005.0335} {arXiv:1005.0335 [hep-ph]} \BibitemShut {NoStop}%
\bibitem [{\citenamefont {Chen}\ \emph {et~al.}(2020)\citenamefont {Chen}, \citenamefont {Chen}, \citenamefont {Dong},\ and\ \citenamefont {Su}}]{Chen:2020aos}%
  \BibitemOpen
  \bibfield  {author} {\bibinfo {author} {\bibfnamefont {H.-X.}\ \bibnamefont {Chen}}, \bibinfo {author} {\bibfnamefont {W.}~\bibnamefont {Chen}}, \bibinfo {author} {\bibfnamefont {R.-R.}\ \bibnamefont {Dong}},\ and\ \bibinfo {author} {\bibfnamefont {N.}~\bibnamefont {Su}},\ }\bibfield  {title} {\bibinfo {title} {{$X_0$(2900) and $X_1$(2900): Hadronic Molecules or Compact Tetraquarks}},\ }\href {https://doi.org/10.1088/0256-307X/37/10/101201} {\bibfield  {journal} {\bibinfo  {journal} {Chin. Phys. Lett.}\ }\textbf {\bibinfo {volume} {37}},\ \bibinfo {pages} {101201} (\bibinfo {year} {2020})},\ \Eprint {https://arxiv.org/abs/2008.07516} {arXiv:2008.07516 [hep-ph]} \BibitemShut {NoStop}%
\bibitem [{\citenamefont {He}\ and\ \citenamefont {Chen}(2021)}]{He:2020btl}%
  \BibitemOpen
  \bibfield  {author} {\bibinfo {author} {\bibfnamefont {J.}~\bibnamefont {He}}\ and\ \bibinfo {author} {\bibfnamefont {D.-Y.}\ \bibnamefont {Chen}},\ }\bibfield  {title} {\bibinfo {title} {{Molecular picture for $X_0(2900)$ and $X_1(2900)$}},\ }\href {https://doi.org/10.1088/1674-1137/abeda8} {\bibfield  {journal} {\bibinfo  {journal} {Chin. Phys. C}\ }\textbf {\bibinfo {volume} {45}},\ \bibinfo {pages} {063102} (\bibinfo {year} {2021})},\ \Eprint {https://arxiv.org/abs/2008.07782} {arXiv:2008.07782 [hep-ph]} \BibitemShut {NoStop}%
\bibitem [{\citenamefont {Liu}\ \emph {et~al.}(2020)\citenamefont {Liu}, \citenamefont {Xie},\ and\ \citenamefont {Geng}}]{Liu:2020nil}%
  \BibitemOpen
  \bibfield  {author} {\bibinfo {author} {\bibfnamefont {M.-Z.}\ \bibnamefont {Liu}}, \bibinfo {author} {\bibfnamefont {J.-J.}\ \bibnamefont {Xie}},\ and\ \bibinfo {author} {\bibfnamefont {L.-S.}\ \bibnamefont {Geng}},\ }\bibfield  {title} {\bibinfo {title} {{$X_0(2866)$ as a $D^*\bar{K}^*$ molecular state}},\ }\href {https://doi.org/10.1103/PhysRevD.102.091502} {\bibfield  {journal} {\bibinfo  {journal} {Phys. Rev. D}\ }\textbf {\bibinfo {volume} {102}},\ \bibinfo {pages} {091502} (\bibinfo {year} {2020})},\ \Eprint {https://arxiv.org/abs/2008.07389} {arXiv:2008.07389 [hep-ph]} \BibitemShut {NoStop}%
\bibitem [{\citenamefont {Hu}\ \emph {et~al.}(2021)\citenamefont {Hu}, \citenamefont {Lao}, \citenamefont {Ling},\ and\ \citenamefont {Wang}}]{Hu:2020mxp}%
  \BibitemOpen
  \bibfield  {author} {\bibinfo {author} {\bibfnamefont {M.-W.}\ \bibnamefont {Hu}}, \bibinfo {author} {\bibfnamefont {X.-Y.}\ \bibnamefont {Lao}}, \bibinfo {author} {\bibfnamefont {P.}~\bibnamefont {Ling}},\ and\ \bibinfo {author} {\bibfnamefont {Q.}~\bibnamefont {Wang}},\ }\bibfield  {title} {\bibinfo {title} {{$X_0$(2900) and its heavy quark spin partners in molecular picture}},\ }\href {https://doi.org/10.1088/1674-1137/abcfaa} {\bibfield  {journal} {\bibinfo  {journal} {Chin. Phys. C}\ }\textbf {\bibinfo {volume} {45}},\ \bibinfo {pages} {021003} (\bibinfo {year} {2021})},\ \Eprint {https://arxiv.org/abs/2008.06894} {arXiv:2008.06894 [hep-ph]} \BibitemShut {NoStop}%
\bibitem [{\citenamefont {Agaev}\ \emph {et~al.}(2021)\citenamefont {Agaev}, \citenamefont {Azizi},\ and\ \citenamefont {Sundu}}]{Agaev:2020nrc}%
  \BibitemOpen
  \bibfield  {author} {\bibinfo {author} {\bibfnamefont {S.~S.}\ \bibnamefont {Agaev}}, \bibinfo {author} {\bibfnamefont {K.}~\bibnamefont {Azizi}},\ and\ \bibinfo {author} {\bibfnamefont {H.}~\bibnamefont {Sundu}},\ }\bibfield  {title} {\bibinfo {title} {{New scalar resonance X 0(2900) as a molecule: mass and width}},\ }\href {https://doi.org/10.1088/1361-6471/ac0b31} {\bibfield  {journal} {\bibinfo  {journal} {J. Phys. G}\ }\textbf {\bibinfo {volume} {48}},\ \bibinfo {pages} {085012} (\bibinfo {year} {2021})},\ \Eprint {https://arxiv.org/abs/2008.13027} {arXiv:2008.13027 [hep-ph]} \BibitemShut {NoStop}%
\bibitem [{\citenamefont {Wang}\ and\ \citenamefont {Zhu}(2022)}]{Wang:2021lwy}%
  \BibitemOpen
  \bibfield  {author} {\bibinfo {author} {\bibfnamefont {B.}~\bibnamefont {Wang}}\ and\ \bibinfo {author} {\bibfnamefont {S.-L.}\ \bibnamefont {Zhu}},\ }\bibfield  {title} {\bibinfo {title} {{How to understand the X(2900)?}},\ }\href {https://doi.org/10.1140/epjc/s10052-022-10396-9} {\bibfield  {journal} {\bibinfo  {journal} {Eur. Phys. J. C}\ }\textbf {\bibinfo {volume} {82}},\ \bibinfo {pages} {419} (\bibinfo {year} {2022})},\ \Eprint {https://arxiv.org/abs/2107.09275} {arXiv:2107.09275 [hep-ph]} \BibitemShut {NoStop}%
\bibitem [{\citenamefont {Ortega}\ \emph {et~al.}(2023)\citenamefont {Ortega}, \citenamefont {Entem}, \citenamefont {Fernandez},\ and\ \citenamefont {Segovia}}]{Ortega:2023azl}%
  \BibitemOpen
  \bibfield  {author} {\bibinfo {author} {\bibfnamefont {P.~G.}\ \bibnamefont {Ortega}}, \bibinfo {author} {\bibfnamefont {D.~R.}\ \bibnamefont {Entem}}, \bibinfo {author} {\bibfnamefont {F.}~\bibnamefont {Fernandez}},\ and\ \bibinfo {author} {\bibfnamefont {J.}~\bibnamefont {Segovia}},\ }\bibfield  {title} {\bibinfo {title} {{Novel Tcs and Tcs{\textasciimacron} candidates in a constituent-quark-model-based meson-meson coupled-channels calculation}},\ }\href {https://doi.org/10.1103/PhysRevD.108.094035} {\bibfield  {journal} {\bibinfo  {journal} {Phys. Rev. D}\ }\textbf {\bibinfo {volume} {108}},\ \bibinfo {pages} {094035} (\bibinfo {year} {2023})},\ \Eprint {https://arxiv.org/abs/2305.14430} {arXiv:2305.14430 [hep-ph]} \BibitemShut {NoStop}%
\bibitem [{\citenamefont {Wang}\ \emph {et~al.}(2024{\natexlab{a}})\citenamefont {Wang}, \citenamefont {Chen}, \citenamefont {Meng},\ and\ \citenamefont {Zhu}}]{Wang:2023hpp}%
  \BibitemOpen
  \bibfield  {author} {\bibinfo {author} {\bibfnamefont {B.}~\bibnamefont {Wang}}, \bibinfo {author} {\bibfnamefont {K.}~\bibnamefont {Chen}}, \bibinfo {author} {\bibfnamefont {L.}~\bibnamefont {Meng}},\ and\ \bibinfo {author} {\bibfnamefont {S.-L.}\ \bibnamefont {Zhu}},\ }\bibfield  {title} {\bibinfo {title} {{Spectrum of the molecular tetraquarks: Unraveling the Tcs0(2900) and Tcs{\textasciimacron}0a(2900)}},\ }\href {https://doi.org/10.1103/PhysRevD.109.034027} {\bibfield  {journal} {\bibinfo  {journal} {Phys. Rev. D}\ }\textbf {\bibinfo {volume} {109}},\ \bibinfo {pages} {034027} (\bibinfo {year} {2024}{\natexlab{a}})},\ \Eprint {https://arxiv.org/abs/2309.02191} {arXiv:2309.02191 [hep-ph]} \BibitemShut {NoStop}%
\bibitem [{\citenamefont {Tornqvist}(1991)}]{Tornqvist:1991ks}%
  \BibitemOpen
  \bibfield  {author} {\bibinfo {author} {\bibfnamefont {N.~A.}\ \bibnamefont {Tornqvist}},\ }\bibfield  {title} {\bibinfo {title} {{Possible large deuteron - like meson meson states bound by pions}},\ }\href {https://doi.org/10.1103/PhysRevLett.67.556} {\bibfield  {journal} {\bibinfo  {journal} {Phys. Rev. Lett.}\ }\textbf {\bibinfo {volume} {67}},\ \bibinfo {pages} {556} (\bibinfo {year} {1991})}\BibitemShut {NoStop}%
\bibitem [{\citenamefont {Tornqvist}(1994)}]{Tornqvist:1993ng}%
  \BibitemOpen
  \bibfield  {author} {\bibinfo {author} {\bibfnamefont {N.~A.}\ \bibnamefont {Tornqvist}},\ }\bibfield  {title} {\bibinfo {title} {{From the deuteron to deusons, an analysis of deuteron - like meson meson bound states}},\ }\href {https://doi.org/10.1007/BF01413192} {\bibfield  {journal} {\bibinfo  {journal} {Z. Phys. C}\ }\textbf {\bibinfo {volume} {61}},\ \bibinfo {pages} {525} (\bibinfo {year} {1994})},\ \Eprint {https://arxiv.org/abs/hep-ph/9310247} {arXiv:hep-ph/9310247} \BibitemShut {NoStop}%
\bibitem [{\citenamefont {Gamermann}\ \emph {et~al.}(2007)\citenamefont {Gamermann}, \citenamefont {Oset}, \citenamefont {Strottman},\ and\ \citenamefont {Vicente~Vacas}}]{Gamermann:2006nm}%
  \BibitemOpen
  \bibfield  {author} {\bibinfo {author} {\bibfnamefont {D.}~\bibnamefont {Gamermann}}, \bibinfo {author} {\bibfnamefont {E.}~\bibnamefont {Oset}}, \bibinfo {author} {\bibfnamefont {D.}~\bibnamefont {Strottman}},\ and\ \bibinfo {author} {\bibfnamefont {M.~J.}\ \bibnamefont {Vicente~Vacas}},\ }\bibfield  {title} {\bibinfo {title} {{Dynamically generated open and hidden charm meson systems}},\ }\href {https://doi.org/10.1103/PhysRevD.76.074016} {\bibfield  {journal} {\bibinfo  {journal} {Phys. Rev. D}\ }\textbf {\bibinfo {volume} {76}},\ \bibinfo {pages} {074016} (\bibinfo {year} {2007})},\ \Eprint {https://arxiv.org/abs/hep-ph/0612179} {arXiv:hep-ph/0612179} \BibitemShut {NoStop}%
\bibitem [{\citenamefont {Liu}\ \emph {et~al.}(2008{\natexlab{a}})\citenamefont {Liu}, \citenamefont {Liu}, \citenamefont {Deng},\ and\ \citenamefont {Zhu}}]{Liu:2008fh}%
  \BibitemOpen
  \bibfield  {author} {\bibinfo {author} {\bibfnamefont {Y.-R.}\ \bibnamefont {Liu}}, \bibinfo {author} {\bibfnamefont {X.}~\bibnamefont {Liu}}, \bibinfo {author} {\bibfnamefont {W.-Z.}\ \bibnamefont {Deng}},\ and\ \bibinfo {author} {\bibfnamefont {S.-L.}\ \bibnamefont {Zhu}},\ }\bibfield  {title} {\bibinfo {title} {{Is $X(3872) $ Really a Molecular State?}},\ }\href {https://doi.org/10.1140/epjc/s10052-008-0640-4} {\bibfield  {journal} {\bibinfo  {journal} {Eur. Phys. J. C}\ }\textbf {\bibinfo {volume} {56}},\ \bibinfo {pages} {63} (\bibinfo {year} {2008}{\natexlab{a}})},\ \Eprint {https://arxiv.org/abs/0801.3540} {arXiv:0801.3540 [hep-ph]} \BibitemShut {NoStop}%
\bibitem [{\citenamefont {Liu}\ \emph {et~al.}(2008{\natexlab{b}})\citenamefont {Liu}, \citenamefont {Liu}, \citenamefont {Deng},\ and\ \citenamefont {Zhu}}]{Liu:2008xz}%
  \BibitemOpen
  \bibfield  {author} {\bibinfo {author} {\bibfnamefont {X.}~\bibnamefont {Liu}}, \bibinfo {author} {\bibfnamefont {Y.-R.}\ \bibnamefont {Liu}}, \bibinfo {author} {\bibfnamefont {W.-Z.}\ \bibnamefont {Deng}},\ and\ \bibinfo {author} {\bibfnamefont {S.-L.}\ \bibnamefont {Zhu}},\ }\bibfield  {title} {\bibinfo {title} {{Z+(4430) as a D(1)-prime D* (D(1) D*) molecular state}},\ }\href {https://doi.org/10.1103/PhysRevD.77.094015} {\bibfield  {journal} {\bibinfo  {journal} {Phys. Rev. D}\ }\textbf {\bibinfo {volume} {77}},\ \bibinfo {pages} {094015} (\bibinfo {year} {2008}{\natexlab{b}})},\ \Eprint {https://arxiv.org/abs/0803.1295} {arXiv:0803.1295 [hep-ph]} \BibitemShut {NoStop}%
\bibitem [{\citenamefont {Thomas}\ and\ \citenamefont {Close}(2008)}]{Thomas:2008ja}%
  \BibitemOpen
  \bibfield  {author} {\bibinfo {author} {\bibfnamefont {C.~E.}\ \bibnamefont {Thomas}}\ and\ \bibinfo {author} {\bibfnamefont {F.~E.}\ \bibnamefont {Close}},\ }\bibfield  {title} {\bibinfo {title} {{Is X(3872) a molecule?}},\ }\href {https://doi.org/10.1103/PhysRevD.78.034007} {\bibfield  {journal} {\bibinfo  {journal} {Phys. Rev. D}\ }\textbf {\bibinfo {volume} {78}},\ \bibinfo {pages} {034007} (\bibinfo {year} {2008})},\ \Eprint {https://arxiv.org/abs/0805.3653} {arXiv:0805.3653 [hep-ph]} \BibitemShut {NoStop}%
\bibitem [{\citenamefont {Liu}\ \emph {et~al.}(2009)\citenamefont {Liu}, \citenamefont {Luo}, \citenamefont {Liu},\ and\ \citenamefont {Zhu}}]{Liu:2009qhy}%
  \BibitemOpen
  \bibfield  {author} {\bibinfo {author} {\bibfnamefont {X.}~\bibnamefont {Liu}}, \bibinfo {author} {\bibfnamefont {Z.-G.}\ \bibnamefont {Luo}}, \bibinfo {author} {\bibfnamefont {Y.-R.}\ \bibnamefont {Liu}},\ and\ \bibinfo {author} {\bibfnamefont {S.-L.}\ \bibnamefont {Zhu}},\ }\bibfield  {title} {\bibinfo {title} {{X(3872) and Other Possible Heavy Molecular States}},\ }\href {https://doi.org/10.1140/epjc/s10052-009-1020-4} {\bibfield  {journal} {\bibinfo  {journal} {Eur. Phys. J. C}\ }\textbf {\bibinfo {volume} {61}},\ \bibinfo {pages} {411} (\bibinfo {year} {2009})},\ \Eprint {https://arxiv.org/abs/0808.0073} {arXiv:0808.0073 [hep-ph]} \BibitemShut {NoStop}%
\bibitem [{\citenamefont {Ding}\ \emph {et~al.}(2009)\citenamefont {Ding}, \citenamefont {Liu},\ and\ \citenamefont {Yan}}]{Ding:2009vj}%
  \BibitemOpen
  \bibfield  {author} {\bibinfo {author} {\bibfnamefont {G.-J.}\ \bibnamefont {Ding}}, \bibinfo {author} {\bibfnamefont {J.-F.}\ \bibnamefont {Liu}},\ and\ \bibinfo {author} {\bibfnamefont {M.-L.}\ \bibnamefont {Yan}},\ }\bibfield  {title} {\bibinfo {title} {{Dynamics of Hadronic Molecule in One-Boson Exchange Approach and Possible Heavy Flavor Molecules}},\ }\href {https://doi.org/10.1103/PhysRevD.79.054005} {\bibfield  {journal} {\bibinfo  {journal} {Phys. Rev. D}\ }\textbf {\bibinfo {volume} {79}},\ \bibinfo {pages} {054005} (\bibinfo {year} {2009})},\ \Eprint {https://arxiv.org/abs/0901.0426} {arXiv:0901.0426 [hep-ph]} \BibitemShut {NoStop}%
\bibitem [{\citenamefont {Lee}\ \emph {et~al.}(2009)\citenamefont {Lee}, \citenamefont {Faessler}, \citenamefont {Gutsche},\ and\ \citenamefont {Lyubovitskij}}]{Lee:2009hy}%
  \BibitemOpen
  \bibfield  {author} {\bibinfo {author} {\bibfnamefont {I.~W.}\ \bibnamefont {Lee}}, \bibinfo {author} {\bibfnamefont {A.}~\bibnamefont {Faessler}}, \bibinfo {author} {\bibfnamefont {T.}~\bibnamefont {Gutsche}},\ and\ \bibinfo {author} {\bibfnamefont {V.~E.}\ \bibnamefont {Lyubovitskij}},\ }\bibfield  {title} {\bibinfo {title} {{X(3872) as a molecular DD* state in a potential model}},\ }\href {https://doi.org/10.1103/PhysRevD.80.094005} {\bibfield  {journal} {\bibinfo  {journal} {Phys. Rev. D}\ }\textbf {\bibinfo {volume} {80}},\ \bibinfo {pages} {094005} (\bibinfo {year} {2009})},\ \Eprint {https://arxiv.org/abs/0910.1009} {arXiv:0910.1009 [hep-ph]} \BibitemShut {NoStop}%
\bibitem [{\citenamefont {Sun}\ \emph {et~al.}(2011)\citenamefont {Sun}, \citenamefont {He}, \citenamefont {Liu}, \citenamefont {Luo},\ and\ \citenamefont {Zhu}}]{Sun:2011uh}%
  \BibitemOpen
  \bibfield  {author} {\bibinfo {author} {\bibfnamefont {Z.-F.}\ \bibnamefont {Sun}}, \bibinfo {author} {\bibfnamefont {J.}~\bibnamefont {He}}, \bibinfo {author} {\bibfnamefont {X.}~\bibnamefont {Liu}}, \bibinfo {author} {\bibfnamefont {Z.-G.}\ \bibnamefont {Luo}},\ and\ \bibinfo {author} {\bibfnamefont {S.-L.}\ \bibnamefont {Zhu}},\ }\bibfield  {title} {\bibinfo {title} {{$Z_b(10610)^\pm$ and $Z_b(10650)^\pm$ as the $B^*\bar{B}$ and $B^*\bar{B}^{*}$ molecular states}},\ }\href {https://doi.org/10.1103/PhysRevD.84.054002} {\bibfield  {journal} {\bibinfo  {journal} {Phys. Rev. D}\ }\textbf {\bibinfo {volume} {84}},\ \bibinfo {pages} {054002} (\bibinfo {year} {2011})},\ \Eprint {https://arxiv.org/abs/1106.2968} {arXiv:1106.2968 [hep-ph]} \BibitemShut {NoStop}%
\bibitem [{\citenamefont {Li}\ and\ \citenamefont {Zhu}(2012)}]{Li:2012cs}%
  \BibitemOpen
  \bibfield  {author} {\bibinfo {author} {\bibfnamefont {N.}~\bibnamefont {Li}}\ and\ \bibinfo {author} {\bibfnamefont {S.-L.}\ \bibnamefont {Zhu}},\ }\bibfield  {title} {\bibinfo {title} {{Isospin breaking, Coupled-channel effects and Diagnosis of X(3872)}},\ }\href {https://doi.org/10.1103/PhysRevD.86.074022} {\bibfield  {journal} {\bibinfo  {journal} {Phys. Rev. D}\ }\textbf {\bibinfo {volume} {86}},\ \bibinfo {pages} {074022} (\bibinfo {year} {2012})},\ \Eprint {https://arxiv.org/abs/1207.3954} {arXiv:1207.3954 [hep-ph]} \BibitemShut {NoStop}%
\bibitem [{\citenamefont {Li}\ \emph {et~al.}(2013)\citenamefont {Li}, \citenamefont {Sun}, \citenamefont {Liu},\ and\ \citenamefont {Zhu}}]{Li:2012ss}%
  \BibitemOpen
  \bibfield  {author} {\bibinfo {author} {\bibfnamefont {N.}~\bibnamefont {Li}}, \bibinfo {author} {\bibfnamefont {Z.-F.}\ \bibnamefont {Sun}}, \bibinfo {author} {\bibfnamefont {X.}~\bibnamefont {Liu}},\ and\ \bibinfo {author} {\bibfnamefont {S.-L.}\ \bibnamefont {Zhu}},\ }\bibfield  {title} {\bibinfo {title} {{Coupled-channel analysis of the possible $D^{(*)}D^{(*)}, \overline{B}^{(*)}\overline{B}^{(*)}$ and $D^{(*)}\overline{B}^{(*)}$ molecular states}},\ }\href {https://doi.org/10.1103/PhysRevD.88.114008} {\bibfield  {journal} {\bibinfo  {journal} {Phys. Rev. D}\ }\textbf {\bibinfo {volume} {88}},\ \bibinfo {pages} {114008} (\bibinfo {year} {2013})},\ \Eprint {https://arxiv.org/abs/1211.5007} {arXiv:1211.5007 [hep-ph]} \BibitemShut {NoStop}%
\bibitem [{\citenamefont {Chen}\ \emph {et~al.}(2017)\citenamefont {Chen}, \citenamefont {Hosaka},\ and\ \citenamefont {Liu}}]{Chen:2017vai}%
  \BibitemOpen
  \bibfield  {author} {\bibinfo {author} {\bibfnamefont {R.}~\bibnamefont {Chen}}, \bibinfo {author} {\bibfnamefont {A.}~\bibnamefont {Hosaka}},\ and\ \bibinfo {author} {\bibfnamefont {X.}~\bibnamefont {Liu}},\ }\bibfield  {title} {\bibinfo {title} {{Heavy molecules and one-$\sigma/\omega$-exchange model}},\ }\href {https://doi.org/10.1103/PhysRevD.96.116012} {\bibfield  {journal} {\bibinfo  {journal} {Phys. Rev. D}\ }\textbf {\bibinfo {volume} {96}},\ \bibinfo {pages} {116012} (\bibinfo {year} {2017})},\ \Eprint {https://arxiv.org/abs/1707.08306} {arXiv:1707.08306 [hep-ph]} \BibitemShut {NoStop}%
\bibitem [{\citenamefont {Liu}\ \emph {et~al.}(2019{\natexlab{b}})\citenamefont {Liu}, \citenamefont {Wu}, \citenamefont {Pavon~Valderrama}, \citenamefont {Xie},\ and\ \citenamefont {Geng}}]{Liu:2019stu}%
  \BibitemOpen
  \bibfield  {author} {\bibinfo {author} {\bibfnamefont {M.-Z.}\ \bibnamefont {Liu}}, \bibinfo {author} {\bibfnamefont {T.-W.}\ \bibnamefont {Wu}}, \bibinfo {author} {\bibfnamefont {M.}~\bibnamefont {Pavon~Valderrama}}, \bibinfo {author} {\bibfnamefont {J.-J.}\ \bibnamefont {Xie}},\ and\ \bibinfo {author} {\bibfnamefont {L.-S.}\ \bibnamefont {Geng}},\ }\bibfield  {title} {\bibinfo {title} {{Heavy-quark spin and flavor symmetry partners of the X(3872) revisited: What can we learn from the one boson exchange model?}},\ }\href {https://doi.org/10.1103/PhysRevD.99.094018} {\bibfield  {journal} {\bibinfo  {journal} {Phys. Rev. D}\ }\textbf {\bibinfo {volume} {99}},\ \bibinfo {pages} {094018} (\bibinfo {year} {2019}{\natexlab{b}})},\ \Eprint {https://arxiv.org/abs/1902.03044} {arXiv:1902.03044 [hep-ph]} \BibitemShut {NoStop}%
\bibitem [{\citenamefont {Chen}\ and\ \citenamefont {Huang}(2021)}]{Chen:2020yvq}%
  \BibitemOpen
  \bibfield  {author} {\bibinfo {author} {\bibfnamefont {R.}~\bibnamefont {Chen}}\ and\ \bibinfo {author} {\bibfnamefont {Q.}~\bibnamefont {Huang}},\ }\bibfield  {title} {\bibinfo {title} {{$Z_{cs}(3985)^-$: A strange hidden-charm tetraquark resonance or not?}},\ }\href {https://doi.org/10.1103/PhysRevD.103.034008} {\bibfield  {journal} {\bibinfo  {journal} {Phys. Rev. D}\ }\textbf {\bibinfo {volume} {103}},\ \bibinfo {pages} {034008} (\bibinfo {year} {2021})},\ \Eprint {https://arxiv.org/abs/2011.09156} {arXiv:2011.09156 [hep-ph]} \BibitemShut {NoStop}%
\bibitem [{\citenamefont {Chen}\ \emph {et~al.}(2021)\citenamefont {Chen}, \citenamefont {Huang}, \citenamefont {Liu},\ and\ \citenamefont {Zhu}}]{Chen:2021vhg}%
  \BibitemOpen
  \bibfield  {author} {\bibinfo {author} {\bibfnamefont {R.}~\bibnamefont {Chen}}, \bibinfo {author} {\bibfnamefont {Q.}~\bibnamefont {Huang}}, \bibinfo {author} {\bibfnamefont {X.}~\bibnamefont {Liu}},\ and\ \bibinfo {author} {\bibfnamefont {S.-L.}\ \bibnamefont {Zhu}},\ }\bibfield  {title} {\bibinfo {title} {{Predicting another doubly charmed molecular resonance Tcc'+ (3876)}},\ }\href {https://doi.org/10.1103/PhysRevD.104.114042} {\bibfield  {journal} {\bibinfo  {journal} {Phys. Rev. D}\ }\textbf {\bibinfo {volume} {104}},\ \bibinfo {pages} {114042} (\bibinfo {year} {2021})},\ \Eprint {https://arxiv.org/abs/2108.01911} {arXiv:2108.01911 [hep-ph]} \BibitemShut {NoStop}%
\bibitem [{\citenamefont {Dong}\ \emph {et~al.}(2021{\natexlab{a}})\citenamefont {Dong}, \citenamefont {Guo},\ and\ \citenamefont {Zou}}]{Dong:2021bvy}%
  \BibitemOpen
  \bibfield  {author} {\bibinfo {author} {\bibfnamefont {X.-K.}\ \bibnamefont {Dong}}, \bibinfo {author} {\bibfnamefont {F.-K.}\ \bibnamefont {Guo}},\ and\ \bibinfo {author} {\bibfnamefont {B.-S.}\ \bibnamefont {Zou}},\ }\bibfield  {title} {\bibinfo {title} {{A survey of heavy{\textendash}heavy hadronic molecules}},\ }\href {https://doi.org/10.1088/1572-9494/ac27a2} {\bibfield  {journal} {\bibinfo  {journal} {Commun. Theor. Phys.}\ }\textbf {\bibinfo {volume} {73}},\ \bibinfo {pages} {125201} (\bibinfo {year} {2021}{\natexlab{a}})},\ \Eprint {https://arxiv.org/abs/2108.02673} {arXiv:2108.02673 [hep-ph]} \BibitemShut {NoStop}%
\bibitem [{\citenamefont {Dong}\ \emph {et~al.}(2021{\natexlab{b}})\citenamefont {Dong}, \citenamefont {Guo},\ and\ \citenamefont {Zou}}]{Dong:2021juy}%
  \BibitemOpen
  \bibfield  {author} {\bibinfo {author} {\bibfnamefont {X.-K.}\ \bibnamefont {Dong}}, \bibinfo {author} {\bibfnamefont {F.-K.}\ \bibnamefont {Guo}},\ and\ \bibinfo {author} {\bibfnamefont {B.-S.}\ \bibnamefont {Zou}},\ }\bibfield  {title} {\bibinfo {title} {{A survey of heavy-antiheavy hadronic molecules}},\ }\href {https://doi.org/10.13725/j.cnki.pip.2021.02.001} {\bibfield  {journal} {\bibinfo  {journal} {Progr. Phys.}\ }\textbf {\bibinfo {volume} {41}},\ \bibinfo {pages} {65} (\bibinfo {year} {2021}{\natexlab{b}})},\ \Eprint {https://arxiv.org/abs/2101.01021} {arXiv:2101.01021 [hep-ph]} \BibitemShut {NoStop}%
\bibitem [{\citenamefont {Lin}\ \emph {et~al.}(2024{\natexlab{a}})\citenamefont {Lin}, \citenamefont {Cheng},\ and\ \citenamefont {Zhu}}]{Lin:2022wmj}%
  \BibitemOpen
  \bibfield  {author} {\bibinfo {author} {\bibfnamefont {Z.-Y.}\ \bibnamefont {Lin}}, \bibinfo {author} {\bibfnamefont {J.-B.}\ \bibnamefont {Cheng}},\ and\ \bibinfo {author} {\bibfnamefont {S.-L.}\ \bibnamefont {Zhu}},\ }\bibfield  {title} {\bibinfo {title} {{Tcc+ and {\ensuremath{\chi}}c1(3872) with the complex scaling method and DD(D{\textasciimacron}){\ensuremath{\pi}} three-body effect}},\ }\href {https://doi.org/10.1103/PhysRevD.110.054008} {\bibfield  {journal} {\bibinfo  {journal} {Phys. Rev. D}\ }\textbf {\bibinfo {volume} {110}},\ \bibinfo {pages} {054008} (\bibinfo {year} {2024}{\natexlab{a}})},\ \Eprint {https://arxiv.org/abs/2205.14628} {arXiv:2205.14628 [hep-ph]} \BibitemShut {NoStop}%
\bibitem [{\citenamefont {Cheng}\ \emph {et~al.}(2022)\citenamefont {Cheng}, \citenamefont {Lin},\ and\ \citenamefont {Zhu}}]{Cheng:2022qcm}%
  \BibitemOpen
  \bibfield  {author} {\bibinfo {author} {\bibfnamefont {J.-B.}\ \bibnamefont {Cheng}}, \bibinfo {author} {\bibfnamefont {Z.-Y.}\ \bibnamefont {Lin}},\ and\ \bibinfo {author} {\bibfnamefont {S.-L.}\ \bibnamefont {Zhu}},\ }\bibfield  {title} {\bibinfo {title} {{Double-charm tetraquark under the complex scaling method}},\ }\href {https://doi.org/10.1103/PhysRevD.106.016012} {\bibfield  {journal} {\bibinfo  {journal} {Phys. Rev. D}\ }\textbf {\bibinfo {volume} {106}},\ \bibinfo {pages} {016012} (\bibinfo {year} {2022})},\ \Eprint {https://arxiv.org/abs/2205.13354} {arXiv:2205.13354 [hep-ph]} \BibitemShut {NoStop}%
\bibitem [{\citenamefont {Peng}\ \emph {et~al.}(2023)\citenamefont {Peng}, \citenamefont {Yan},\ and\ \citenamefont {Pavon~Valderrama}}]{Peng:2023lfw}%
  \BibitemOpen
  \bibfield  {author} {\bibinfo {author} {\bibfnamefont {F.-Z.}\ \bibnamefont {Peng}}, \bibinfo {author} {\bibfnamefont {M.-J.}\ \bibnamefont {Yan}},\ and\ \bibinfo {author} {\bibfnamefont {M.}~\bibnamefont {Pavon~Valderrama}},\ }\bibfield  {title} {\bibinfo {title} {{Heavy- and light-flavor symmetry partners of the Tcc+(3875), the X(3872), and the X(3960) from light-meson exchange saturation}},\ }\href {https://doi.org/10.1103/PhysRevD.108.114001} {\bibfield  {journal} {\bibinfo  {journal} {Phys. Rev. D}\ }\textbf {\bibinfo {volume} {108}},\ \bibinfo {pages} {114001} (\bibinfo {year} {2023})},\ \Eprint {https://arxiv.org/abs/2304.13515} {arXiv:2304.13515 [hep-ph]} \BibitemShut {NoStop}%
\bibitem [{\citenamefont {Lin}\ \emph {et~al.}(2024{\natexlab{b}})\citenamefont {Lin}, \citenamefont {Wang}, \citenamefont {Cheng}, \citenamefont {Meng},\ and\ \citenamefont {Zhu}}]{Lin:2024qcq}%
  \BibitemOpen
  \bibfield  {author} {\bibinfo {author} {\bibfnamefont {Z.-Y.}\ \bibnamefont {Lin}}, \bibinfo {author} {\bibfnamefont {J.-Z.}\ \bibnamefont {Wang}}, \bibinfo {author} {\bibfnamefont {J.-B.}\ \bibnamefont {Cheng}}, \bibinfo {author} {\bibfnamefont {L.}~\bibnamefont {Meng}},\ and\ \bibinfo {author} {\bibfnamefont {S.-L.}\ \bibnamefont {Zhu}},\ }\bibfield  {title} {\bibinfo {title} {{Identification of the $G(3900)$ as the P-wave $D\bar{D}^*/\bar{D}D^*$ resonance}},\ }\href {https://doi.org/10.1103/PhysRevLett.133.241903} {\bibfield  {journal} {\bibinfo  {journal} {Phys. Rev. Lett.}\ }\textbf {\bibinfo {volume} {133}},\ \bibinfo {pages} {241903} (\bibinfo {year} {2024}{\natexlab{b}})},\ \Eprint {https://arxiv.org/abs/2403.01727} {arXiv:2403.01727 [hep-ph]} \BibitemShut {NoStop}%
\bibitem [{\citenamefont {Wang}\ \emph {et~al.}(2024{\natexlab{b}})\citenamefont {Wang}, \citenamefont {Lin}, \citenamefont {Wang}, \citenamefont {Meng},\ and\ \citenamefont {Zhu}}]{Wang:2024ukc}%
  \BibitemOpen
  \bibfield  {author} {\bibinfo {author} {\bibfnamefont {J.-Z.}\ \bibnamefont {Wang}}, \bibinfo {author} {\bibfnamefont {Z.-Y.}\ \bibnamefont {Lin}}, \bibinfo {author} {\bibfnamefont {B.}~\bibnamefont {Wang}}, \bibinfo {author} {\bibfnamefont {L.}~\bibnamefont {Meng}},\ and\ \bibinfo {author} {\bibfnamefont {S.-L.}\ \bibnamefont {Zhu}},\ }\bibfield  {title} {\bibinfo {title} {{Double pole structures of X1(2900) as the P-wave D{\textasciimacron}*K* resonances}},\ }\href {https://doi.org/10.1103/PhysRevD.110.114003} {\bibfield  {journal} {\bibinfo  {journal} {Phys. Rev. D}\ }\textbf {\bibinfo {volume} {110}},\ \bibinfo {pages} {114003} (\bibinfo {year} {2024}{\natexlab{b}})},\ \Eprint {https://arxiv.org/abs/2408.08965} {arXiv:2408.08965 [hep-ph]} \BibitemShut {NoStop}%
\bibitem [{\citenamefont {Wang}\ \emph {et~al.}(2025)\citenamefont {Wang}, \citenamefont {Lin}, \citenamefont {Cheng}, \citenamefont {Meng},\ and\ \citenamefont {Zhu}}]{Wang:2025kpm}%
  \BibitemOpen
  \bibfield  {author} {\bibinfo {author} {\bibfnamefont {J.-Z.}\ \bibnamefont {Wang}}, \bibinfo {author} {\bibfnamefont {Z.-Y.}\ \bibnamefont {Lin}}, \bibinfo {author} {\bibfnamefont {J.-B.}\ \bibnamefont {Cheng}}, \bibinfo {author} {\bibfnamefont {L.}~\bibnamefont {Meng}},\ and\ \bibinfo {author} {\bibfnamefont {S.-L.}\ \bibnamefont {Zhu}},\ }\bibfield  {title} {\bibinfo {title} {{Emergence of new heavy quarkoniumlike states: $Y(10600)$ and $Y(10650)$}},\ }\href@noop {} {\  (\bibinfo {year} {2025})},\ \Eprint {https://arxiv.org/abs/2505.02742} {arXiv:2505.02742 [hep-ph]} \BibitemShut {NoStop}%
\bibitem [{\citenamefont {Cheng}\ \emph {et~al.}(2026)\citenamefont {Cheng}, \citenamefont {Lin}, \citenamefont {Wang},\ and\ \citenamefont {Zhu}}]{Cheng:2026cgo}%
  \BibitemOpen
  \bibfield  {author} {\bibinfo {author} {\bibfnamefont {J.-B.}\ \bibnamefont {Cheng}}, \bibinfo {author} {\bibfnamefont {Z.-Y.}\ \bibnamefont {Lin}}, \bibinfo {author} {\bibfnamefont {J.-Z.}\ \bibnamefont {Wang}},\ and\ \bibinfo {author} {\bibfnamefont {S.-L.}\ \bibnamefont {Zhu}},\ }\bibfield  {title} {\bibinfo {title} {{Decoding $Z_c(4430)$ and $Z_c(4200)$: The role of $P$-wave charmed mesons}},\ }\href@noop {} {\  (\bibinfo {year} {2026})},\ \Eprint {https://arxiv.org/abs/2601.20740} {arXiv:2601.20740 [hep-ph]} \BibitemShut {NoStop}%
\bibitem [{\citenamefont {Zhu}\ \emph {et~al.}(2025)\citenamefont {Zhu}, \citenamefont {Meng}, \citenamefont {Ma}, \citenamefont {Li}, \citenamefont {Chen},\ and\ \citenamefont {Zhu}}]{Zhu:2024hgm}%
  \BibitemOpen
  \bibfield  {author} {\bibinfo {author} {\bibfnamefont {H.-X.}\ \bibnamefont {Zhu}}, \bibinfo {author} {\bibfnamefont {L.}~\bibnamefont {Meng}}, \bibinfo {author} {\bibfnamefont {Y.}~\bibnamefont {Ma}}, \bibinfo {author} {\bibfnamefont {N.}~\bibnamefont {Li}}, \bibinfo {author} {\bibfnamefont {W.}~\bibnamefont {Chen}},\ and\ \bibinfo {author} {\bibfnamefont {S.-L.}\ \bibnamefont {Zhu}},\ }\bibfield  {title} {\bibinfo {title} {{Constraining the DDD* three-body bound state via the Zc(3900) pole}},\ }\href {https://doi.org/10.1103/PhysRevD.111.094022} {\bibfield  {journal} {\bibinfo  {journal} {Phys. Rev. D}\ }\textbf {\bibinfo {volume} {111}},\ \bibinfo {pages} {094022} (\bibinfo {year} {2025})},\ \Eprint {https://arxiv.org/abs/2412.12816} {arXiv:2412.12816 [hep-ph]} \BibitemShut {NoStop}%
\bibitem [{\citenamefont {Hiyama}\ \emph {et~al.}(2003)\citenamefont {Hiyama}, \citenamefont {Kino},\ and\ \citenamefont {Kamimura}}]{Hiyama:2003cu}%
  \BibitemOpen
  \bibfield  {author} {\bibinfo {author} {\bibfnamefont {E.}~\bibnamefont {Hiyama}}, \bibinfo {author} {\bibfnamefont {Y.}~\bibnamefont {Kino}},\ and\ \bibinfo {author} {\bibfnamefont {M.}~\bibnamefont {Kamimura}},\ }\bibfield  {title} {\bibinfo {title} {{Gaussian expansion method for few-body systems}},\ }\href {https://doi.org/10.1016/S0146-6410(03)90015-9} {\bibfield  {journal} {\bibinfo  {journal} {Prog. Part. Nucl. Phys.}\ }\textbf {\bibinfo {volume} {51}},\ \bibinfo {pages} {223} (\bibinfo {year} {2003})}\BibitemShut {NoStop}%
\bibitem [{\citenamefont {Aguilar}\ and\ \citenamefont {Combes}(1971)}]{Aguilar:1971ve}%
  \BibitemOpen
  \bibfield  {author} {\bibinfo {author} {\bibfnamefont {J.}~\bibnamefont {Aguilar}}\ and\ \bibinfo {author} {\bibfnamefont {J.~M.}\ \bibnamefont {Combes}},\ }\bibfield  {title} {\bibinfo {title} {{A class of analytic perturbations for one-body schroedinger hamiltonians}},\ }\href {https://doi.org/10.1007/BF01877510} {\bibfield  {journal} {\bibinfo  {journal} {Commun. Math. Phys.}\ }\textbf {\bibinfo {volume} {22}},\ \bibinfo {pages} {269} (\bibinfo {year} {1971})}\BibitemShut {NoStop}%
\bibitem [{\citenamefont {Balslev}\ and\ \citenamefont {Combes}(1971)}]{Balslev:1971vb}%
  \BibitemOpen
  \bibfield  {author} {\bibinfo {author} {\bibfnamefont {E.}~\bibnamefont {Balslev}}\ and\ \bibinfo {author} {\bibfnamefont {J.~M.}\ \bibnamefont {Combes}},\ }\bibfield  {title} {\bibinfo {title} {{Spectral properties of many-body schroedinger operators with dilatation-analytic interactions}},\ }\href {https://doi.org/10.1007/BF01877511} {\bibfield  {journal} {\bibinfo  {journal} {Commun. Math. Phys.}\ }\textbf {\bibinfo {volume} {22}},\ \bibinfo {pages} {280} (\bibinfo {year} {1971})}\BibitemShut {NoStop}%
\bibitem [{\citenamefont {Moiseyev}(1998)}]{Moiseyev:1998gjp}%
  \BibitemOpen
  \bibfield  {author} {\bibinfo {author} {\bibfnamefont {N.}~\bibnamefont {Moiseyev}},\ }\bibfield  {title} {\bibinfo {title} {{Quantum theory of resonances: calculating energies, widths and cross-sections by complex scaling}},\ }\href {https://doi.org/10.1016/S0370-1573(98)00002-7} {\bibfield  {journal} {\bibinfo  {journal} {Phys. Rept.}\ }\textbf {\bibinfo {volume} {302}},\ \bibinfo {pages} {212} (\bibinfo {year} {1998})}\BibitemShut {NoStop}%
\bibitem [{\citenamefont {Aoyama}\ \emph {et~al.}(2006)\citenamefont {Aoyama}, \citenamefont {Myo}, \citenamefont {Kat{\={o}}},\ and\ \citenamefont {Ikeda}}]{Aoyama:2006hrz}%
  \BibitemOpen
  \bibfield  {author} {\bibinfo {author} {\bibfnamefont {S.}~\bibnamefont {Aoyama}}, \bibinfo {author} {\bibfnamefont {T.}~\bibnamefont {Myo}}, \bibinfo {author} {\bibfnamefont {K.}~\bibnamefont {Kat{\={o}}}},\ and\ \bibinfo {author} {\bibfnamefont {K.}~\bibnamefont {Ikeda}},\ }\bibfield  {title} {\bibinfo {title} {{The Complex Scaling Method for Many-Body Resonances and Its Applications to Three-Body Resonances}},\ }\href {https://doi.org/10.1143/ptp.116.1} {\bibfield  {journal} {\bibinfo  {journal} {Prog. Theor. Phys.}\ }\textbf {\bibinfo {volume} {116}},\ \bibinfo {pages} {1} (\bibinfo {year} {2006})}\BibitemShut {NoStop}%
\bibitem [{\citenamefont {Navas}\ \emph {et~al.}(2024)\citenamefont {Navas} \emph {et~al.}}]{ParticleDataGroup:2024cfk}%
  \BibitemOpen
  \bibfield  {author} {\bibinfo {author} {\bibfnamefont {S.}~\bibnamefont {Navas}} \emph {et~al.} (\bibinfo {collaboration} {Particle Data Group}),\ }\bibfield  {title} {\bibinfo {title} {{Review of particle physics}},\ }\href {https://doi.org/10.1103/PhysRevD.110.030001} {\bibfield  {journal} {\bibinfo  {journal} {Phys. Rev. D}\ }\textbf {\bibinfo {volume} {110}},\ \bibinfo {pages} {030001} (\bibinfo {year} {2024})}\BibitemShut {NoStop}%
\bibitem [{\citenamefont {Georgi}(1991)}]{Georgi:1990cx}%
  \BibitemOpen
  \bibfield  {author} {\bibinfo {author} {\bibfnamefont {H.}~\bibnamefont {Georgi}},\ }\bibfield  {title} {\bibinfo {title} {{Comment on heavy baryon weak form-factors}},\ }\href {https://doi.org/10.1016/0550-3213(91)90519-4} {\bibfield  {journal} {\bibinfo  {journal} {Nucl. Phys. B}\ }\textbf {\bibinfo {volume} {348}},\ \bibinfo {pages} {293} (\bibinfo {year} {1991})}\BibitemShut {NoStop}%
\bibitem [{\citenamefont {Mannel}\ \emph {et~al.}(1991)\citenamefont {Mannel}, \citenamefont {Roberts},\ and\ \citenamefont {Ryzak}}]{Mannel:1990vg}%
  \BibitemOpen
  \bibfield  {author} {\bibinfo {author} {\bibfnamefont {T.}~\bibnamefont {Mannel}}, \bibinfo {author} {\bibfnamefont {W.}~\bibnamefont {Roberts}},\ and\ \bibinfo {author} {\bibfnamefont {Z.}~\bibnamefont {Ryzak}},\ }\bibfield  {title} {\bibinfo {title} {{Baryons in the heavy quark effective theory}},\ }\href {https://doi.org/10.1016/0550-3213(91)90301-D} {\bibfield  {journal} {\bibinfo  {journal} {Nucl. Phys. B}\ }\textbf {\bibinfo {volume} {355}},\ \bibinfo {pages} {38} (\bibinfo {year} {1991})}\BibitemShut {NoStop}%
\bibitem [{\citenamefont {Falk}(1992)}]{Falk:1991nq}%
  \BibitemOpen
  \bibfield  {author} {\bibinfo {author} {\bibfnamefont {A.~F.}\ \bibnamefont {Falk}},\ }\bibfield  {title} {\bibinfo {title} {{Hadrons of arbitrary spin in the heavy quark effective theory}},\ }\href {https://doi.org/10.1016/0550-3213(92)90004-U} {\bibfield  {journal} {\bibinfo  {journal} {Nucl. Phys. B}\ }\textbf {\bibinfo {volume} {378}},\ \bibinfo {pages} {79} (\bibinfo {year} {1992})}\BibitemShut {NoStop}%
\bibitem [{\citenamefont {Wise}(1992)}]{Wise:1992hn}%
  \BibitemOpen
  \bibfield  {author} {\bibinfo {author} {\bibfnamefont {M.~B.}\ \bibnamefont {Wise}},\ }\bibfield  {title} {\bibinfo {title} {{Chiral perturbation theory for hadrons containing a heavy quark}},\ }\href {https://doi.org/10.1103/PhysRevD.45.R2188} {\bibfield  {journal} {\bibinfo  {journal} {Phys. Rev. D}\ }\textbf {\bibinfo {volume} {45}},\ \bibinfo {pages} {R2188} (\bibinfo {year} {1992})}\BibitemShut {NoStop}%
\bibitem [{\citenamefont {Yan}\ \emph {et~al.}(1992)\citenamefont {Yan}, \citenamefont {Cheng}, \citenamefont {Cheung}, \citenamefont {Lin}, \citenamefont {Lin},\ and\ \citenamefont {Yu}}]{Yan:1992gz}%
  \BibitemOpen
  \bibfield  {author} {\bibinfo {author} {\bibfnamefont {T.-M.}\ \bibnamefont {Yan}}, \bibinfo {author} {\bibfnamefont {H.-Y.}\ \bibnamefont {Cheng}}, \bibinfo {author} {\bibfnamefont {C.-Y.}\ \bibnamefont {Cheung}}, \bibinfo {author} {\bibfnamefont {G.-L.}\ \bibnamefont {Lin}}, \bibinfo {author} {\bibfnamefont {Y.~C.}\ \bibnamefont {Lin}},\ and\ \bibinfo {author} {\bibfnamefont {H.-L.}\ \bibnamefont {Yu}},\ }\bibfield  {title} {\bibinfo {title} {{Heavy quark symmetry and chiral dynamics}},\ }\href {https://doi.org/10.1103/PhysRevD.46.1148} {\bibfield  {journal} {\bibinfo  {journal} {Phys. Rev. D}\ }\textbf {\bibinfo {volume} {46}},\ \bibinfo {pages} {1148} (\bibinfo {year} {1992})},\ \bibinfo {note} {[Erratum: Phys.Rev.D 55, 5851 (1997)]}\BibitemShut {NoStop}%
\bibitem [{\citenamefont {Casalbuoni}\ \emph {et~al.}(1997)\citenamefont {Casalbuoni}, \citenamefont {Deandrea}, \citenamefont {Di~Bartolomeo}, \citenamefont {Gatto}, \citenamefont {Feruglio},\ and\ \citenamefont {Nardulli}}]{Casalbuoni:1996pg}%
  \BibitemOpen
  \bibfield  {author} {\bibinfo {author} {\bibfnamefont {R.}~\bibnamefont {Casalbuoni}}, \bibinfo {author} {\bibfnamefont {A.}~\bibnamefont {Deandrea}}, \bibinfo {author} {\bibfnamefont {N.}~\bibnamefont {Di~Bartolomeo}}, \bibinfo {author} {\bibfnamefont {R.}~\bibnamefont {Gatto}}, \bibinfo {author} {\bibfnamefont {F.}~\bibnamefont {Feruglio}},\ and\ \bibinfo {author} {\bibfnamefont {G.}~\bibnamefont {Nardulli}},\ }\bibfield  {title} {\bibinfo {title} {{Phenomenology of heavy meson chiral Lagrangians}},\ }\href {https://doi.org/10.1016/S0370-1573(96)00027-0} {\bibfield  {journal} {\bibinfo  {journal} {Phys. Rept.}\ }\textbf {\bibinfo {volume} {281}},\ \bibinfo {pages} {145} (\bibinfo {year} {1997})},\ \Eprint {https://arxiv.org/abs/hep-ph/9605342} {arXiv:hep-ph/9605342} \BibitemShut {NoStop}%
\bibitem [{\citenamefont {Isola}\ \emph {et~al.}(2003)\citenamefont {Isola}, \citenamefont {Ladisa}, \citenamefont {Nardulli},\ and\ \citenamefont {Santorelli}}]{Isola:2003fh}%
  \BibitemOpen
  \bibfield  {author} {\bibinfo {author} {\bibfnamefont {C.}~\bibnamefont {Isola}}, \bibinfo {author} {\bibfnamefont {M.}~\bibnamefont {Ladisa}}, \bibinfo {author} {\bibfnamefont {G.}~\bibnamefont {Nardulli}},\ and\ \bibinfo {author} {\bibfnamefont {P.}~\bibnamefont {Santorelli}},\ }\bibfield  {title} {\bibinfo {title} {{Charming penguins in B ---{\ensuremath{>}} K* pi, K(rho, omega, phi) decays}},\ }\href {https://doi.org/10.1103/PhysRevD.68.114001} {\bibfield  {journal} {\bibinfo  {journal} {Phys. Rev. D}\ }\textbf {\bibinfo {volume} {68}},\ \bibinfo {pages} {114001} (\bibinfo {year} {2003})},\ \Eprint {https://arxiv.org/abs/hep-ph/0307367} {arXiv:hep-ph/0307367} \BibitemShut {NoStop}%
\bibitem [{\citenamefont {Albaladejo}\ \emph {et~al.}(2016)\citenamefont {Albaladejo}, \citenamefont {Guo}, \citenamefont {Hidalgo-Duque},\ and\ \citenamefont {Nieves}}]{Albaladejo:2015lob}%
  \BibitemOpen
  \bibfield  {author} {\bibinfo {author} {\bibfnamefont {M.}~\bibnamefont {Albaladejo}}, \bibinfo {author} {\bibfnamefont {F.-K.}\ \bibnamefont {Guo}}, \bibinfo {author} {\bibfnamefont {C.}~\bibnamefont {Hidalgo-Duque}},\ and\ \bibinfo {author} {\bibfnamefont {J.}~\bibnamefont {Nieves}},\ }\bibfield  {title} {\bibinfo {title} {{$Z_c(3900)$: What has been really seen?}},\ }\href {https://doi.org/10.1016/j.physletb.2016.02.025} {\bibfield  {journal} {\bibinfo  {journal} {Phys. Lett. B}\ }\textbf {\bibinfo {volume} {755}},\ \bibinfo {pages} {337} (\bibinfo {year} {2016})},\ \Eprint {https://arxiv.org/abs/1512.03638} {arXiv:1512.03638 [hep-ph]} \BibitemShut {NoStop}%
\bibitem [{\citenamefont {Pilloni}\ \emph {et~al.}(2017)\citenamefont {Pilloni}, \citenamefont {Fernandez-Ramirez}, \citenamefont {Jackura}, \citenamefont {Mathieu}, \citenamefont {Mikhasenko}, \citenamefont {Nys},\ and\ \citenamefont {Szczepaniak}}]{Pilloni:2016obd}%
  \BibitemOpen
  \bibfield  {author} {\bibinfo {author} {\bibfnamefont {A.}~\bibnamefont {Pilloni}}, \bibinfo {author} {\bibfnamefont {C.}~\bibnamefont {Fernandez-Ramirez}}, \bibinfo {author} {\bibfnamefont {A.}~\bibnamefont {Jackura}}, \bibinfo {author} {\bibfnamefont {V.}~\bibnamefont {Mathieu}}, \bibinfo {author} {\bibfnamefont {M.}~\bibnamefont {Mikhasenko}}, \bibinfo {author} {\bibfnamefont {J.}~\bibnamefont {Nys}},\ and\ \bibinfo {author} {\bibfnamefont {A.~P.}\ \bibnamefont {Szczepaniak}} (\bibinfo {collaboration} {JPAC}),\ }\bibfield  {title} {\bibinfo {title} {{Amplitude analysis and the nature of the Z$_c$(3900)}},\ }\href {https://doi.org/10.1016/j.physletb.2017.06.030} {\bibfield  {journal} {\bibinfo  {journal} {Phys. Lett. B}\ }\textbf {\bibinfo {volume} {772}},\ \bibinfo {pages} {200} (\bibinfo {year} {2017})},\ \Eprint {https://arxiv.org/abs/1612.06490} {arXiv:1612.06490 [hep-ph]} \BibitemShut {NoStop}%
\bibitem [{\citenamefont {Nakamura}\ \emph {et~al.}(2025)\citenamefont {Nakamura}, \citenamefont {Li}, \citenamefont {Peng}, \citenamefont {Sun},\ and\ \citenamefont {Zhou}}]{Nakamura:2023obk}%
  \BibitemOpen
  \bibfield  {author} {\bibinfo {author} {\bibfnamefont {S.~X.}\ \bibnamefont {Nakamura}}, \bibinfo {author} {\bibfnamefont {X.~H.}\ \bibnamefont {Li}}, \bibinfo {author} {\bibfnamefont {H.~P.}\ \bibnamefont {Peng}}, \bibinfo {author} {\bibfnamefont {Z.~T.}\ \bibnamefont {Sun}},\ and\ \bibinfo {author} {\bibfnamefont {X.~R.}\ \bibnamefont {Zhou}},\ }\bibfield  {title} {\bibinfo {title} {{Global coupled-channel analysis of e+e-{\textrightarrow}cc{\textasciimacron} processes in s=3.75{\,}to{\,}4.7{\,}{\,}GeV}},\ }\href {https://doi.org/10.1103/fch8-xwb8} {\bibfield  {journal} {\bibinfo  {journal} {Phys. Rev. D}\ }\textbf {\bibinfo {volume} {112}},\ \bibinfo {pages} {054027} (\bibinfo {year} {2025})},\ \Eprint {https://arxiv.org/abs/2312.17658} {arXiv:2312.17658 [hep-ph]} \BibitemShut {NoStop}%
\bibitem [{\citenamefont {Yu}\ \emph {et~al.}(2024)\citenamefont {Yu}, \citenamefont {Wang}, \citenamefont {Wu},\ and\ \citenamefont {Yang}}]{Yu:2024sqv}%
  \BibitemOpen
  \bibfield  {author} {\bibinfo {author} {\bibfnamefont {K.}~\bibnamefont {Yu}}, \bibinfo {author} {\bibfnamefont {G.-J.}\ \bibnamefont {Wang}}, \bibinfo {author} {\bibfnamefont {J.-J.}\ \bibnamefont {Wu}},\ and\ \bibinfo {author} {\bibfnamefont {Z.}~\bibnamefont {Yang}},\ }\bibfield  {title} {\bibinfo {title} {{Three-coupled-channel analysis of Zc(3900) involving DD{\textasciimacron}*, {\ensuremath{\pi}}J/{\ensuremath{\psi}}, and {\ensuremath{\rho}}{\ensuremath{\eta}}c}},\ }\href {https://doi.org/10.1103/PhysRevD.110.114029} {\bibfield  {journal} {\bibinfo  {journal} {Phys. Rev. D}\ }\textbf {\bibinfo {volume} {110}},\ \bibinfo {pages} {114029} (\bibinfo {year} {2024})},\ \Eprint {https://arxiv.org/abs/2409.10865} {arXiv:2409.10865 [hep-ph]} \BibitemShut {NoStop}%
\bibitem [{\citenamefont {Chen}\ \emph {et~al.}(2024{\natexlab{a}})\citenamefont {Chen}, \citenamefont {Du},\ and\ \citenamefont {Guo}}]{Chen:2023def}%
  \BibitemOpen
  \bibfield  {author} {\bibinfo {author} {\bibfnamefont {Y.-H.}\ \bibnamefont {Chen}}, \bibinfo {author} {\bibfnamefont {M.-L.}\ \bibnamefont {Du}},\ and\ \bibinfo {author} {\bibfnamefont {F.-K.}\ \bibnamefont {Guo}},\ }\bibfield  {title} {\bibinfo {title} {{Precise determination of the pole position of the exotic Z$_{c}$(3900)}},\ }\href {https://doi.org/10.1007/s11433-023-2408-1} {\bibfield  {journal} {\bibinfo  {journal} {Sci. China Phys. Mech. Astron.}\ }\textbf {\bibinfo {volume} {67}},\ \bibinfo {pages} {291011} (\bibinfo {year} {2024}{\natexlab{a}})},\ \Eprint {https://arxiv.org/abs/2310.15965} {arXiv:2310.15965 [hep-ph]} \BibitemShut {NoStop}%
\bibitem [{\citenamefont {Kalantar-Nayestanaki}\ \emph {et~al.}(2012)\citenamefont {Kalantar-Nayestanaki}, \citenamefont {Epelbaum}, \citenamefont {Messchendorp},\ and\ \citenamefont {Nogga}}]{Kalantar-Nayestanaki:2011rzs}%
  \BibitemOpen
  \bibfield  {author} {\bibinfo {author} {\bibfnamefont {N.}~\bibnamefont {Kalantar-Nayestanaki}}, \bibinfo {author} {\bibfnamefont {E.}~\bibnamefont {Epelbaum}}, \bibinfo {author} {\bibfnamefont {J.~G.}\ \bibnamefont {Messchendorp}},\ and\ \bibinfo {author} {\bibfnamefont {A.}~\bibnamefont {Nogga}},\ }\bibfield  {title} {\bibinfo {title} {{Signatures of three-nucleon interactions in few-nucleon systems}},\ }\href {https://doi.org/10.1088/0034-4885/75/1/016301} {\bibfield  {journal} {\bibinfo  {journal} {Rept. Prog. Phys.}\ }\textbf {\bibinfo {volume} {75}},\ \bibinfo {pages} {016301} (\bibinfo {year} {2012})},\ \Eprint {https://arxiv.org/abs/1108.1227} {arXiv:1108.1227 [nucl-th]} \BibitemShut {NoStop}%
\bibitem [{\citenamefont {Epelbaum}\ and\ \citenamefont {Gegelia}(2009)}]{Epelbaum:2009sd}%
  \BibitemOpen
  \bibfield  {author} {\bibinfo {author} {\bibfnamefont {E.}~\bibnamefont {Epelbaum}}\ and\ \bibinfo {author} {\bibfnamefont {J.}~\bibnamefont {Gegelia}},\ }\bibfield  {title} {\bibinfo {title} {{Regularization, renormalization and 'peratization' in effective field theory for two nucleons}},\ }\href {https://doi.org/10.1140/epja/i2009-10833-3} {\bibfield  {journal} {\bibinfo  {journal} {Eur. Phys. J. A}\ }\textbf {\bibinfo {volume} {41}},\ \bibinfo {pages} {341} (\bibinfo {year} {2009})},\ \Eprint {https://arxiv.org/abs/0906.3822} {arXiv:0906.3822 [nucl-th]} \BibitemShut {NoStop}%
\bibitem [{\citenamefont {Machleidt}\ and\ \citenamefont {Entem}(2011)}]{Machleidt:2011zz}%
  \BibitemOpen
  \bibfield  {author} {\bibinfo {author} {\bibfnamefont {R.}~\bibnamefont {Machleidt}}\ and\ \bibinfo {author} {\bibfnamefont {D.~R.}\ \bibnamefont {Entem}},\ }\bibfield  {title} {\bibinfo {title} {{Chiral effective field theory and nuclear forces}},\ }\href {https://doi.org/10.1016/j.physrep.2011.02.001} {\bibfield  {journal} {\bibinfo  {journal} {Phys. Rept.}\ }\textbf {\bibinfo {volume} {503}},\ \bibinfo {pages} {1} (\bibinfo {year} {2011})},\ \Eprint {https://arxiv.org/abs/1105.2919} {arXiv:1105.2919 [nucl-th]} \BibitemShut {NoStop}%
\bibitem [{\citenamefont {van Kolck}(1994)}]{vanKolck:1994yi}%
  \BibitemOpen
  \bibfield  {author} {\bibinfo {author} {\bibfnamefont {U.}~\bibnamefont {van Kolck}},\ }\bibfield  {title} {\bibinfo {title} {{Few nucleon forces from chiral Lagrangians}},\ }\href {https://doi.org/10.1103/PhysRevC.49.2932} {\bibfield  {journal} {\bibinfo  {journal} {Phys. Rev. C}\ }\textbf {\bibinfo {volume} {49}},\ \bibinfo {pages} {2932} (\bibinfo {year} {1994})}\BibitemShut {NoStop}%
\bibitem [{\citenamefont {Hammer}\ \emph {et~al.}(2013)\citenamefont {Hammer}, \citenamefont {Nogga},\ and\ \citenamefont {Schwenk}}]{Hammer:2012id}%
  \BibitemOpen
  \bibfield  {author} {\bibinfo {author} {\bibfnamefont {H.-W.}\ \bibnamefont {Hammer}}, \bibinfo {author} {\bibfnamefont {A.}~\bibnamefont {Nogga}},\ and\ \bibinfo {author} {\bibfnamefont {A.}~\bibnamefont {Schwenk}},\ }\bibfield  {title} {\bibinfo {title} {{Three-body forces: From cold atoms to nuclei}},\ }\href {https://doi.org/10.1103/RevModPhys.85.197} {\bibfield  {journal} {\bibinfo  {journal} {Rev. Mod. Phys.}\ }\textbf {\bibinfo {volume} {85}},\ \bibinfo {pages} {197} (\bibinfo {year} {2013})},\ \Eprint {https://arxiv.org/abs/1210.4273} {arXiv:1210.4273 [nucl-th]} \BibitemShut {NoStop}%
\bibitem [{\citenamefont {Pan}\ \emph {et~al.}(2026)\citenamefont {Pan}, \citenamefont {Liu},\ and\ \citenamefont {Geng}}]{Pan:2025vyg}%
  \BibitemOpen
  \bibfield  {author} {\bibinfo {author} {\bibfnamefont {Y.-W.}\ \bibnamefont {Pan}}, \bibinfo {author} {\bibfnamefont {M.-Z.}\ \bibnamefont {Liu}},\ and\ \bibinfo {author} {\bibfnamefont {L.-S.}\ \bibnamefont {Geng}},\ }\bibfield  {title} {\bibinfo {title} {{Probing the Three-Body Force in Hadronic Systems with Specific Charge Parity}},\ }\href {https://doi.org/10.1103/vl2y-8xdn} {\bibfield  {journal} {\bibinfo  {journal} {Phys. Rev. Lett.}\ }\textbf {\bibinfo {volume} {136}},\ \bibinfo {pages} {151901} (\bibinfo {year} {2026})},\ \Eprint {https://arxiv.org/abs/2512.01468} {arXiv:2512.01468 [nucl-th]} \BibitemShut {NoStop}%
\bibitem [{\citenamefont {Meng}\ \emph {et~al.}(2021)\citenamefont {Meng}, \citenamefont {Hiyama}, \citenamefont {Hosaka}, \citenamefont {Oka}, \citenamefont {Gubler}, \citenamefont {Can}, \citenamefont {Takahashi},\ and\ \citenamefont {Zong}}]{Meng:2020knc}%
  \BibitemOpen
  \bibfield  {author} {\bibinfo {author} {\bibfnamefont {Q.}~\bibnamefont {Meng}}, \bibinfo {author} {\bibfnamefont {E.}~\bibnamefont {Hiyama}}, \bibinfo {author} {\bibfnamefont {A.}~\bibnamefont {Hosaka}}, \bibinfo {author} {\bibfnamefont {M.}~\bibnamefont {Oka}}, \bibinfo {author} {\bibfnamefont {P.}~\bibnamefont {Gubler}}, \bibinfo {author} {\bibfnamefont {K.~U.}\ \bibnamefont {Can}}, \bibinfo {author} {\bibfnamefont {T.~T.}\ \bibnamefont {Takahashi}},\ and\ \bibinfo {author} {\bibfnamefont {H.~S.}\ \bibnamefont {Zong}},\ }\bibfield  {title} {\bibinfo {title} {{Stable double-heavy tetraquarks: spectrum and structure}},\ }\href {https://doi.org/10.1016/j.physletb.2021.136095} {\bibfield  {journal} {\bibinfo  {journal} {Phys. Lett. B}\ }\textbf {\bibinfo {volume} {814}},\ \bibinfo {pages} {136095} (\bibinfo {year} {2021})},\ \Eprint {https://arxiv.org/abs/2009.14493} {arXiv:2009.14493 [nucl-th]} \BibitemShut {NoStop}%
\bibitem [{\citenamefont {Meng}\ \emph {et~al.}(2023{\natexlab{b}})\citenamefont {Meng}, \citenamefont {Chen}, \citenamefont {Ma},\ and\ \citenamefont {Zhu}}]{Meng:2023jqk}%
  \BibitemOpen
  \bibfield  {author} {\bibinfo {author} {\bibfnamefont {L.}~\bibnamefont {Meng}}, \bibinfo {author} {\bibfnamefont {Y.-K.}\ \bibnamefont {Chen}}, \bibinfo {author} {\bibfnamefont {Y.}~\bibnamefont {Ma}},\ and\ \bibinfo {author} {\bibfnamefont {S.-L.}\ \bibnamefont {Zhu}},\ }\bibfield  {title} {\bibinfo {title} {{Tetraquark bound states in constituent quark models: Benchmark test calculations}},\ }\href {https://doi.org/10.1103/PhysRevD.108.114016} {\bibfield  {journal} {\bibinfo  {journal} {Phys. Rev. D}\ }\textbf {\bibinfo {volume} {108}},\ \bibinfo {pages} {114016} (\bibinfo {year} {2023}{\natexlab{b}})},\ \Eprint {https://arxiv.org/abs/2310.13354} {arXiv:2310.13354 [hep-ph]} \BibitemShut {NoStop}%
\bibitem [{\citenamefont {Carbonell}\ \emph {et~al.}(2014)\citenamefont {Carbonell}, \citenamefont {Deltuva}, \citenamefont {Fonseca},\ and\ \citenamefont {Lazauskas}}]{Carbonell:2013ywa}%
  \BibitemOpen
  \bibfield  {author} {\bibinfo {author} {\bibfnamefont {J.}~\bibnamefont {Carbonell}}, \bibinfo {author} {\bibfnamefont {A.}~\bibnamefont {Deltuva}}, \bibinfo {author} {\bibfnamefont {A.~C.}\ \bibnamefont {Fonseca}},\ and\ \bibinfo {author} {\bibfnamefont {R.}~\bibnamefont {Lazauskas}},\ }\bibfield  {title} {\bibinfo {title} {{Bound state techniques to solve the multiparticle scattering problem}},\ }\href {https://doi.org/10.1016/j.ppnp.2013.10.003} {\bibfield  {journal} {\bibinfo  {journal} {Prog. Part. Nucl. Phys.}\ }\textbf {\bibinfo {volume} {74}},\ \bibinfo {pages} {55} (\bibinfo {year} {2014})},\ \Eprint {https://arxiv.org/abs/1310.6631} {arXiv:1310.6631 [nucl-th]} \BibitemShut {NoStop}%
\bibitem [{\citenamefont {Hiyama}\ \emph {et~al.}(2016)\citenamefont {Hiyama}, \citenamefont {Lazauskas}, \citenamefont {Carbonell},\ and\ \citenamefont {Kamimura}}]{Hiyama:2016nwn}%
  \BibitemOpen
  \bibfield  {author} {\bibinfo {author} {\bibfnamefont {E.}~\bibnamefont {Hiyama}}, \bibinfo {author} {\bibfnamefont {R.}~\bibnamefont {Lazauskas}}, \bibinfo {author} {\bibfnamefont {J.}~\bibnamefont {Carbonell}},\ and\ \bibinfo {author} {\bibfnamefont {M.}~\bibnamefont {Kamimura}},\ }\bibfield  {title} {\bibinfo {title} {{Possibility of generating a 4-neutron resonance with a $T=3/2$ isospin 3-neutron force}},\ }\href {https://doi.org/10.1103/PhysRevC.93.044004} {\bibfield  {journal} {\bibinfo  {journal} {Phys. Rev. C}\ }\textbf {\bibinfo {volume} {93}},\ \bibinfo {pages} {044004} (\bibinfo {year} {2016})},\ \Eprint {https://arxiv.org/abs/1604.04363} {arXiv:1604.04363 [nucl-th]} \BibitemShut {NoStop}%
\bibitem [{\citenamefont {Dot{\'e}}\ \emph {et~al.}(2018)\citenamefont {Dot{\'e}}, \citenamefont {Inoue},\ and\ \citenamefont {Myo}}]{Dote:2017wkk}%
  \BibitemOpen
  \bibfield  {author} {\bibinfo {author} {\bibfnamefont {A.}~\bibnamefont {Dot{\'e}}}, \bibinfo {author} {\bibfnamefont {T.}~\bibnamefont {Inoue}},\ and\ \bibinfo {author} {\bibfnamefont {T.}~\bibnamefont {Myo}},\ }\bibfield  {title} {\bibinfo {title} {{Fully coupled-channel study of $K^-pp$ resonance in a chiral SU(3)-based $\bar{K}N$ potential}},\ }\href {https://doi.org/10.1016/j.physletb.2018.08.029} {\bibfield  {journal} {\bibinfo  {journal} {Phys. Lett. B}\ }\textbf {\bibinfo {volume} {784}},\ \bibinfo {pages} {405} (\bibinfo {year} {2018})},\ \Eprint {https://arxiv.org/abs/1710.07589} {arXiv:1710.07589 [nucl-th]} \BibitemShut {NoStop}%
\bibitem [{\citenamefont {Happ}\ \emph {et~al.}(2024)\citenamefont {Happ}, \citenamefont {Naidon},\ and\ \citenamefont {Hiyama}}]{Happ:2023kcc}%
  \BibitemOpen
  \bibfield  {author} {\bibinfo {author} {\bibfnamefont {L.}~\bibnamefont {Happ}}, \bibinfo {author} {\bibfnamefont {P.}~\bibnamefont {Naidon}},\ and\ \bibinfo {author} {\bibfnamefont {E.}~\bibnamefont {Hiyama}},\ }\bibfield  {title} {\bibinfo {title} {{Mass Ratio Dependence of Three-Body Resonance Lifetimes in 1D and 3D}},\ }\href {https://doi.org/10.1007/s00601-024-01900-w} {\bibfield  {journal} {\bibinfo  {journal} {Few Body Syst.}\ }\textbf {\bibinfo {volume} {65}},\ \bibinfo {pages} {38} (\bibinfo {year} {2024})},\ \Eprint {https://arxiv.org/abs/2312.04080} {arXiv:2312.04080 [quant-ph]} \BibitemShut {NoStop}%
\bibitem [{\citenamefont {Chen}\ \emph {et~al.}(2024{\natexlab{b}})\citenamefont {Chen}, \citenamefont {Meng}, \citenamefont {Lin},\ and\ \citenamefont {Zhu}}]{Chen:2023eri}%
  \BibitemOpen
  \bibfield  {author} {\bibinfo {author} {\bibfnamefont {Y.-K.}\ \bibnamefont {Chen}}, \bibinfo {author} {\bibfnamefont {L.}~\bibnamefont {Meng}}, \bibinfo {author} {\bibfnamefont {Z.-Y.}\ \bibnamefont {Lin}},\ and\ \bibinfo {author} {\bibfnamefont {S.-L.}\ \bibnamefont {Zhu}},\ }\bibfield  {title} {\bibinfo {title} {{Virtual states in the coupled-channel problems with an improved complex scaling method}},\ }\href {https://doi.org/10.1103/PhysRevD.109.034006} {\bibfield  {journal} {\bibinfo  {journal} {Phys. Rev. D}\ }\textbf {\bibinfo {volume} {109}},\ \bibinfo {pages} {034006} (\bibinfo {year} {2024}{\natexlab{b}})},\ \Eprint {https://arxiv.org/abs/2308.12424} {arXiv:2308.12424 [hep-ph]} \BibitemShut {NoStop}%
\bibitem [{\citenamefont {Chen}\ \emph {et~al.}(2024{\natexlab{c}})\citenamefont {Chen}, \citenamefont {Wu}, \citenamefont {Meng},\ and\ \citenamefont {Zhu}}]{Chen:2023syh}%
  \BibitemOpen
  \bibfield  {author} {\bibinfo {author} {\bibfnamefont {Y.-K.}\ \bibnamefont {Chen}}, \bibinfo {author} {\bibfnamefont {W.-L.}\ \bibnamefont {Wu}}, \bibinfo {author} {\bibfnamefont {L.}~\bibnamefont {Meng}},\ and\ \bibinfo {author} {\bibfnamefont {S.-L.}\ \bibnamefont {Zhu}},\ }\bibfield  {title} {\bibinfo {title} {{Unified description of the Qsq{\textasciimacron}q{\textasciimacron} molecular bound states, molecular resonances, and compact tetraquark states in the quark potential model}},\ }\href {https://doi.org/10.1103/PhysRevD.109.014010} {\bibfield  {journal} {\bibinfo  {journal} {Phys. Rev. D}\ }\textbf {\bibinfo {volume} {109}},\ \bibinfo {pages} {014010} (\bibinfo {year} {2024}{\natexlab{c}})},\ \Eprint {https://arxiv.org/abs/2310.14597} {arXiv:2310.14597 [hep-ph]} \BibitemShut {NoStop}%
\bibitem [{\citenamefont {Meng}\ \emph {et~al.}(2025)\citenamefont {Meng}, \citenamefont {Wang},\ and\ \citenamefont {Oka}}]{Meng:2024yhu}%
  \BibitemOpen
  \bibfield  {author} {\bibinfo {author} {\bibfnamefont {Q.}~\bibnamefont {Meng}}, \bibinfo {author} {\bibfnamefont {G.-J.}\ \bibnamefont {Wang}},\ and\ \bibinfo {author} {\bibfnamefont {M.}~\bibnamefont {Oka}},\ }\bibfield  {title} {\bibinfo {title} {{Mass spectra of full-heavy and double-heavy tetraquark states in the conventional quark model}},\ }\href {https://doi.org/10.1103/PhysRevD.111.014018} {\bibfield  {journal} {\bibinfo  {journal} {Phys. Rev. D}\ }\textbf {\bibinfo {volume} {111}},\ \bibinfo {pages} {014018} (\bibinfo {year} {2025})},\ \Eprint {https://arxiv.org/abs/2404.01238} {arXiv:2404.01238 [hep-ph]} \BibitemShut {NoStop}%
\bibitem [{\citenamefont {Ma}\ \emph {et~al.}(2024)\citenamefont {Ma}, \citenamefont {Wu}, \citenamefont {Meng}, \citenamefont {Chen},\ and\ \citenamefont {Zhu}}]{Ma:2024vsi}%
  \BibitemOpen
  \bibfield  {author} {\bibinfo {author} {\bibfnamefont {Y.}~\bibnamefont {Ma}}, \bibinfo {author} {\bibfnamefont {W.-L.}\ \bibnamefont {Wu}}, \bibinfo {author} {\bibfnamefont {L.}~\bibnamefont {Meng}}, \bibinfo {author} {\bibfnamefont {Y.-K.}\ \bibnamefont {Chen}},\ and\ \bibinfo {author} {\bibfnamefont {S.-L.}\ \bibnamefont {Zhu}},\ }\bibfield  {title} {\bibinfo {title} {{Fully strange tetraquark resonant states as the cousins of X(6900)}},\ }\href {https://doi.org/10.1103/PhysRevD.110.074026} {\bibfield  {journal} {\bibinfo  {journal} {Phys. Rev. D}\ }\textbf {\bibinfo {volume} {110}},\ \bibinfo {pages} {074026} (\bibinfo {year} {2024})},\ \Eprint {https://arxiv.org/abs/2408.00503} {arXiv:2408.00503 [hep-ph]} \BibitemShut {NoStop}%
\bibitem [{\citenamefont {Wu}\ \emph {et~al.}(2024{\natexlab{a}})\citenamefont {Wu}, \citenamefont {Chen}, \citenamefont {Meng},\ and\ \citenamefont {Zhu}}]{Wu:2024euj}%
  \BibitemOpen
  \bibfield  {author} {\bibinfo {author} {\bibfnamefont {W.-L.}\ \bibnamefont {Wu}}, \bibinfo {author} {\bibfnamefont {Y.-K.}\ \bibnamefont {Chen}}, \bibinfo {author} {\bibfnamefont {L.}~\bibnamefont {Meng}},\ and\ \bibinfo {author} {\bibfnamefont {S.-L.}\ \bibnamefont {Zhu}},\ }\bibfield  {title} {\bibinfo {title} {{Benchmark calculations of fully heavy compact and molecular tetraquark states}},\ }\href {https://doi.org/10.1103/PhysRevD.109.054034} {\bibfield  {journal} {\bibinfo  {journal} {Phys. Rev. D}\ }\textbf {\bibinfo {volume} {109}},\ \bibinfo {pages} {054034} (\bibinfo {year} {2024}{\natexlab{a}})},\ \Eprint {https://arxiv.org/abs/2401.14899} {arXiv:2401.14899 [hep-ph]} \BibitemShut {NoStop}%
\bibitem [{\citenamefont {Wu}\ \emph {et~al.}(2024{\natexlab{b}})\citenamefont {Wu}, \citenamefont {Ma}, \citenamefont {Chen}, \citenamefont {Meng},\ and\ \citenamefont {Zhu}}]{Wu:2024zbx}%
  \BibitemOpen
  \bibfield  {author} {\bibinfo {author} {\bibfnamefont {W.-L.}\ \bibnamefont {Wu}}, \bibinfo {author} {\bibfnamefont {Y.}~\bibnamefont {Ma}}, \bibinfo {author} {\bibfnamefont {Y.-K.}\ \bibnamefont {Chen}}, \bibinfo {author} {\bibfnamefont {L.}~\bibnamefont {Meng}},\ and\ \bibinfo {author} {\bibfnamefont {S.-L.}\ \bibnamefont {Zhu}},\ }\bibfield  {title} {\bibinfo {title} {{Doubly heavy tetraquark bound and resonant states}},\ }\href {https://doi.org/10.1103/PhysRevD.110.094041} {\bibfield  {journal} {\bibinfo  {journal} {Phys. Rev. D}\ }\textbf {\bibinfo {volume} {110}},\ \bibinfo {pages} {094041} (\bibinfo {year} {2024}{\natexlab{b}})},\ \Eprint {https://arxiv.org/abs/2409.03373} {arXiv:2409.03373 [hep-ph]} \BibitemShut {NoStop}%
\end{thebibliography}%
\end{document}